\documentclass{aa}  

\usepackage{graphicx}
\usepackage{xcolor}

\usepackage{txfonts}
\usepackage{booktabs}
\usepackage{multirow}

\usepackage{makecell}

\usepackage{nicefrac}

\usepackage[colorlinks=true,linkcolor=blue,allcolors=blue]{hyperref}

\definecolor{mycolor}{rgb}{0.05, 0.65, 0.2}

\newcommand{\Halpha}{{\text{H}\ensuremath{\alpha}}}
\newcommand{\Hbeta}{{\text{H}\ensuremath{\beta}}}
\newcommand{\oiii}{{\text{[\ion{O}{iii}]}}}

\usepackage{comment}

\begin{document}

   \title{The Engine and its Flows: Little Red Dot spectra are shaped by the column densities of their gas envelopes
}
    \titlerunning{Balmer Line Profiles of LRDs}

\authorrunning{Matthee et al.}

\newcommand{\orcid}[1]{} 

   \author{
Jorryt~Matthee\orcid{0000-0003-2871-127X}\inst{\ref{inst:ista}}\thanks{jorryt.matthee@ista.ac.at}
\and Alberto~Torralba\orcid{0000-0001-5586-6950}\inst{\ref{inst:ista}}
\and Gabriele Pezzulli\orcid{0000-0003-0736-7879}\inst{\ref{inst:kapteyn}}
\and Rohan P. Naidu\orcid{0000-0003-3997-5705}\inst{\ref{inst:MIT_kavli}}\thanks{Pappalardo Fellow}
\and John Chisholm  \orcid{0000-0002-0302-2577}\inst{\ref{inst:texas}}
\and Sara Mascia\orcid{0000-0002-9572-7813}\inst{\ref{inst:ista}}
\and Jenny E. Greene\orcid{0000-0002-5612-3427} \inst{\ref{inst:princeton}}
\and Yuzo Ishikawa\orcid{0000-0001-7572-5231} \inst{\ref{inst:MIT_kavli}}
\and Max Gronke\orcid{0000-0003-2491-060X} \inst{\ref{inst:ari}}
\and Stijn Wuyts\orcid{0000-0003-3735-1931}\inst{\ref{inst:bath}} 
\and Rongmon Bordoloi\orcid{0000-0002-3120-7173} \inst{\ref{inst:northcarolina}}
\and Gabriel Brammer\orcid{0000-0003-2680-005X} \inst{\ref{inst:dawn}}
\and Seok-Jun Chang\orcid{0000-0002-0112-5900}\inst{\ref{inst:mpa}}
\and Anna-Christina Eilers\orcid{0000-0003-2895-6218} \inst{\ref{inst:MIT_kavli}}
\and Anna de Graaff\orcid{0000-0002-2380-9801}\inst{\ref{inst:MPIA},\ref{inst:harvard}}\thanks{Clay Fellow} 
\and Raphael E. Hviding\orcid{0000-0002-4684-9005} \inst{\ref{inst:MPIA}} 
\and Edoardo Iani\orcid{0000-0001-8386-3546} \inst{\ref{inst:ista}}
\and Garth Illingworth\orcid{0000-0002-8096-2837}\inst{\ref{inst:santacruz}}
\and Daichi Kashino \orcid{0000-0001-9044-1747}\inst{\ref{inst:nagoya}}
\and Ivo Labbe\orcid{0000-0002-2057-5376}\inst{\ref{inst:swinburne}}
\and Yilun Ma\orcid{0000-0002-0463-9528}\inst{\ref{inst:princeton}}
\and Michael V. Maseda\orcid{0000-0003-0695-4414}\inst{\ref{inst:madison}}
\and Romain Meyer\orcid{0000-0001-5492-4522}\inst{\ref{inst:geneva}}
\and Erica Nelson\orcid{0000-0002-7524-374X}\inst{\ref{inst:boulder}}
\and Pascal Oesch\orcid{0000-0001-5851-6649} \inst{\ref{inst:geneva}}
\and Mengyuan Xiao\orcid{0000-0003-1207-5344}\inst{\ref{inst:geneva}}
}

   % }

   \institute{
    Institute of Science and Technology Austria (ISTA), Am Campus 1, 3400 Klosterneuburg, Austria.\label{inst:ista}
    \and Kapteyn Astronomical Institute, University of Groningen, Landleven 12, NL-9747 AD Groningen, the Netherlands.\label{inst:kapteyn}
    \and MIT Kavli Institute for Astrophysics and Space Research, Massachusetts Institute of Technology, Cambridge, MA 02139, USA.\label{inst:MIT_kavli}
    \and Department of Astronomy, The University of Texas at Austin, Austin, TX 78712, USA. \label{inst:texas}
    \and Department of Astrophysical Sciences, Princeton University, Princeton, NJ 08544, USA.\label{inst:princeton}
    \and Astronomisches Rechen-Institut, Zentrum f\"{u}r Astronomie, Universit\"{a}t Heidelberg, Mönchhofstra$\beta$e 12-14, 69120 Heidelberg, Germany \label{inst:ari} 
    \and Department of Physics, University of Bath, Claverton Down, Bath, BA2 7AY, UK. \label{inst:bath}
    \and Department of Physics and Astronomy, North Carolina State University, Raleigh, 27695, North Carolina, USA. \label{inst:northcarolina}
    \and Cosmic Dawn Center (DAWN), Niels Bohr Institute, University of Copenhagen, Jagtvej 128, K\o benhavn N, DK-2200, Denmark. \label{inst:dawn}
    \and Max-Planck-Institut f\"{u}r Astrophysik, Karl-Schwarzschild-Stra$\beta$e 1, 85748 Garching b. M\"{u}nchen, Germany. \label{inst:mpa}   
    \and Max-Planck-Institut f{\"u}r Astronomie, K{\"o}nigstuhl 17, D-69117 Heidelberg, Germany. \label{inst:MPIA}    
    \and Center for Astrophysics $|$ Harvard \& Smithsonian, 60 Garden St., Cambridge MA 02138 USA. \label{inst:harvard}
    \and Department of Astronomy and Astrophysics, University of California, Santa Cruz, CA 95064, USA. \label{inst:santacruz}  
    \and Centre for Astrophysics and Supercomputing, Swinburne University of Technology, Melbourne, VIC 3122, Australia. \label{inst:swinburne}
        \and Department of Astronomy, University of Wisconsin-Madison, 475 N. Charter St., Madison, WI 53706 USA. \label{inst:madison}
    \and Institute for Advanced Research, Nagoya University, Nagoya 464-8601, Japan. \label{inst:nagoya}     
    \and Department of Astronomy, University of Geneva, Chemin Pegasi 51, 1290 Versoix, Switzerland. \label{inst:geneva}
    \and Department for Astrophysical and Planetary Science, University of Colorado, Boulder, CO 80309, USA. \label{inst:boulder}    
    }

 %  \date{Accepted XXX. Received YYY; in original form ZZZ}

  \abstract{
JWST data have enabled the abundant identification of compact broad Balmer line sources nicknamed the Little Red Dots. While they share broad lines with active galactic nuclei, they are unusually X-ray and infrared weak. We investigate the origin of the Balmer line profiles based on an empirical analysis of 18 broad H$\alpha$-selected sources with high quality spectra at $z\approx3-7$. The H$\alpha$ line profiles vary systematically with Balmer break strength: sources with blue UV to optical colors show a narrow core profile, redder sources with Balmer breaks a blue shifted absorption (P~Cygni shape), and the reddest sources display absorption-dominated cores. All H$\alpha$ lines have symmetric exponential wings, which are more dominant and slightly broader in red sources. Balmer absorption is present in $\sim60$ \% of the sample, with H$\beta$ showing relatively stronger absorption. Drawing upon empirical analogies with stellar phenomena, we interpret these trends as being due to radiative processes that depend on variations in the optical depth, ionisation state and column density of a clumpy, partially ionised envelope. We unveil a correlation between the absorber velocity and Balmer break strength, with the densest absorbers inflowing and bluer sources having faster outflows. This indicates viewing angle or evolutionary effects where optically thick gas is inflowing, as suggested in models of super-Eddington accretion, and the engine can more easily drive outflows in directions with lower column densities. This new understanding of Balmer line profiles as tracing gas properties rather than dynamical broadening helps resolve tensions associated with high inferred black hole masses from standard virial calibrations, and reveals the complex gas environment around the hot central engine.
}

   \maketitle

%%%%%%%%%%%%%%%%%%%%%%%%%%%%%%%%%%%%%%%%%%%%%%%%%%%%%%%%%%%%%%%%%%%%
\section{Introduction}\label{sec:introduction}
Thanks to its sensitive imaging and spectroscopic capabilities in the (near-)infrared, {\it JWST} has enabled the identification of large numbers of broad Balmer line emitters at high-redshift ($z\sim3-9$; e.g. \citealt{Ubler23,Harikane2023,Matthee24,Lin24,Greene24,Kocevski24,Hviding25,Taylor25z9}). Given their broad line widths and luminosities, these sources are typically interpreted as being powered by active galactic nuclei (AGNs). Compared to known AGNs and quasars, they extend the parameter space of AGNs to several magnitudes fainter luminosities and higher redshifts \citep[e.g.][]{Scholtz23}. Among these are compact sources with red UV to optical colors (so-called V-shaped spectra; e.g. \citealt{kokorev2024a}) and compact morphology  -- the `little red dots' (LRDs) -- whose nature has been the subject of significant debate \citep[see][for a recent review]{InayoshiHo25}.

Generally, JWST-identified broad-line emitters tend to be misfits to the AGN population, with unusually faint X-Ray \citep{Ananna24,Yue24} and faint IR luminosities \citep{Setton2025,Xiao25} given their broad-line luminosity and regardless of their UV to optical colors \citep{Brazzini26}. They generally also lack broad components in the UV lines or the strong, broad permitted Fe{\sc ii} features usually seen in previously identified AGNs. The LRDs are well known to show strong Balmer breaks that are highly unusual for AGNs \citep{Wang24_RUBIES,Setton24b,Naidu25,deGraaff25} and lack UV variability \citep{Furtak25,Kokubo25,Tee25}. When applying standard calibrations between the H$\alpha$ luminosities, line-widths and the black hole (BH) masses, the BHs in these AGNs appear to be highly overly massive compared to their stellar masses \citep[e.g.][]{Pacucci23,maiolino2024a,Li24,Matthee25clustering}. Combined with the number densities of $\approx10^{-5}$ cMpc$^{-3}$, these BH masses also give rise to tensions with the evolution of the BH mass function \citep{Luberto25,Roberts26}, as little evolution would be permitted after $z\approx5$ to match continuity constraints from measurements in the later Universe. Further, given the X-Ray and IR weakness, bolometric luminosities may be much lower \citep{Greene26}, implying that the majority of sources would be sub-Eddington in case virial BH mass indicators would apply \citep[e.g.][]{Juodzbalis25Direct}. This is at odds with expectations from selection effects \citep{Lauer07,Schulze14,Li24} and from the shape of luminosity functions \citep{YMa25b} that promote detection of super-Eddington sources.

High columns of dense gas covering the central engine have emerged as a key feature explaining the peculiarities of LRDs. Signatures of dense gas were first identified as narrow absorption in the Balmer series and in HeI \citep[e.g.][]{Matthee24,Juodzbalis24b,Kocevski24,NaiduALT24,BWang24a,Deugenio25}. Balmer absorption lines are rarely detected in AGN spectra (with reported fractions being $\lesssim0.1$ \%; \citealt{Aoki06,Izotov08,Schulze18}) as it requires high column densities of H{\sc i} gas with a populated $n=2$ level, even though the associated densities are fairly common for densities in broad line regions \citep[e.g.][]{Peterson06}. Observationally, such narrow absorption lines are difficult to detect due to the need for high sensitivity and resolution and initial estimates of their occurrence rate are $\sim10-20$ \% \citep{Matthee24,Lin24} based on NIRCam grism observations that yield complete luminosity-limited samples to estimate such statistics.

The dense gas surrounding the engines of the LRDs also plays a critical role in the interpretation of the broad-band spectrum of LRDs. As pointed out by \cite{Inayoshi24}, the gas that causes Balmer absorption should also cause a Balmer break in the spectrum. Indeed, the unusually strong and smooth Balmer breaks observed in the spectra of LRDs can be matched with photoionization models where an AGN spectrum is surrounded by a slab of dense gas, provided the gas has very high (column) densities and turbulence \citep{Ji25,Naidu25,deGraaff25}. The dense gas may be Compton thick to any emitted X-rays \citep{Kocevski24,Maiolino24}, it causes resonant scattering effects that alter line profiles and Balmer decrements \citep{Chang26} and it reddens the spectrum without the need for significant dust attenuation \citep{deGraaff25}, generally yielding lower bolometric luminosities \citep{Greene26}. Moreover, along with partially excited H{\sc i} gas there is likely a high column of free electrons \citep{Torralba25b,Sneppen26}, which broadens emission-lines through Thomson scattering \citep{Rusakov25,Chang26}, explaining the exponential shapes of the broad wings without the presence of overly massive black holes \citep[cf.][]{Brazzini25}.

Various theoretical explanations for the nature of LRDs have been explored, ranging from late-stage quasi-stars \citep{Begelman25}, black hole envelopes \citep{Kido25}, optically thick accretion flows \citep{Liu25,Chen26} to binary black holes \citep{InayoshiBinary} and supermassive stars \citep{Zwick25,Nandal26,Chisholm26}. Yet, many questions remain. What is the engine powering LRDs? What is the geometry and origin of the dense gas? Where does the continuum and line emission originate? What causes variation across samples of broad line emitters? How do we estimate their (black hole) masses? 

Emission-line spectra that carry imprints from dense (hydrogen) gas are routinely found in stellar environments, most relevantly in spectra of Type IIn supernovae \citep[e.g.][]{Xu92,Dessart09} and outbursts from luminous blue variables  \citep[LBVs, e.g.][]{Humphreys94,Mehner13}. These generally show Balmer line profiles with broad wings and narrow absorption features akin to LRD spectra that are attributed to properties of their (optically thick) winds. While LRDs are not transients \citep{Kokubo25} and they are much more luminous than LBVs, the similarity of (parts of) LRD spectra to stellar atmospheres has been noted in the literature \citep[e.g.][]{Naidu25,deGraaff25b,Barro25,Zhang25var,Liu26}. In this work, we draw guidance from these empirical similarities in terms of how we analyse line-profiles and interpret the trends identified.

In this paper, we investigate the physical conditions in LRDs by focusing on their Balmer line profiles. As detailed in Section $\ref{sec:data}$, we compile broad H$\alpha$-selected sources at $z=3-7$ with high quality spectra. In Section $\ref{sec:measurements}$, we show that the shapes of the Balmer line profiles correlate with the broad-band spectrum, with blue sources having narrow Balmer line-cores on top of exponential wings, redder sources showing central P~Cygni features and the reddest sources show central Balmer absorption. We also present our modeling of the line-profiles and the correlations among the line-shapes and global properties of the LRD spectra. We interpret the results in Section $\ref{sec:interpretation}$, where we argue the spectral variation can be understood in the context of dense envelopes with a varying optical depth. In Section $\ref{sec:discussion}$, we discuss the implications that our results have on our understanding of the powering mechanism and origin of LRDs. The paper is summarized in Section $\ref{sec:summary}$. Throughout this work we use a $\Lambda$CDM cosmology as described by \citet{Planck18}, with $\Omega_\Lambda=0.69$, $\Omega_\text{M}=0.31$, and $H_0=67.7$ km\,s$^{-1}$\,Mpc$^{-1}$.

%%%%%%%%%%%%%%%%%%%%%%%%%%%%%%%%%%%%%%%%%%%%%%%%%%%%%%%%%%%%%%%%%%%%

\begin{table*}
\centering
\caption{The broad H$\alpha$ line emitter sample used in this work. Coordinates are in the J2000 system. Systemic redshifts are obtained from the [O{\sc iii}]$_{4960,5008}$ doublet. PID refers to the JWST program ID as part of which the grating data were taken. We note "{\it the Cliff}" is the commonly used nickname for RUBIES-UDS-154183. } \label{tab:sample}
\begin{tabular}{lcccccc}
ID & R.A. & Dec. & {$z_{\rm [OIII]}$}  & Grating setup & PID \\ \hline
  FRESCO-GN-9771 & 189.2810 & 62.2473 & 5.535  & IFU G395H & 5664  \\
  FRESCO-GN-12839 & 189.3443 & 62.2634 & 5.241 & IFU G395H & 5664\\
  FRESCO-GN-15498 & 189.2855 & 62.2807  & 5.085 & IFU G395H & 5664 \\
  FRESCO-GN-16813 & 189.1793 & 62.2925 & 5.358 & IFU G395H & 5664 \\
  FRESCO-GS-13971 &  53.1386 & 27.7903 & 5.482 & IFU G395H & 5664 \\
  JADES-GN-68797 & 189.2291 & 62.1462 & 5.039  & MSA G395M & 1181 \\
  JADES-GN-38147 & 189.2707 & 62.1484 & 5.867  & MSA G395M & 1181 \\
  JADES-GN-73488 & 189.1974& 62.1772 & 4.133  & MSA G235M & 1181 \\
  RUBIES-EGS-42046 & 214.7954 & 52.7888 & 5.276  & MSA G235M & 4233 \\
  RUBIES-EGS-50052 & 214.8235 & 52.8303 & 5.239 & MSA G235M & 4233 \\
  RUBIES-EGS-55604 & 214.9830 & 52.9560 & 6.983  & MSA G235M & 4233 \\
  RUBIES-EGS-49140 & 214.8922 & 52.8774 & 6.685  & MSA G235M & 4233 \\
  RUBIES-UDS-47509 & 34.2646 & -5.2326  & 5.672  & MSA G235M & 4233 \\
  RUBIES-UDS-182791 & 34.2138 & -5.0870 & 4.715  & MSA G235M & 4233\\
 {\it the Cliff} & 34.4107 & -5.1297  & 3.548  & IFU G235H & 9433 \\
 UNCOVER-A2744-45924 & 3.5848 &-30.3436 & 4.464  & MSA G395M & 8204 \\
 GS-3073 & 53.0789 & -27.8842 & 5.553  & IFU G395H & 1216 \\
 ALT-69688 & 3.5694 & -30.3482 & 4.307  & MSA G235M & 8204 \\
\hline\end{tabular}
\end{table*}

\section{Data} \label{sec:data}
\subsection{Sample}
Our goal is to investigate the variations among Balmer line profiles of broad-line emitters and their correlations with the UV to optical spectral energy distribution (SED). Therefore, we focus on a sample at $z\sim3-7$, where JWST NIRSpec PRISM spectroscopy enables us to cover the full rest-frame UV to optical range albeit at limited spectral resolution, and we select sources with deep NIRSpec grating spectra covering both the H$\alpha$ and H$\beta$ lines as well as the [O{\sc iii}] doublet with spectral resolution $R\sim1000-2700$. The selection function of this sample is not uniform and primarily limited by the available data (from a combination of new and archival data) that yielded a sample with H$\alpha$ S/N $>20$ and with broad H$\alpha$ emission with significant line-flux extending beyond $\pm1000$ km s$^{-1}$ from line-center. Our sample constitutes 18 sources spanning $z=3.55-6.98$ ($\langle z\rangle= 5.23$ on average). Table $\ref{tab:sample}$ lists the IDs and redshifts of our sample. The FRESCO, UNCOVER and ALT sources were initially selected by their broad H$\alpha$ emission \citep{Matthee24,Matthee25clustering}. The RUBIES sources were initially selected based on their red NIRCam colors \citep{deGraaff24_survey}, JADES sources were selected based on their redshift (i.e. without prior compactness or redness criterion; \citealt{Eisenstein26}), whereas GS-3073 was initially selected as a known AGN candidate with strong nitrogen emission \citep{Vanzella10c}. While these selections appear very different, we note that they yield strongly overlapping samples \citep{Hviding25}. All sources share characteristics as lack of UV variability, X-Ray faintness and lack of broad UV or permitted Fe{\sc ii} emission lines common for quasars and Type II AGN. Seven sources were observed with the NIRSpec IFU and 11 with the micro-shutter array (MSA). Various sources were observed through multiple programs and with multiple modes and we select the highest resolution data wherever available.

To assess how our sample compares to representative samples of LRDs, we show in Figure $\ref{fig:sample}$ the distribution of H$\alpha$ luminosities and Balmer break strengths of our sample compared to the recently published compilation of LRDs from \cite{deGraaff25b}. Generally, most of the luminous broad H$\alpha$ emission-line sources have relatively strong Balmer breaks, with GS-3073 being the luminous blue exception \citep{Ubler23,Brazzini26}. The Balmer break strengths show larger variation among the fainter end of the sample. Our sample is somewhat skewed towards the most luminous H$\alpha$ lines, but mostly captures the distribution in Balmer break strengths. Among the fainter sources, our sample is slightly skewed towards bluer sources. Over the range of H$\alpha$ luminosities probed by our sample, the corresponding number densities are $\approx10^{-5}$ cMpc$^{-3}$ at $z\sim5$, increasing slightly for fainter sources \citep{Matthee24,Lin24,Zhang25}.

\subsection{NIRSpec IFU sample}
We use IFU PRISM and G395H data for five sources initially identified in the FRESCO NIRCam grism survey \citep{Matthee24}. These were observed through program 5664 (PI: Matthee) with total exposure times 6.5 ks (PRISM) and 18.2 ks (G395H) as detailed in \cite{Torralba25b}. Among the parent sample, these targets were selected to span the range in observed H$\alpha$ profiles and in broad-band colors. We complement these data with similar PRISM+G395H IFU observations of GS-3073 (program 1216, PI: L\"utzgendorf) that were analysed in \cite{Ubler23,Ji24,Brazzini26}. This is a nitrogen-selected \citep{Vanzella10c} broad-line AGN whose spectrum resembles the bluer end of the FRESCO-selected sources. We also include recent IFU G235H observations (program 9433, PI: Maiolino) of RUBIES-UDS-154183, better known as "{\it the Cliff}" \citep{deGraaff25} at $z=3.55$ as this supersedes the existing RUBIES G395M spectrum. We use a MSA PRISM spectrum from the RUBIES program for this object. Similar to \cite{Torralba25b}, NIRSpec IFU data were reduced with MSAexp \citep{Brammer23}. In this paper, we analyse 1D spectra that are optimally extracted based on a (wavelength-dependent) point-source aperture, centered at the peak position of the broad H$\alpha$ emission line. Apart from GS-13971, the H$\alpha$ maps of these sources are consistent with point sources (Ishikawa et al. in prep.). 

\subsection{NIRSpec MSA sample}
The IFU sample is complemented with sources that have deep PRISM and NIRspec G235M or G395M grating spectra obtained with the MSA. In addition to {\it the Cliff}, this includes 6 objects observed by the RUBIES program (program 4233, PIs: De Graaff \& Brammer; \citealt{deGraaff24_survey}), three sources observed through the JADES survey (program 1181, PI: Eisenstein; \citealt{Eisenstein26}) and two sources from a recent deep program behind the Abell 2744 cluster (program 8204, PIs: Greene \& Labbe). This sample includes well-studied, red sources such as RUBIES-EGS-49140 and UNCOVER-A2744-45924 \citep[see e.g.][]{Wang24_RUBIES,Labbe2024,Deugenio2025-irony,Lambrides25,Torralba25b}, but also blue broad-line emitters such as ALT-69688 \citep{Matthee25clustering} and RUBIES-EGS-50052 (CEERS-2782; \citealt{Kocevski23}). NIRSpec MSA data were homogeneously reduced using MSAexp and acquired from the DAWN JWST Archive \citep{deGraaff24_survey,Heintz24}\footnote{\url{https://dawn-cph.github.io/dja/}}.

\begin{figure}
    \centering
    \includegraphics[width=\linewidth]{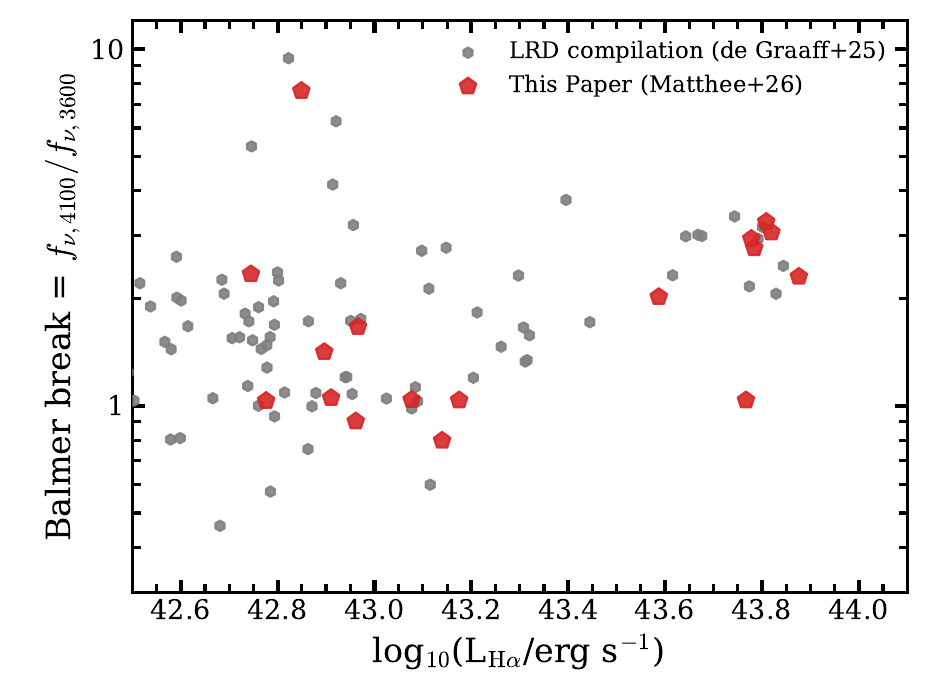}
    \caption{{\bf Demographics and comparison of our sample to the literature.} The H$\alpha$ luminosity and Balmer break strength of the sample included in this work (red), compared to a recent LRD compilation (\citealt{deGraaff25b}) in grey. Our sample is skewed towards higher H$\alpha$ luminosity as these have preferentially been targeted by deep grating follow up. No strong biases are present in terms of the Balmer break strengths.}
    \label{fig:sample}
\end{figure}

\section{Balmer line profiles and SED shapes} \label{sec:measurements} %may needs better name

\subsection{The main empirical trends} \label{sec:maintrends}
\begin{figure*}
    \centering
    \begin{tabular}{ccc}
\hspace{-0.4cm} \includegraphics[height=4.2cm]{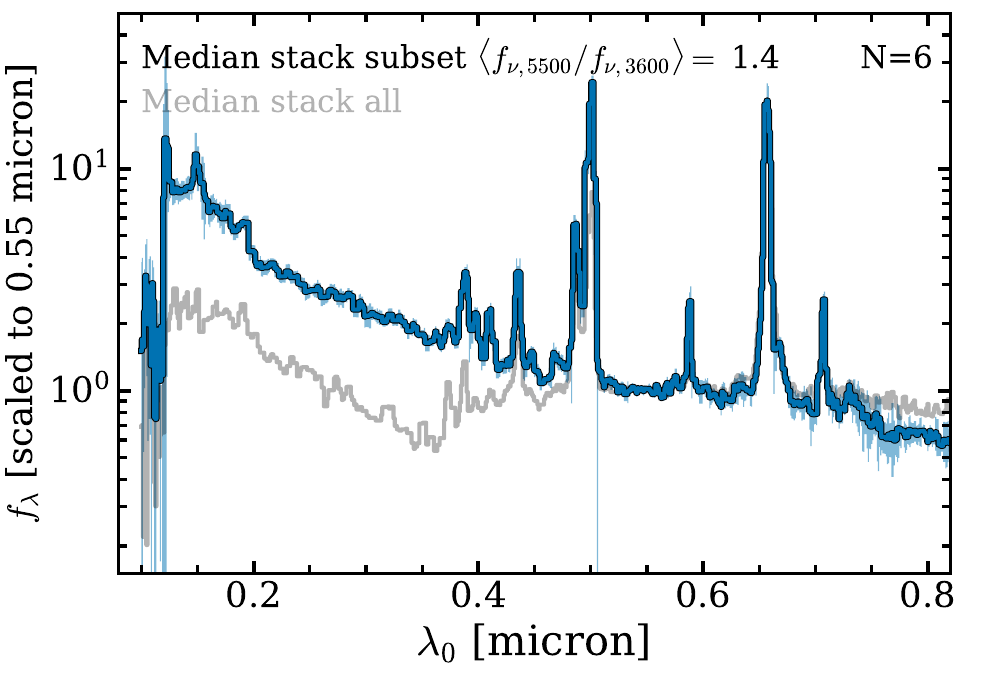} &   
  \hspace{-0.5cm}  \includegraphics[height=4.2cm]{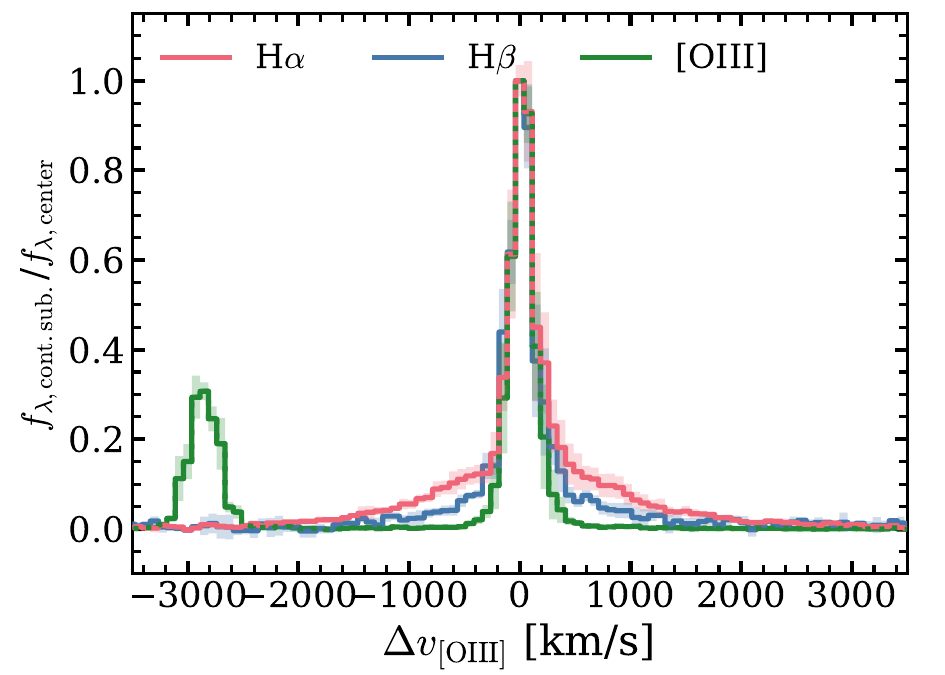} &
   \hspace{-0.5cm}  \includegraphics[height=4.2cm]{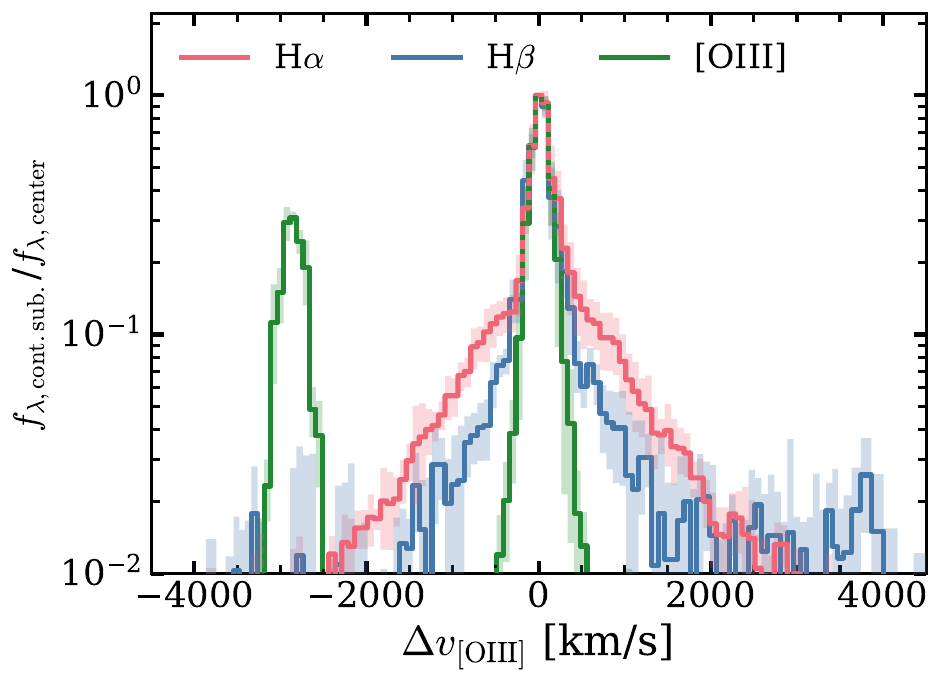} \\

   \hspace{-0.4cm} \includegraphics[height=4.2cm]{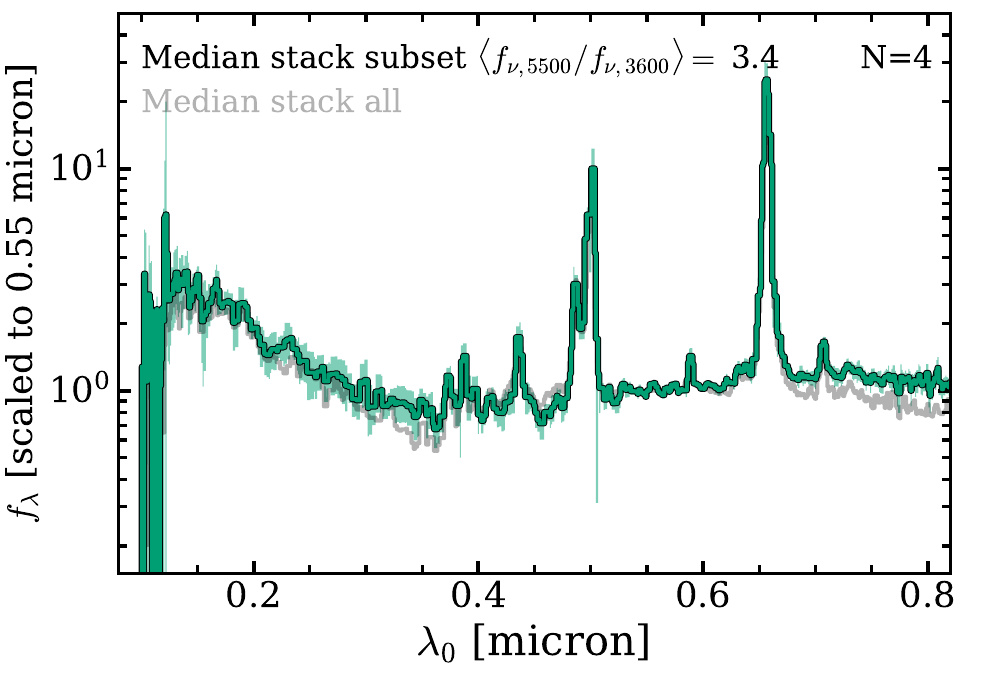} &   
  \hspace{-0.5cm}  \includegraphics[height=4.2cm]{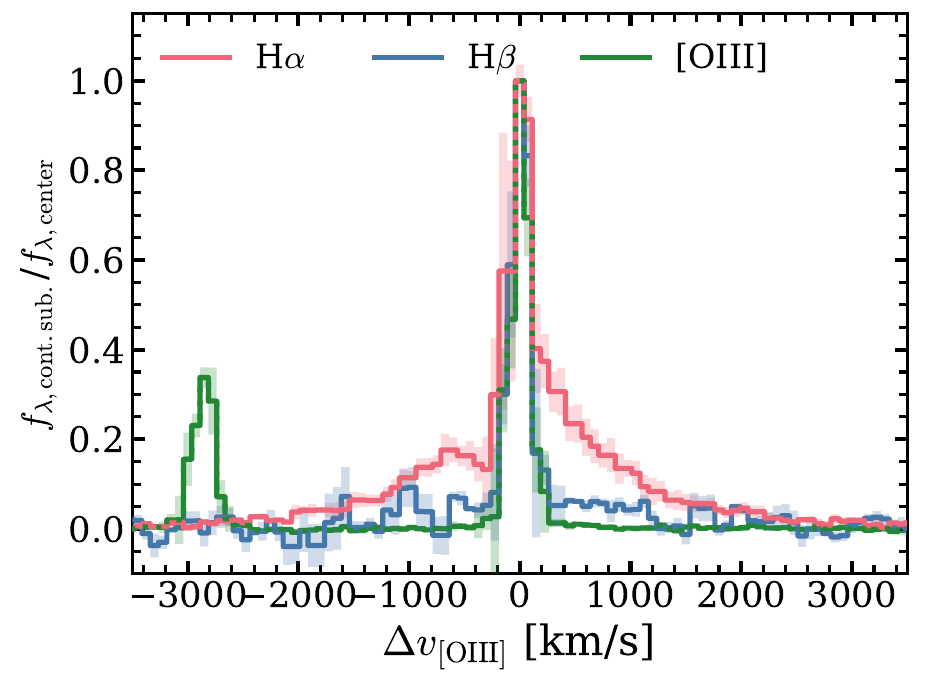} &
   \hspace{-0.5cm}  \includegraphics[height=4.2cm]{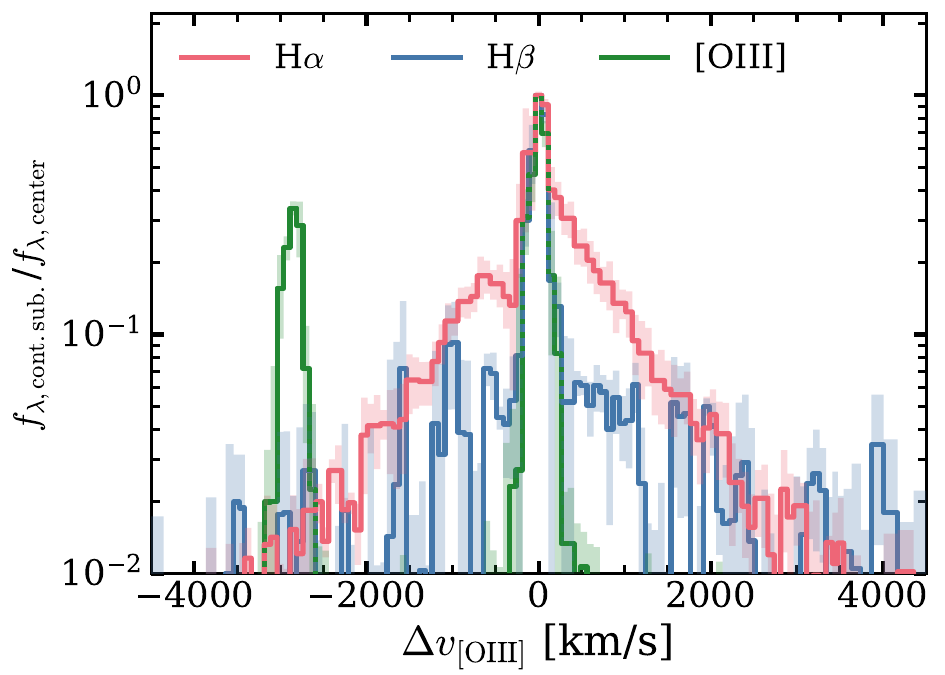} \\

   \hspace{-0.4cm} \includegraphics[height=4.2cm]{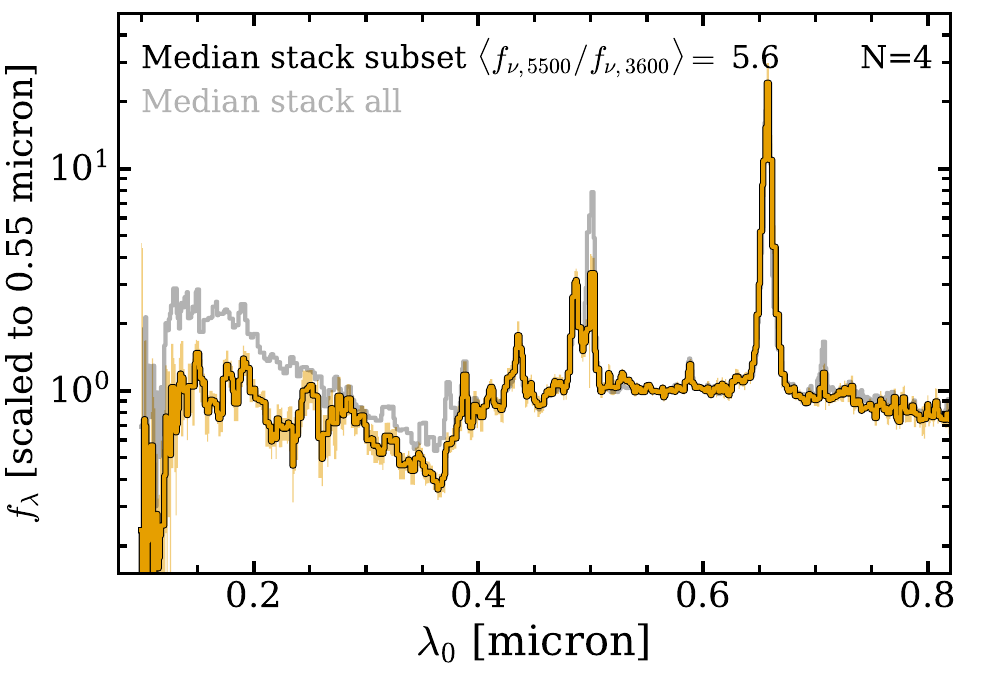} &   
  \hspace{-0.5cm}  \includegraphics[height=4.2cm]{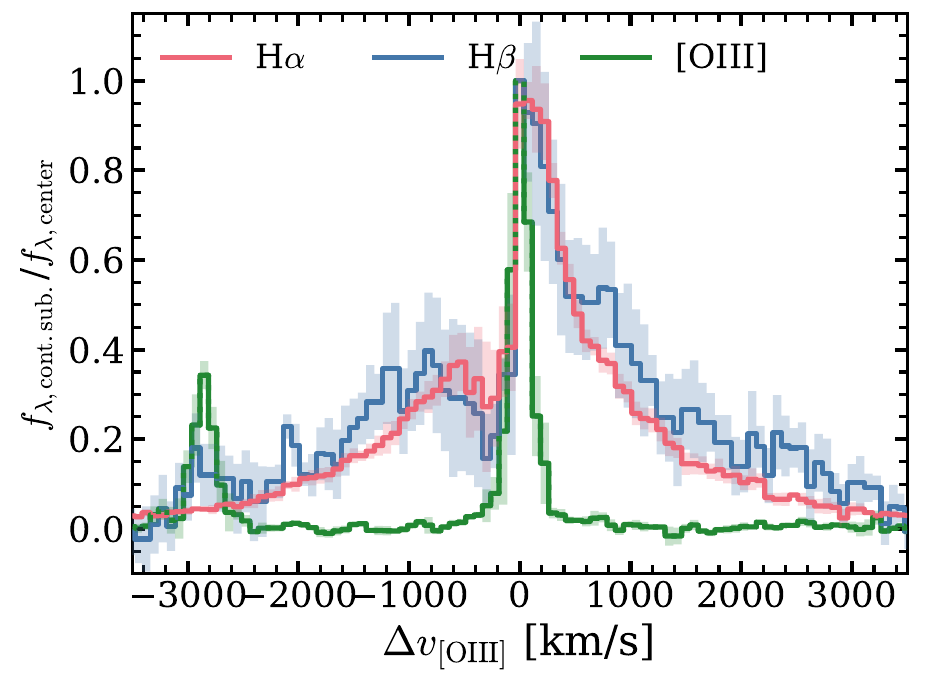} &
   \hspace{-0.5cm}  \includegraphics[height=4.2cm]{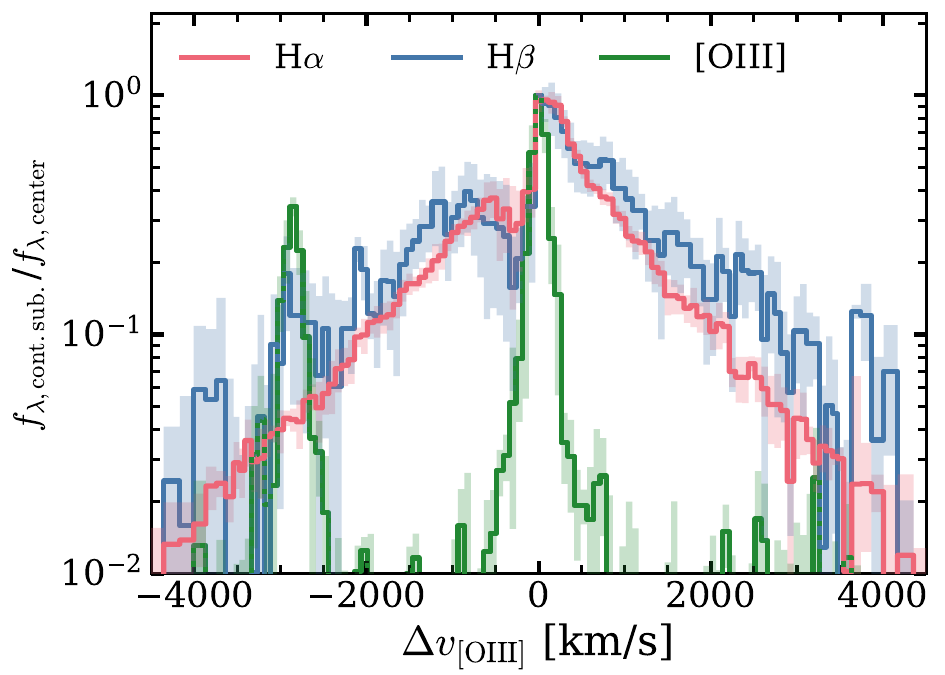} \\

   \hspace{-0.4cm} \includegraphics[height=4.2cm]{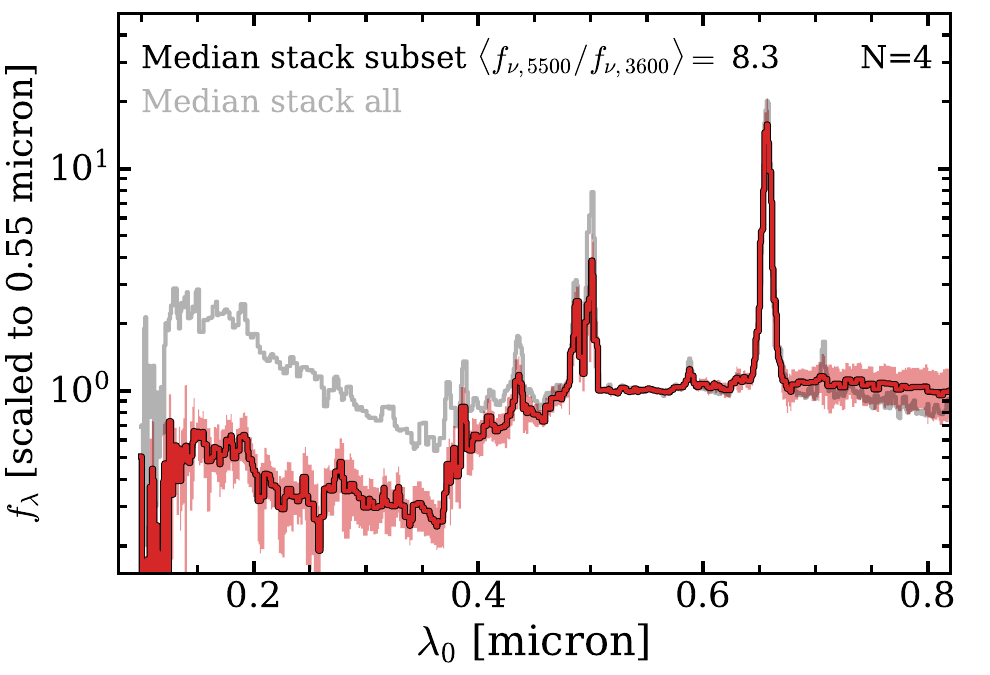} &   
  \hspace{-0.5cm}  \includegraphics[height=4.2cm]{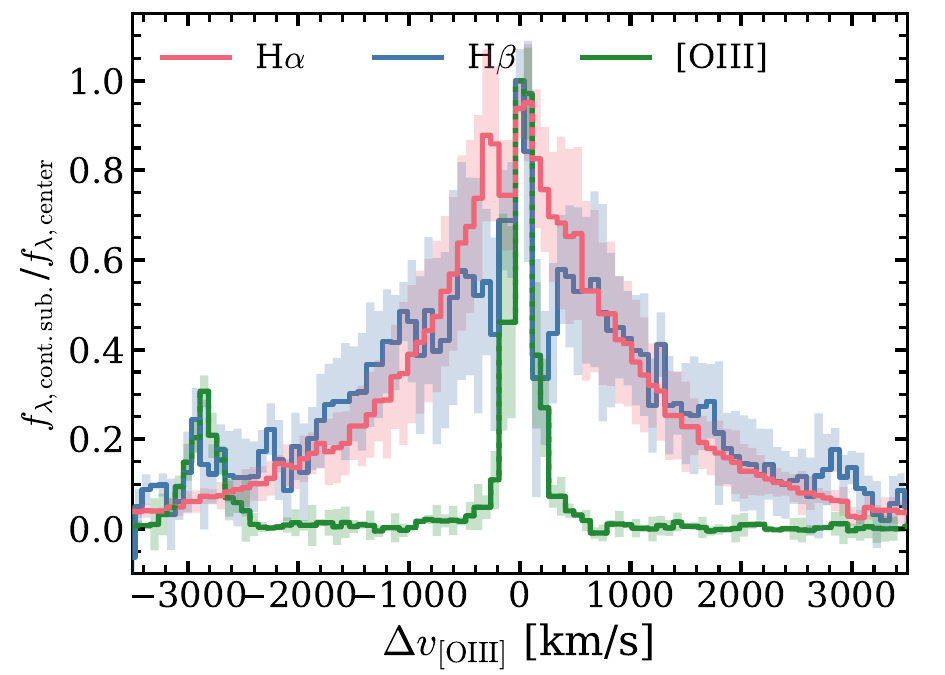} &
   \hspace{-0.5cm}  \includegraphics[height=4.2cm]{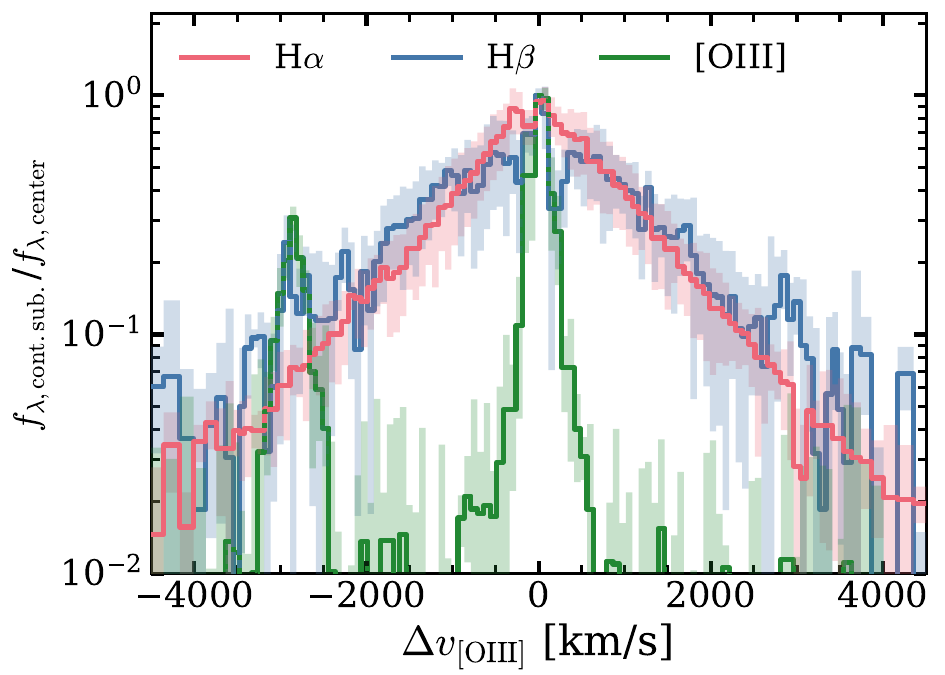} \\
    \end{tabular}
    \caption{{\bf The Balmer line profiles are correlated with the spectral shape of the broad-band spectrum.} Each row shows the median stacked spectrum of a sub-set split by their UV to optical continuum color, increasingly red from top to bottom. Shaded regions illustrate the variation within each stack based on bootstrap resamples. In the left column, we show the stacked broad-band spectrum of the bin (normalised to the continuum level at 5500 {\AA} rest-frame) in color, while the median stack of our full sample is shown in grey. The middle and right (log scale) panels show the H$\alpha$, H$\beta$ and [O{\sc iii}] profiles, in pink, blue and green, respectively. The profiles are normalised to the peak flux at the systemic velocity. The linear scale in the middle column highlights variations in the line-center, while the logarithmic scale in the right column highlights variations in the broad wings. The Balmer lines of redder sources are increasingly more dominated by broad, exponential emission. The central parts of the line-profile changes from narrow and compact to blue-shifted absorption with red-shifted emission (like P~Cygni), to broad absorption with narrow central emission, from blue to red sources.} 
    \label{fig:prism_sample}
\end{figure*}

Here we summarize the main empirical findings that underpin the analysis in this paper. We quantify the broadband spectral shape based on the continuum flux density ratio $f_{\nu, 5500}/f_{\nu, 3600}$, which traces the UV to optical color (see Section $\ref{sec:definitions}$ for a more detailed discussion), and we split our sample in four bins of this color. These bins are defined as $f_{\nu, 5500}/f_{\nu, 3600} = 1.3 - 2.5$, $f_{\nu, 5500}/f_{\nu, 3600} = 2.5 - 5$, $f_{\nu, 5500}/f_{\nu, 3600} = 5 - 7.5$, $f_{\nu, 5500}/f_{\nu, 3600} = 7.5 - 30$. The separation of these wavelength ranges is somewhat wider than definitions used in other recent studies \citep[e.g.][]{Wang24_RUBIES} such that we can account for possible smooth color transitions. Generally, we find that the rank ordering is conserved when using $\lambda \approx4100$ {\AA} instead of $\lambda \approx5500$ {\AA}. The bins contain 6, 4, 4 and 4 sources, with median values 1.4, 3.4, 5.6 and 8.3, respectively. The average redshifts of the sources in the bins are similar, but the two reddest bins have H$\alpha$ luminosities that are on average $\approx4$ times higher than the H$\alpha$ luminosities of the bluest subsets. In Figure $\ref{fig:prism_sample}$, we show the median stacked prism spectra (normalised by the rest-frame optical flux density at 0.55 micron) of these four sub-sets, compared to the median stack of the full sample. Bootstrap resampling is used to quantify the variation within the sub-sets. We also show the median stacked H$\alpha$, H$\beta$ and [O{\sc iii}] spectra, normalised to the peak of the emission since that is most easily measurable without any modeling assumptions. The main trends identified are not sensitive to these normalisation choices. The stacked spectra are shown in linear and logarithmic scale to highlight the cores and the wings of the profiles, respectively. We note that we sample the PRISM stacks to a relatively sparse wavelength grid to  correct for the fact that the spectral resolution varies across sources. However, we also note these stacks are only used for qualitative comparisons. For the stacks on grating data, we smooth all the H grating spectra to the M grating resolution before stacking.\footnote{The stacked spectra will be made available in electronic format upon publication.}

The key result from Figure $\ref{fig:prism_sample}$ is that the shape of the Balmer emission lines correlates with the UV to optical color. The reddest sources have Balmer lines that are most dominated by the broad wings and that show absorption features close to the systemic redshift \citep[e.g.][]{Matthee24,Deugenio25}. The reddest sub-set also shows narrow H$\beta$ line-emission at the line center with an identical profile as the [O{\sc iii}] line. The bluest sources show Balmer lines that are dominated by strong, narrow emission at the line center. Among the intermediate sub-sets, the broad components become increasingly dominant the redder the broad-band spectrum is, the absorption features increasingly strong and shifted closer to the systemic redshift. The Balmer lines show exponential wings \citep[see also][]{Rusakov25} with similar FWHM $\approx1400$ km s$^{-1}$. While the broad-to-total flux ratio generally increases for red sources, there are subtle differences between H$\alpha$ and H$\beta$. In the bluer sub-sets, the H$\alpha$ line has relatively stronger wings than H$\beta$. For the redder sub-sets, it is the opposite case where the wings of the H$\beta$ lines contain a higher fraction of the flux, compared to H$\alpha$. Generally, there are little variations among the [O{\sc iii}] profiles that are in most cases very narrow.

Apart from the emission-line profiles, there are various other spectral features that vary with the UV to optical colors. In the rest-frame optical, the equivalent widths (EWs) of the [O{\sc iii}]$_{4960,5008}$ and HeI$_{5870, 7065}$ lines are significantly stronger in the bluest sources. The H$\alpha$ EWs show much less variation with the UV to optical color and it is usually around 1500 {\AA} \citep[see also][for similar results]{deGraaff25b}. The [S{\sc ii}] doublet is absent across all stacks, unlike in the typical star-forming galaxies at these redshifts \citep[e.g.][]{RobertsBorsani24,Hayes25}, likely due to its low critical density. 

\begin{figure}
    \centering
    \includegraphics[width=0.95\linewidth]{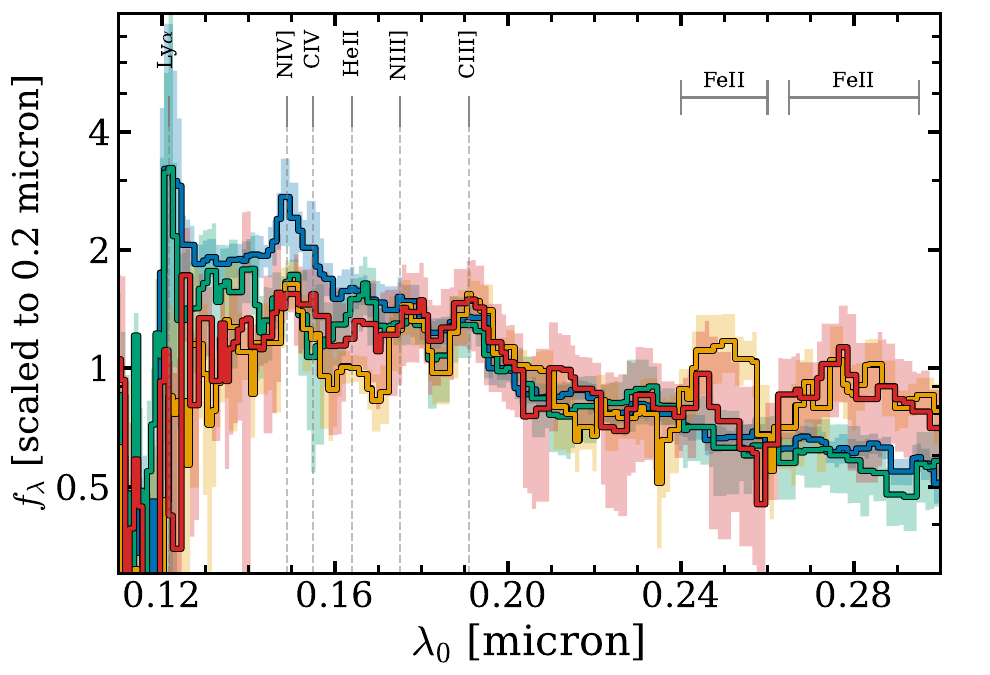}\\
    \includegraphics[width=0.95\linewidth]{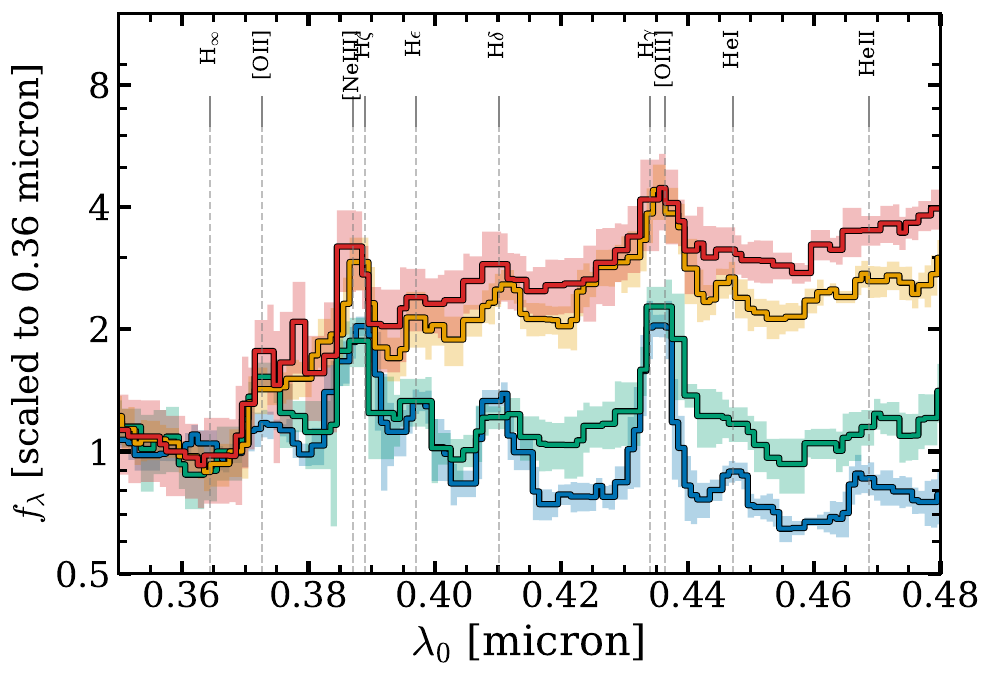}
    \caption{{\bf The variation in faint emission-line strengths across the range of UV to optical colors.} Stacked PRISM spectra (as in the left column in Fig. $\ref{fig:prism_sample}$) in our four sub-sets of UV to optical color, with blue corresponding to the bluest subset, and red to the reddest. Shaded regions illustrate the variation within the sub-sets based on bootstrap resampling. In the top panel, we highlight emission features in the rest-frame UV, while we show faint features in the optical regime in the bottom panel. Differences that correlate with the UV to optical redness are the weakness of N{\sc iv}] emission and the strength of Fe{\sc ii} features. Redder sources show lower EWs in the higher-order Balmer lines and are absent in He{\sc ii}. [Ne{\sc iii}] appears relatively strong across the sub-sets.   }
    \label{fig:prism_zooms}
\end{figure}

In Figure $\ref{fig:prism_zooms}$ we zoom in on various faint emission-line features in the rest-frame UV and the bluer part of the rest-frame optical of the stacked PRISM spectra. While PRISM spectra are not ideal to identify faint features with low EW, grating data are not uniformly available over these wavelength ranges for our sample. In the optical, the forbidden [Ne{\sc iii}]$_{3869}$ line with an excitation potential of 40.9 eV also appears in emission in all subsets, with relatively little variation with UV to optical color. The He{\sc ii}$_{4686}$ line, ionized by even more energetic photons, on the other hand, appears stronger for bluer sources. The Balmer decrements are significantly steeper in the redder sources, with relatively weak higher order Balmer lines such as H$\delta$, compared to the bluest subset. These faint optical lines in the blue sources are studied in more detail in Mascia et al. in prep.

In the rest-frame UV part of the spectrum our analysis is severely challenged by the limited resolution of the PRISM data. However various differences are still apparent. The UV continuum in the bluer sub-sets is power-law like and shows Ly$\alpha$ emission. As illustrated by the shaded regions that highlight the diversity within the subsets, the strength of Ly$\alpha$ emission shows significant scatter within sub-sets. The bluest subset shows indications of strong N{\sc iv}]$_{1488}$, C{\sc iv}$_{1550}$, N{\sc iii}]$_{1750}$ and C{\sc iii}]$_{1909}$ emission. Among individual sources in the stack, the N{\sc iv}] is most often detected, followed by C{\sc iii}] and Lyman-$\alpha$ (see Mascia et al. in prep for a more detailed analysis). Strong nitrogen lines have been reported in broad-line emitters before \citep[e.g.][]{Ubler23,Treiber24,Isobe25}. The strength of C{\sc iv} varies significantly within the blue subset which could be related to its resonant nature. The redder objects tend to be very faint in the UV and appear to have weaker lines. Some hints of N{\sc iv}], N{\sc iii}] and C{\sc iii}] emission with lower EWs are visible, but these need to be confirmed by deep grating spectroscopy. The UV continuum in the redder stacks is flatter and shows broad (composite) features around 2600 {\AA} that are likely related to Fe{\sc ii} transitions. The nature of these transitions is challenging to determine at this resolution \citep{Labbe2024,Torralba25b,Deugenio2025-irony,Tripodi25} as they could be due to combinations of emission and absorption depending on the specific spectral term \citep[e.g.][]{Baldwin04}. As shown in Fig. $\ref{fig:prism_sample}$ ,we note that an inflection or "V-shape" is easily seen in most subsets. In the bluest subset, the data prefer a double power law over a single power law, with a UV continuum slope of $\beta_{\rm UV}\sim-1.8$ (significantly steeper than quasars; \citealt[e.g.][]{Shull25}), whereas $\beta_{\rm opt}\sim-1$ (see Mascia et al. in prep. for a more detailed investigation). For other sub-sets, the optical continuum has a somewhat shallower slope $\beta_{\rm opt}\sim0$ and there is a stronger (Balmer) break.

\begin{figure*}[htbp]
\centering
\renewcommand{\arraystretch}{0.05} 

\begin{tabular}{@{}ccc@{}}
\includegraphics[width=0.32\linewidth]{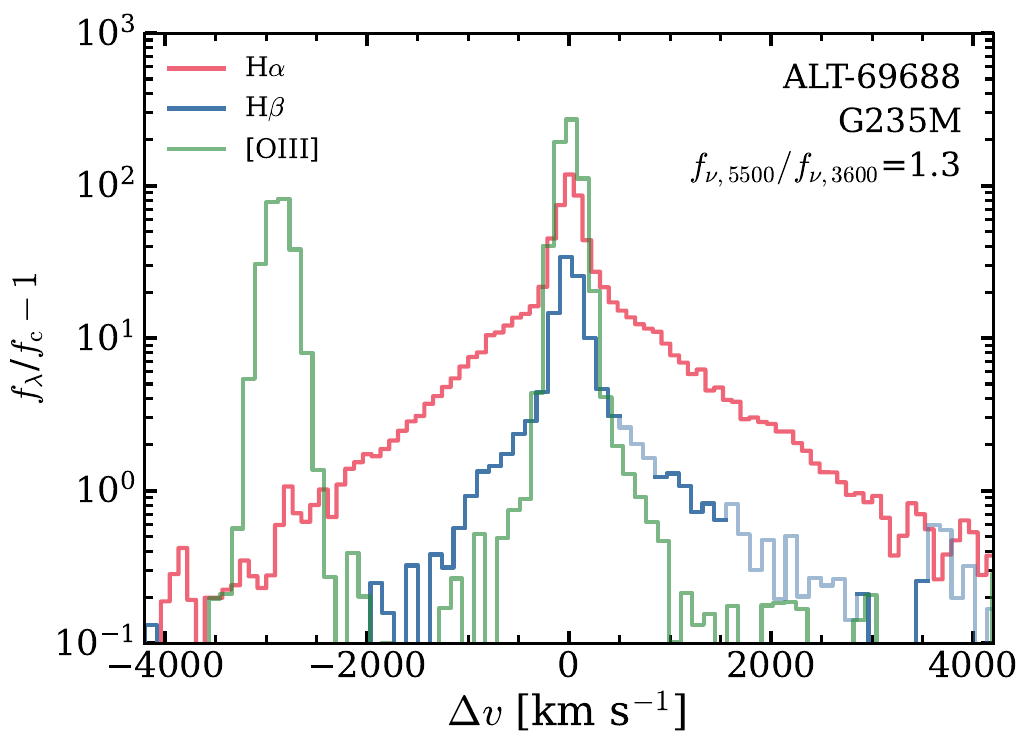} &   
\includegraphics[width=0.32\linewidth]{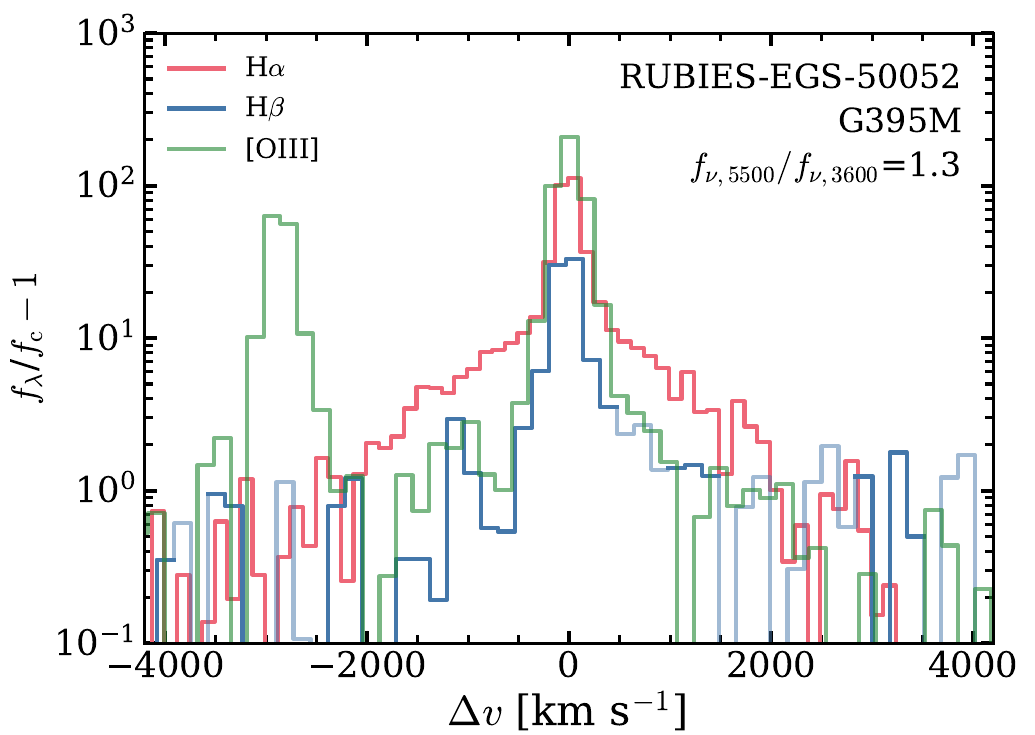} & 
\includegraphics[width=0.32\linewidth]{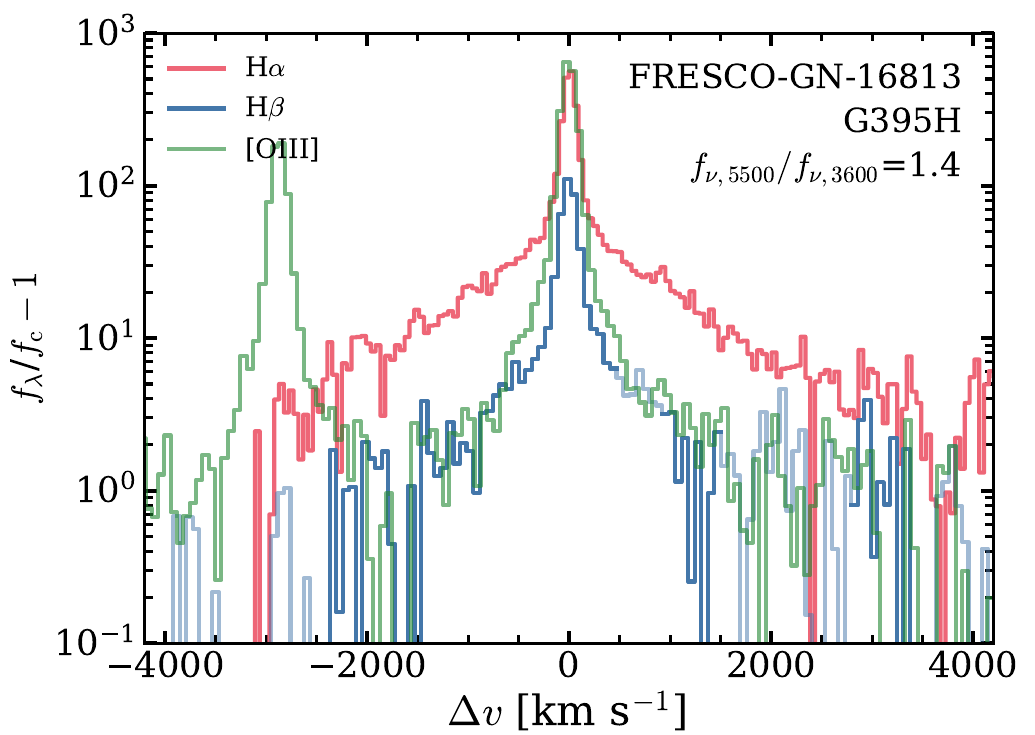} \\[-9pt]

\includegraphics[width=0.32\linewidth]{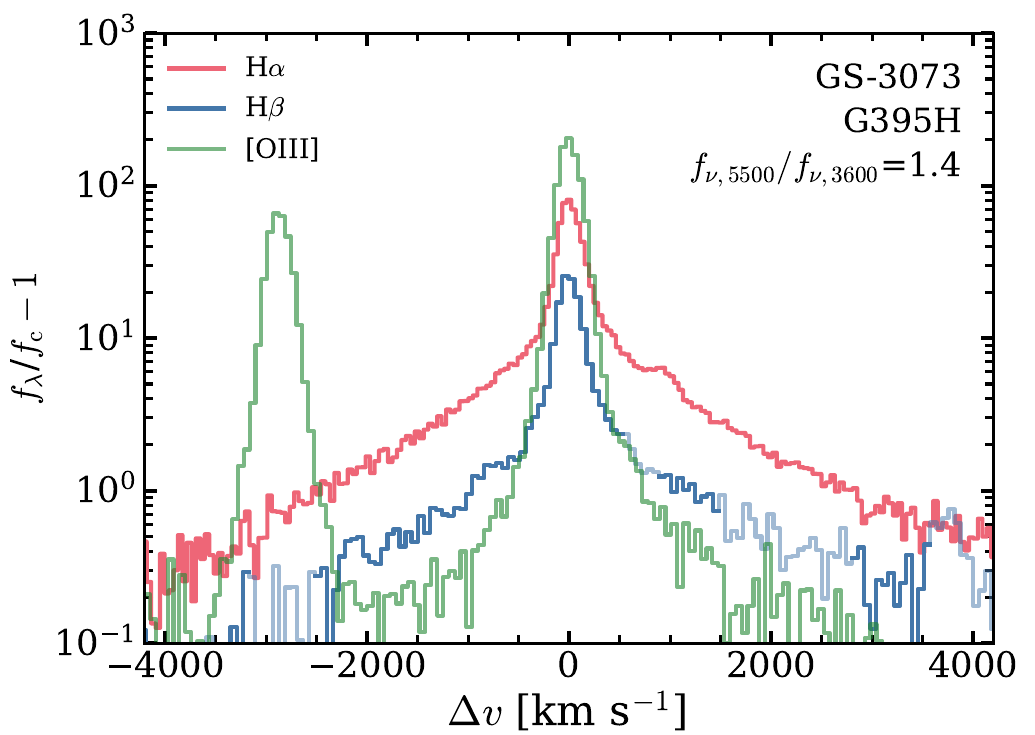} &
\includegraphics[width=0.32\linewidth]{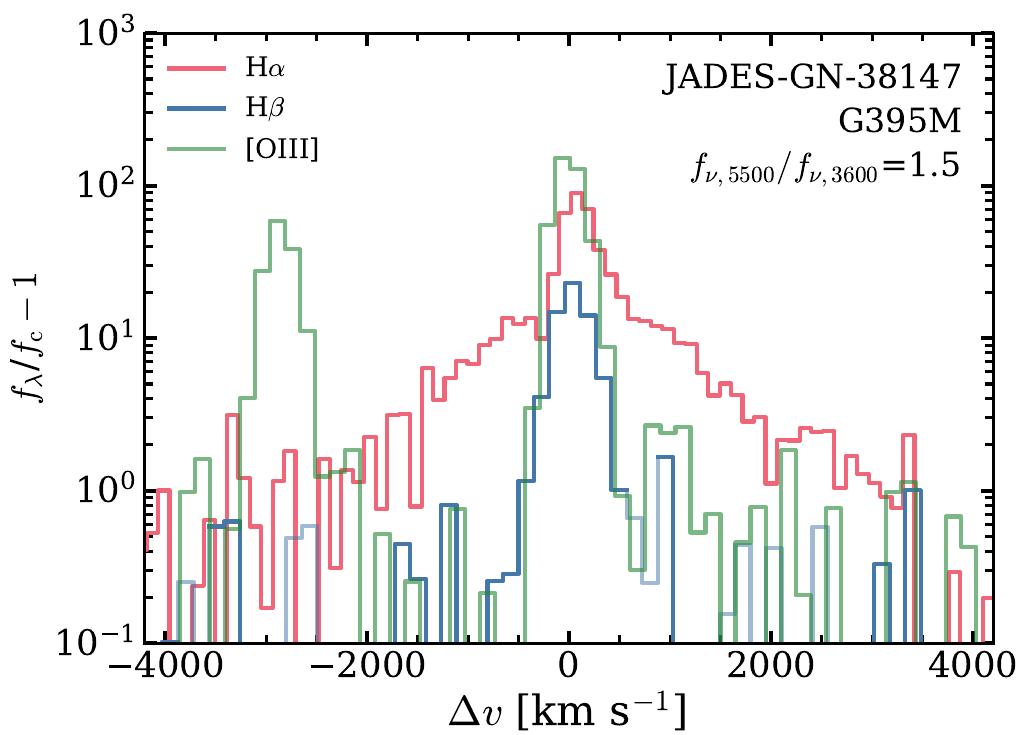} &
\includegraphics[width=0.32\linewidth]{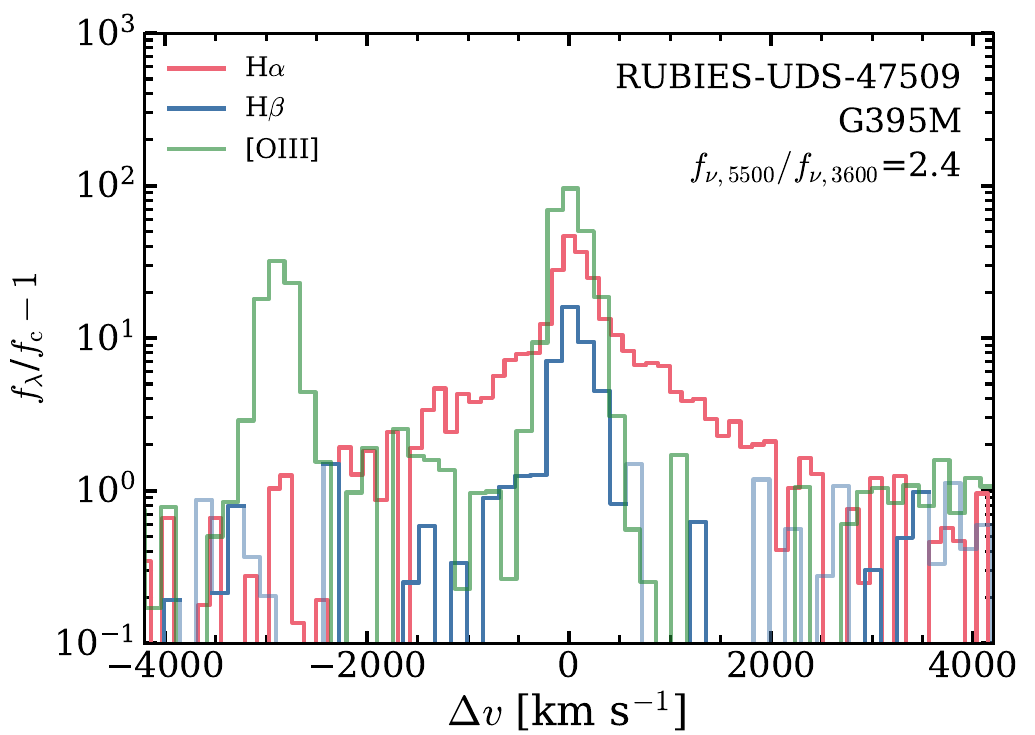} \\[-9pt]

\includegraphics[width=0.32\linewidth]{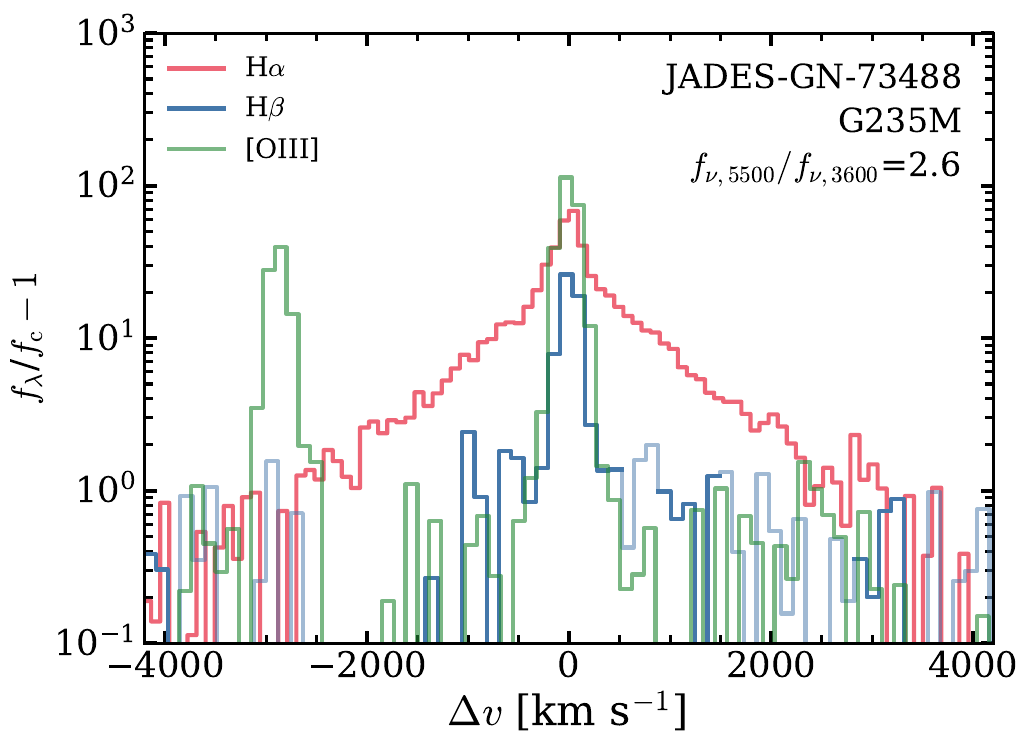} &
\includegraphics[width=0.32\linewidth]{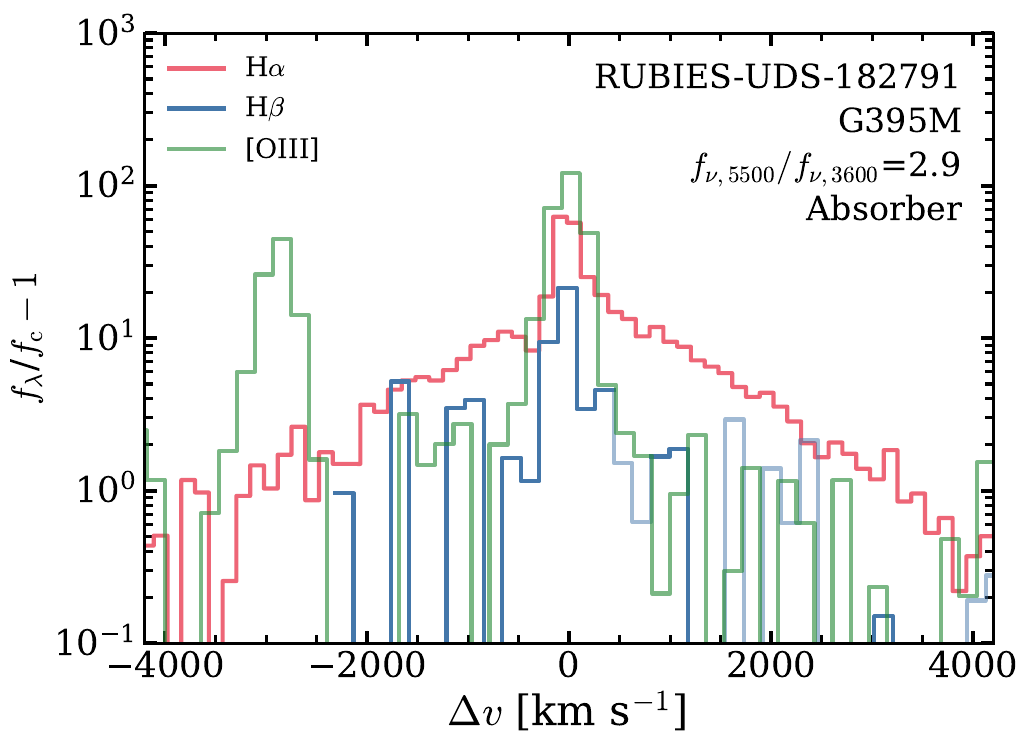} &
\includegraphics[width=0.32\linewidth]{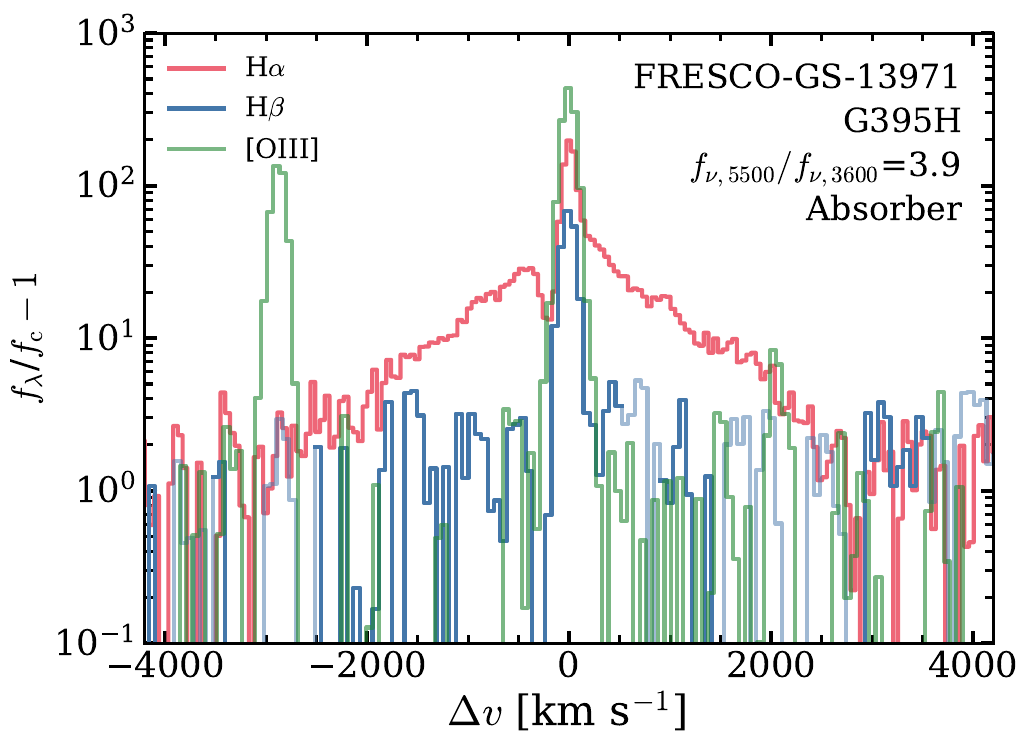} \\[-1pt]

\includegraphics[width=0.32\linewidth]{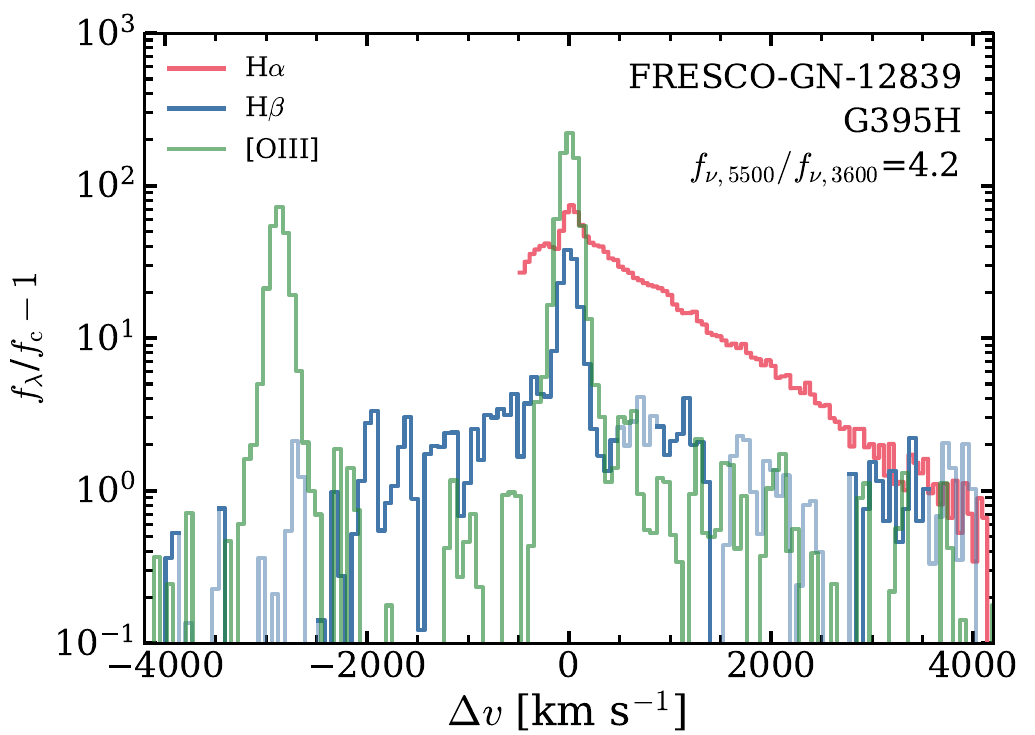} & 
\includegraphics[width=0.32\linewidth]{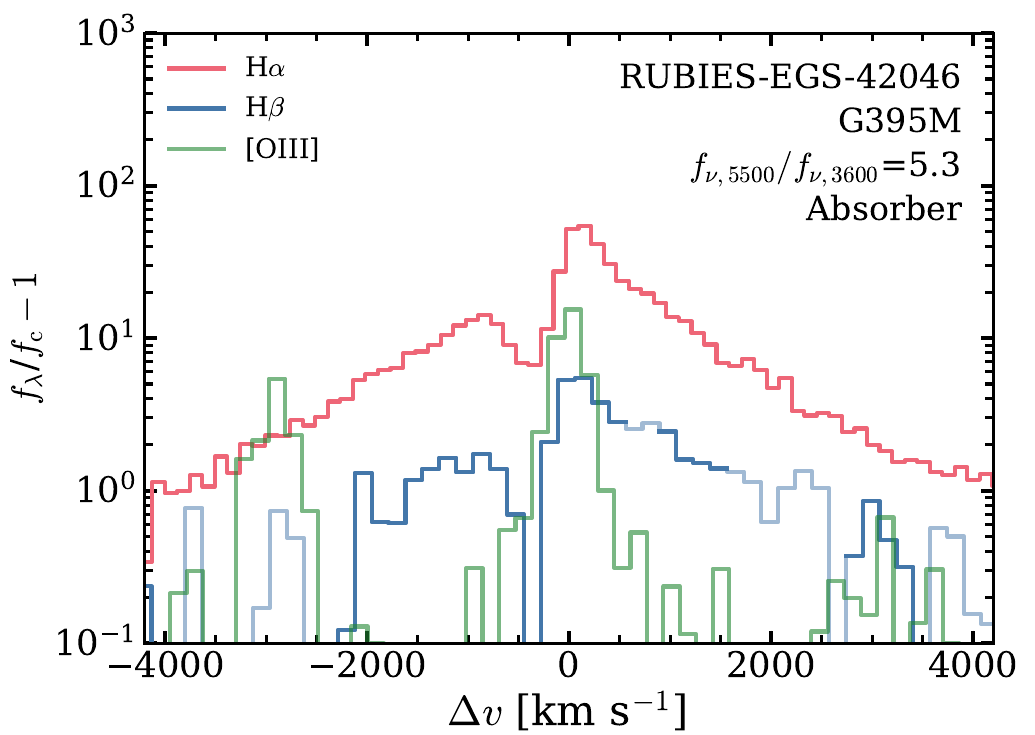} &
\includegraphics[width=0.32\linewidth]{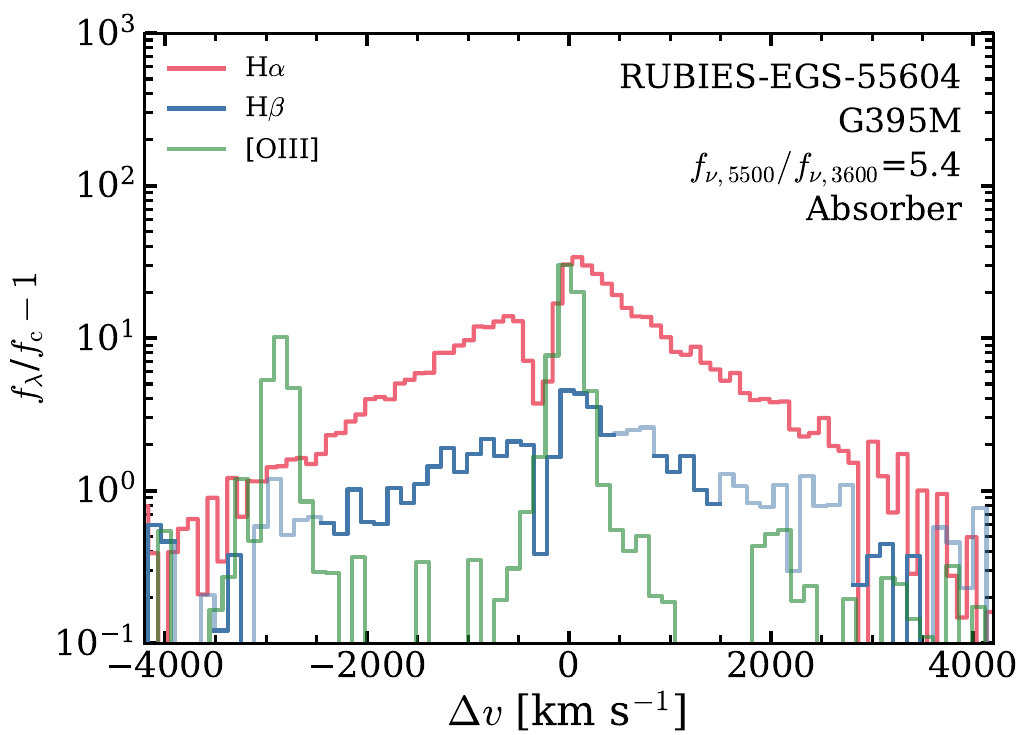} \\[-9pt]       

\includegraphics[width=0.32\linewidth]{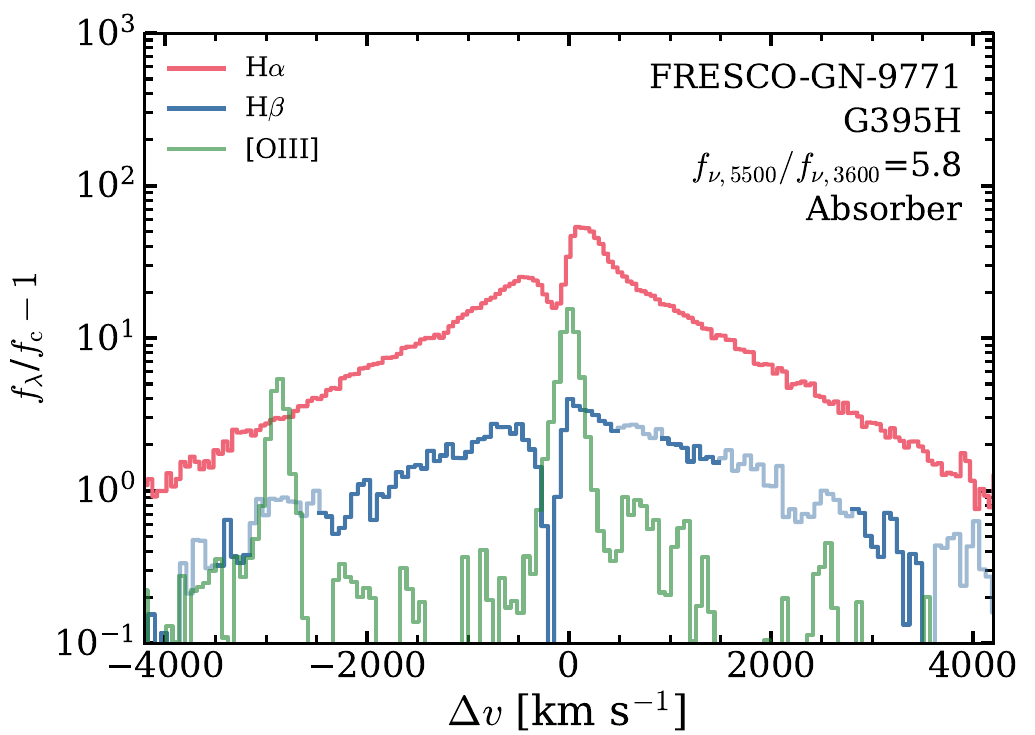} & 
\includegraphics[width=0.32\linewidth]{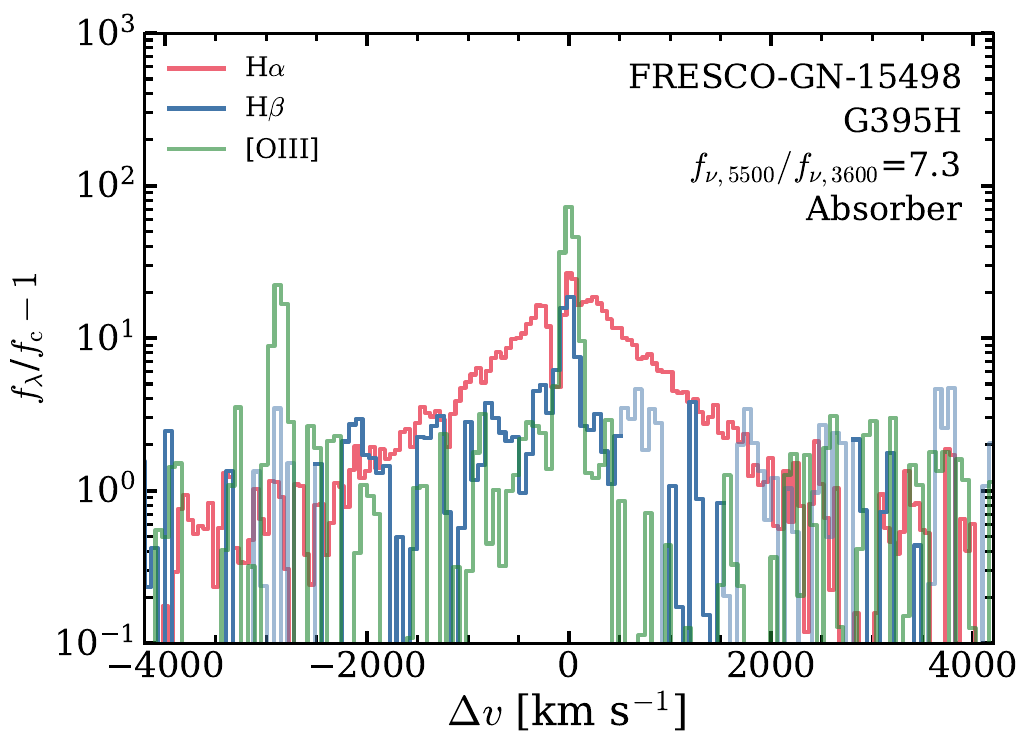} &
\includegraphics[width=0.32\linewidth]{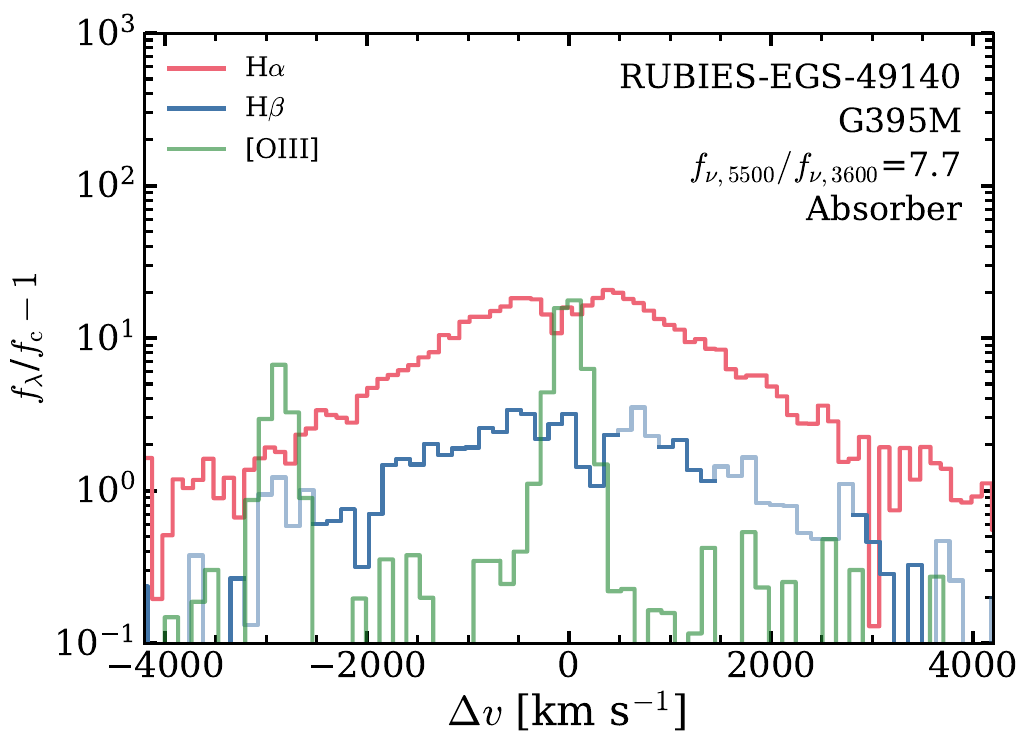} \\[-9pt]      

\includegraphics[width=0.32\linewidth]{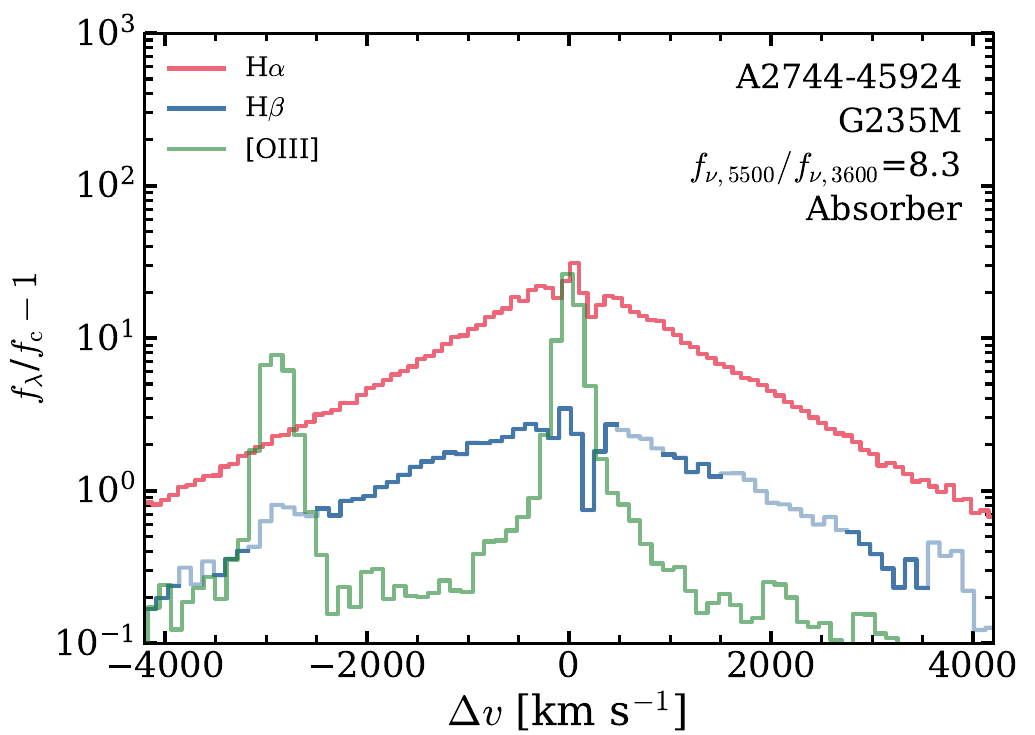} &
\includegraphics[width=0.32\linewidth]{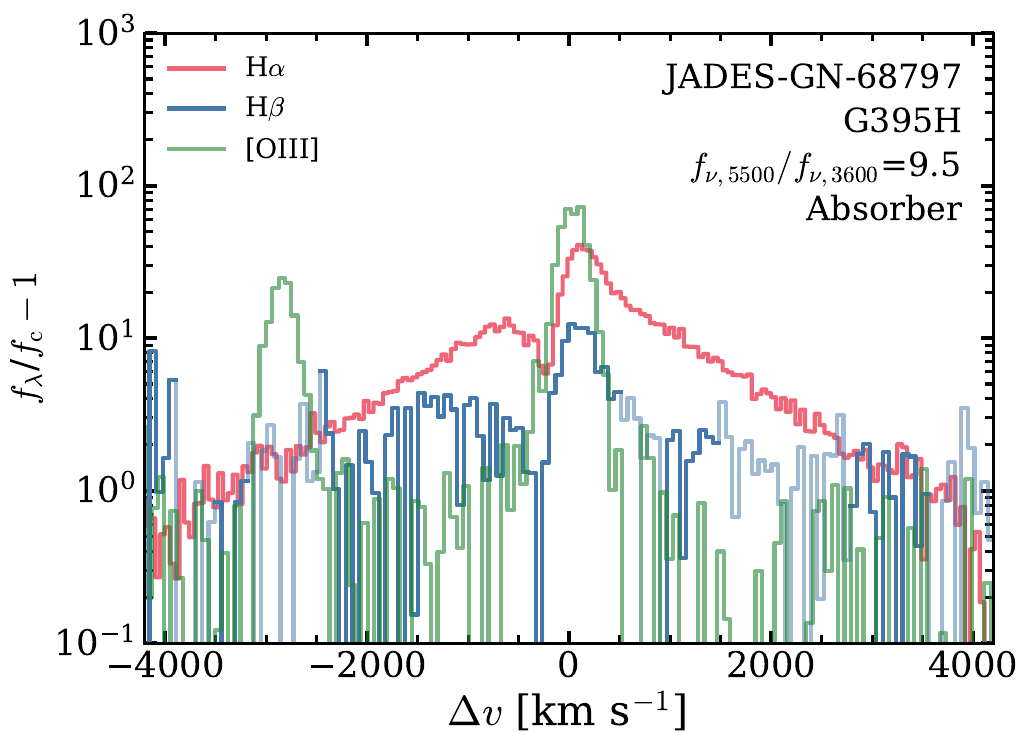} &  
\includegraphics[width=0.32\linewidth]{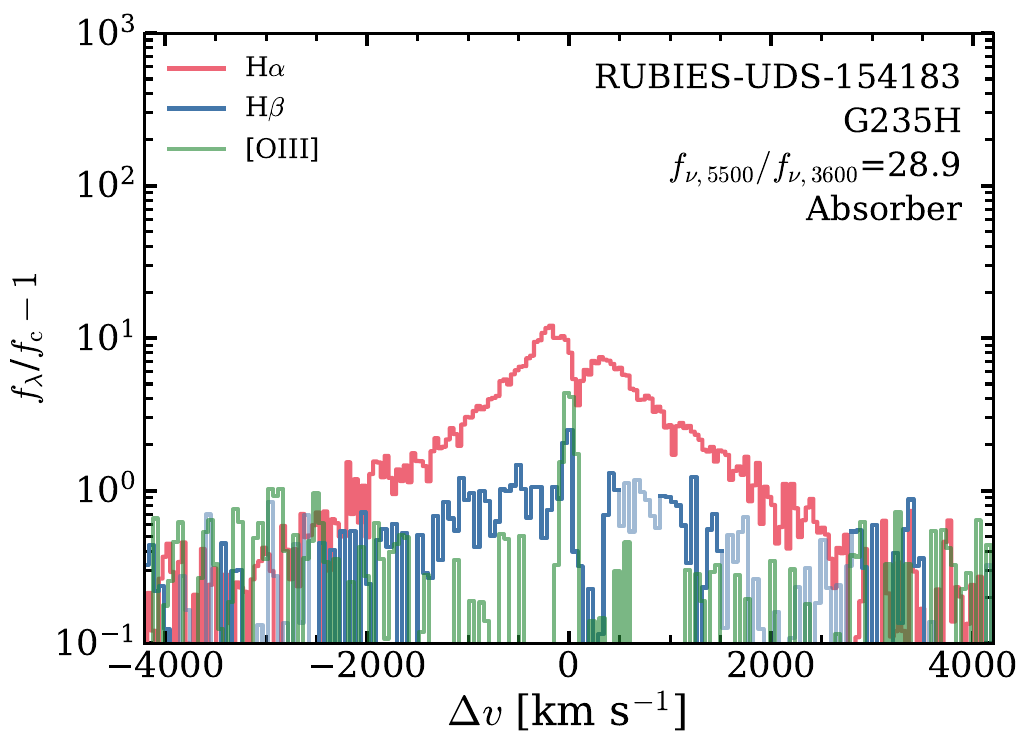} \\ 
\end{tabular}
    \caption{{\bf The similarity and variety among emission-line spectra of broad H$\alpha$ line emitters.} Objects are sorted by the UV to optical color such that blue sources are shown on top-left and the reddest are in the bottom-right. To highlight variations in equivalent width of the various lines, lines are normalised to their corresponding continuum flux density. We highlight sources with significant detection of Balmer absorption lines.}
    \label{fig:overview}
\end{figure*}

Figure $\ref{fig:overview}$ shows the individual line-profiles of the sources, sorted by UV to optical color, highlighting the smooth transition from the trends that are illustrated in our stacked sub-sets presented above. By selection, broad components are detected in all the Balmer lines (apart from some H$\beta$ detections with very low signal-to-noise). The [O{\sc iii}]$_{4960,5008}$ doublet is usually very narrow, except in ALT-69688, GS-3073, A2744-45924 and FRESCO-GN-16813 where a faint broad component can be clearly identified at the location of the point-source. JADES-GN-68797 has a complex, clumpy morphology even within the NIRSpec shutter and we identify two separate [O{\sc iii}] emitting components in the 2D spectrum, mimicking a broader [O{\sc iii}] line. In the next section, we focus on quantifying the line-profile variations and their correlations with various observables by fitting parametric models to the line-profiles. We interpret the trends identified in Section $\ref{sec:interpretation}$.

\subsection{Line profile fitting} \label{sec:profiles}
\subsubsection{Model setup}
In our analysis of line-profiles shown in Fig. $\ref{fig:overview}$ and in our quantitative comparison with galaxy spectral features, we exploit simple parameterizations of the Balmer line profiles as well as model fits. Developing upon the model and procedure as outlined in \cite{Torralba25b} for the analysis of FRESCO-GN-9771, we fit the H$\beta$ and H$\alpha$ emission lines of each object separately. The optical continuum is modeled as a polynomial, after masking the most relevant emission lines, and the line fits are performed on continuum-subtracted spectra. In the first place, we measure the systemic redshift and line width of the \oiii{} $\lambda\lambda 4960,5008$ doublet, fitting a double Gaussian with a fixed flux ratio of 2.98 \citep{Osterbrock06}. The fiducial model that we adopt for both the \Hbeta{} and \Halpha{} lines is physically motivated and consists of the combination of four components: A broad exponential component, an additional Gaussian emission component with an intermediate width, an absorber and a narrow component.

The broad wings are modeled as a symmetric exponential profile, convolved with a Gaussian with the same width as the intermediate emission component
\citep[see e.g.][]{Rusakov25}. We have performed tests where the exponential component is instead an (additive) gaussian component. However, we find that an exponential wing is preferred for all H$\alpha$ fits in our sample, with BIC difference larger than 10. For H$\beta$, an exponential fit is significantly preferred 6/18 sources, while a Gaussian fit is preferred for a single source (ALT-69688) and the remaining fits are inconclusive. The detailed fitting statistics are listed in Appendix $\ref{app:fits}$. The intermediate component is modeled with a single Gaussian profile whose width can vary. The absorber profile is allowed greater flexibility to account for possible complex dynamics \citep[e.g.][]{Deugenio2025-irony}. The optical depth of the absorber is modeled empirically as $\tau(\lambda) = V_{\rm abs}(\lambda)\cdot\gamma(\lambda)$, where $V_{\rm abs}$ is a Voigt profile, and $\gamma$ an skewness parameter defined as $\gamma(\lambda) = 1 + {\rm erf}[\gamma_0 (\lambda - \lambda_0)\cdot(\sqrt{2}\sigma_{\rm abs})^{-1}]$, where `erf' is the Gauss error function, $\sigma_{\rm abs}$ the Voigt-Gaussian parameter of the absorber, and $\gamma_0$ a free parameter that sets the strength and direction of the skewness. The absorption is implemented in the model as a multiplicative factor ${e}^{-\tau(\lambda)}$, and we implicitly assume a covering factor of one \citep[e.g.][]{Ji25,Deugenio25}.
The narrow component is modeled as a single Gaussian, with a width and systemic redshift tied to that of our fit to the \oiii{} doublet. Except for this component, the redshifts of all the components are allowed to vary slightly ($\pm 1000$~km s$^{-1}$) from the systemic. Lastly, for the \Halpha{} fit we add a narrow [\ion{N}{ii}] $\lambda\lambda 6548,6583$ doublet modeled as two Gaussians with a fixed theoretical ratio of 3.049. The width and redshift of the [\ion{N}{ii}] components are also fixed to the \oiii{} doublet.
The fiducial model can be summarized in one equation:
\begin{equation}
    f(\lambda) = \left[e^{-\tau_{\rm e}} I(\lambda)  + (1 - e^{-\tau_{\rm e}})(I * E)(\lambda) + C \right] \cdot e^{-\tau(\lambda)} + N(\lambda) \,,\label{eq:ha_model}
\end{equation}
where $I(\lambda)$ and $N(\lambda)$ are the aforementioned intermediate, and narrow Gaussian components; $E(\lambda)$ a symmetric exponential convolution kernel; $C$ a flat continuum component; and $\tau_{\rm e}$ the Thomson optical depth, that regulates the relative intensity between $I$ (unscattered) and $I*E$ (scattered; see e.g., \citealt{Deugenio2025-irony}).
The full model is convolved with the wavelength-dependent line-spread function (LSF) of the corresponding medium- or high-resolution grating\footnote{\url{https://jwst-docs.stsci.edu/jwst-near-infrared-spectrograph/}}, adopting a resolution $1.8\times R(\lambda)$, given that all our candidates are point-like \citep{degraaff2024_Mdyn}. We perform a $\chi^2$ minimization followed by MCMC sampling to estimate parameter uncertainties, using the Python package \texttt{lmfit}, with wide uninformative priors. In each step, the model function is defined in a $10\times$ over-sampled grid and re-binned to the spectral binning of the data. In Table $\ref{tab:fit_averages}$, we list the average and range of results of the fits. In Appendix $\ref{app:fits}$ we tabulate various fitting statistics and fitted parameters that are particularly explored in this work. 

\begin{table}
    \centering
    \caption{The median and range of the values of the best-fit parameters in our Balmer line profile modeling. * For H$\beta$ statistics involving the exponential component we only include sources with H$\beta$ S/N$>20$.  $\dagger$ For absorption statistics, we only use the sources for which the absorption feature was significantly detected (i.e. the BIC value for models without absorbers differs significantly $\Delta$BIC $>10$).  $\ddagger$ The line-cores are defined as core = total - exponential, which could be negative. }    
    \begin{tabular}{lccc}
    Property & Median & Range & Unit \\ \hline
    FWHM [O{\sc iii}] & 230 & 110 -- 360 & km s$^{-1}$ \\ 
    H$\alpha$/H$\beta_{\rm total}$* & 7.8 & 3.6 -- 14.0 \\ %Hb SN>5
    H$\alpha$/H$\beta_{\rm core}$* & 3.0 & $-2.2$ -- +4.5 \\ %Hb SN>5
    H$\alpha$/H$\beta_{\rm exponential}$* & 8.1 & 4.4 -- 13.8 \\ %Hb SN>5
    
     & & & \\

    {\bf H$\alpha$} & & & \\ \hline
     FWHM intermediate & 410 & 217 -- 820 & km s$^{-1}$ \\
     FWHM exponential    & 1360 & 1040 -- 1600  & km s$^{-1}$ \\
     $\sigma$ absorber & 127 & 106 -- 263 & km s$^{-1}$ \\
     $\Delta$v intermediate & 60 & $-40$ -- +150 & km s$^{-1}$ \\
     $\Delta$v absorber$\dagger$  & $-60$ & $-220$ -- +55 & km s$^{-1}$ \\
     $\tau_0$ absorber$\dagger$  & 3.2 & 0.8 -- 7.2 & \\
     narrow/total & 0.06 & 0.0 -- 0.54 & \\
    core/total$\ddagger$ & 0.20 & $-0.11$ -- +0.47 & \\
     & & & \\

    {\bf H$\beta$} & & & \\ \hline
     FWHM intermediate & 430 & 140 -- 1000 & km s$^{-1}$ \\ %check the 1000, that's cliff that hit a boundary, can be fixed 
     FWHM exponential*   & 1680 & 1150 -- 2040  & km s$^{-1}$ \\ %Hb SN>5
     FWHM absorber & 450 & 270 -- 920 & km s$^{-1}$ \\
     $\Delta$v intermediate & 170 & $-330$ -- +370 & km s$^{-1}$ \\
     $\Delta$v absorber$\dagger$  & $-10$ & $-60$ -- +160 & km s$^{-1}$ \\
     $\tau_0$ absorber$\dagger$  & 4.5 & 3.0 -- 7.3 & \\
     narrow/total* & 0.13 & 0.07 -- 0.6 & \\ %Hb SN>5
    core/total* & $-0.06$ & $-0.23$ -- +0.69 & \\ \hline %Hb SN>5
      
    \end{tabular}

    \label{tab:fit_averages}
\end{table}

\begin{figure*}
\centering
\begin{tabular}{cc}
\vspace{-0.18cm}
    \includegraphics[width=0.43\linewidth]{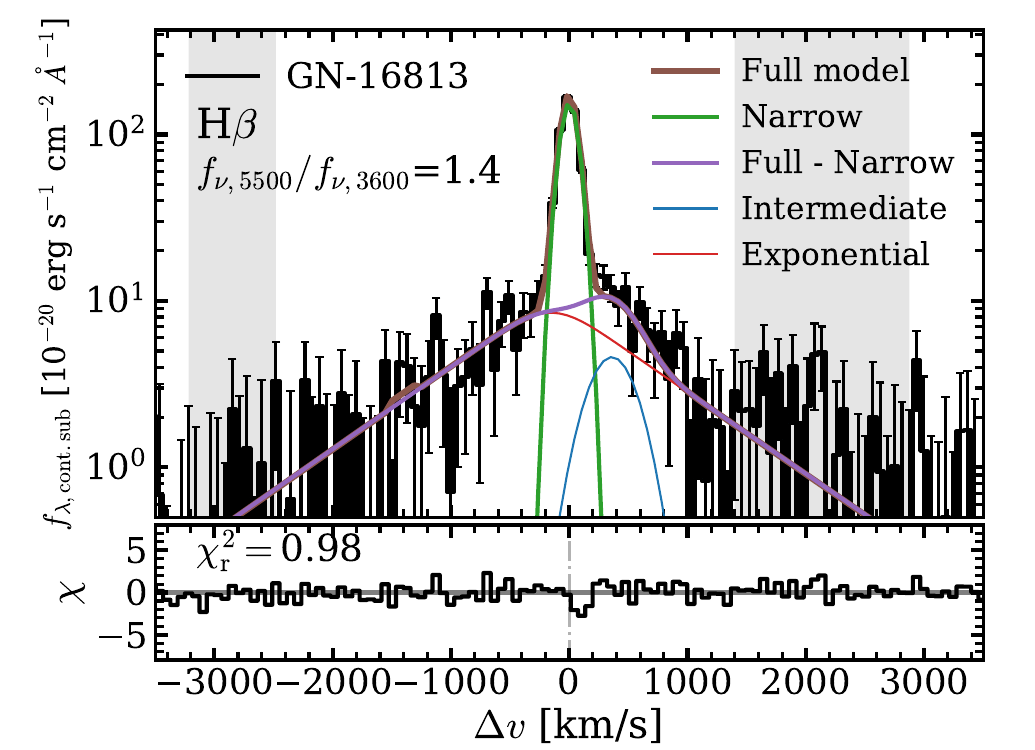} & 
        \includegraphics[width=0.43\linewidth]{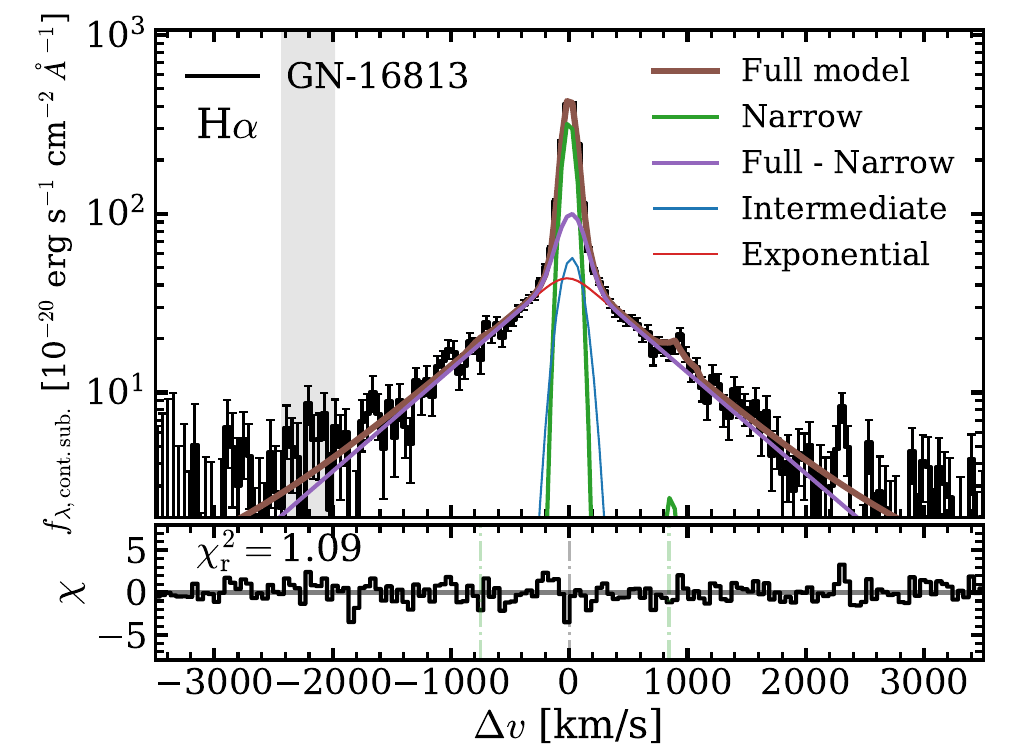}  \\
\vspace{-0.18cm}
    \includegraphics[width=0.43\linewidth]{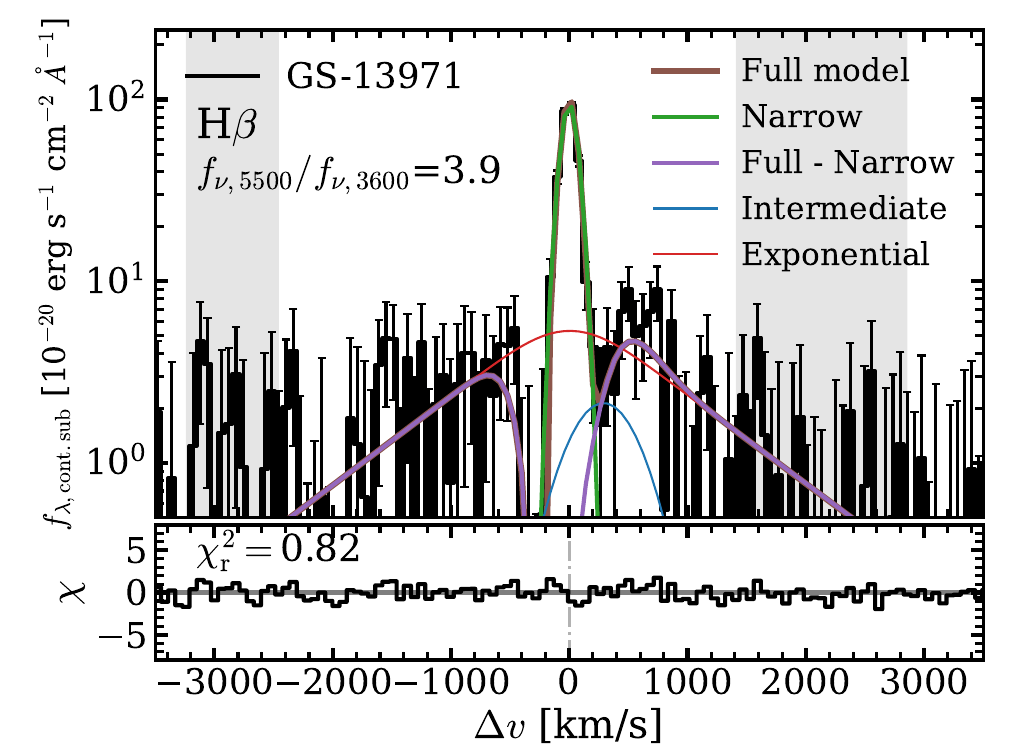} & 
        \includegraphics[width=0.43\linewidth]{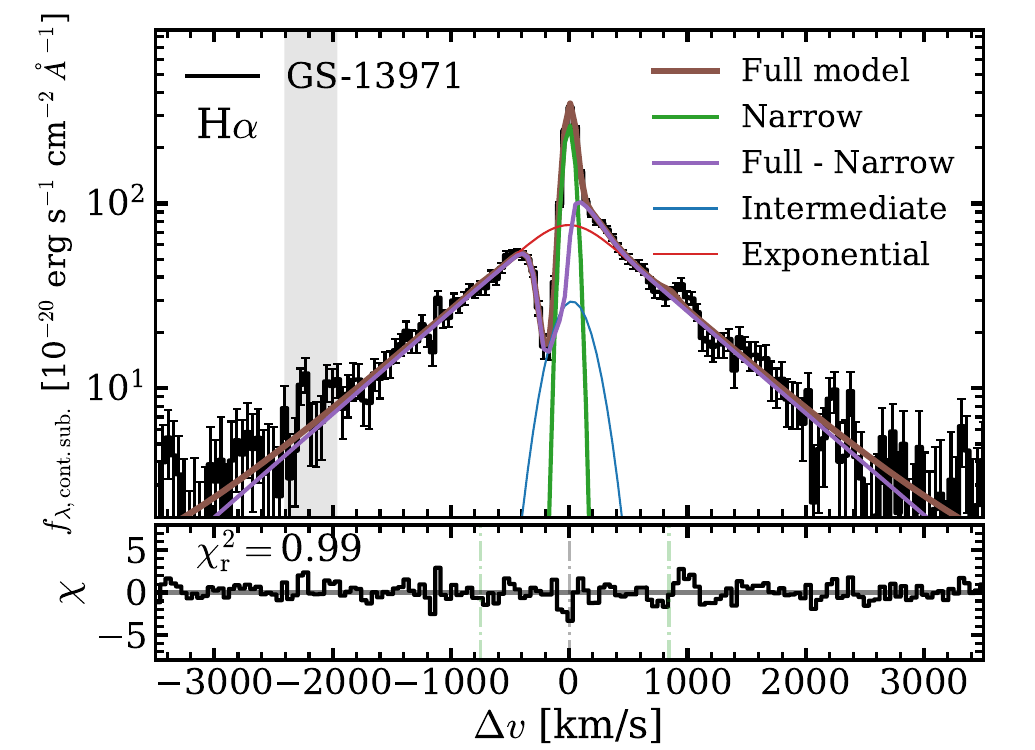}  \\
      \vspace{-0.18cm}  
    \includegraphics[width=0.43\linewidth]{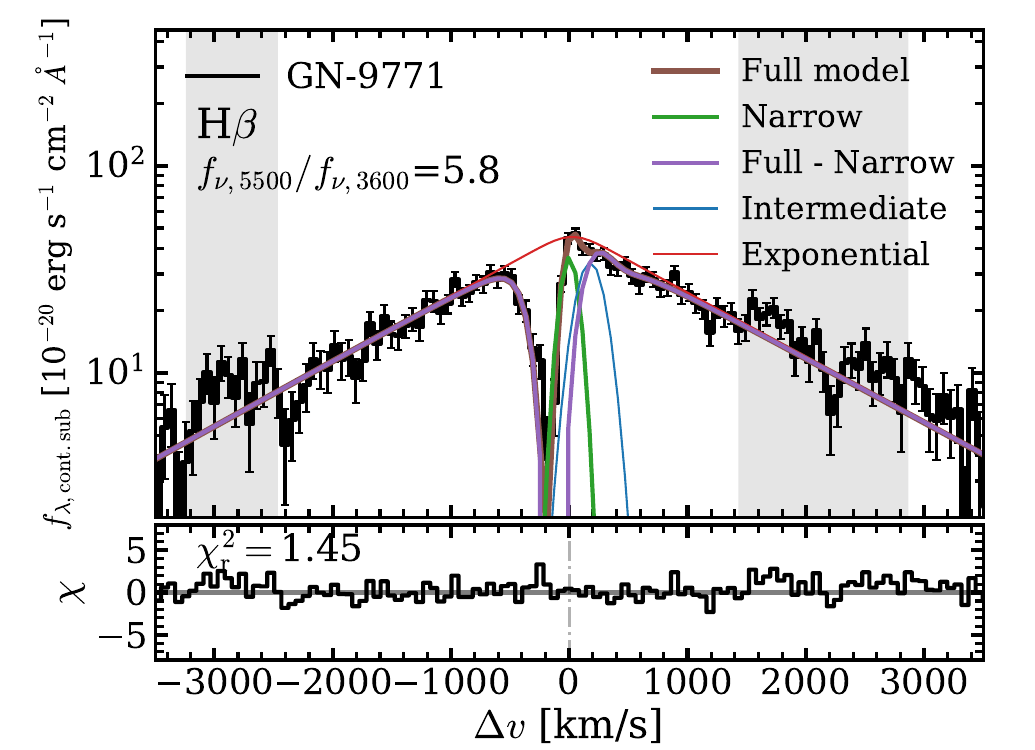} & 
        \includegraphics[width=0.43\linewidth]{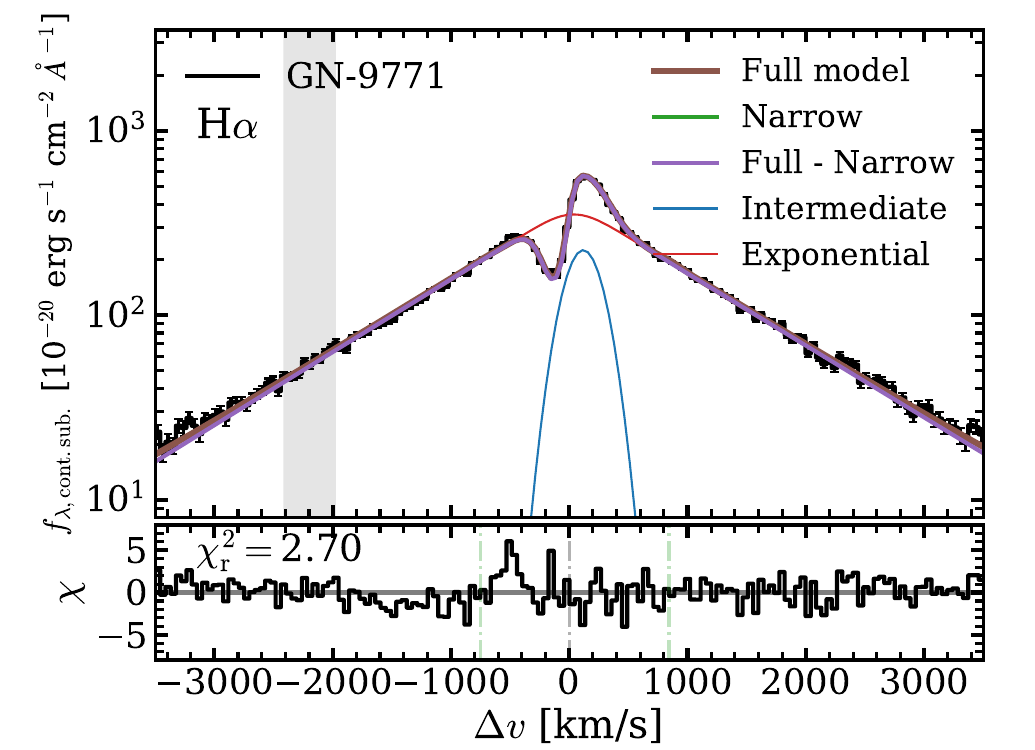}\\  
    \includegraphics[width=0.43\linewidth]{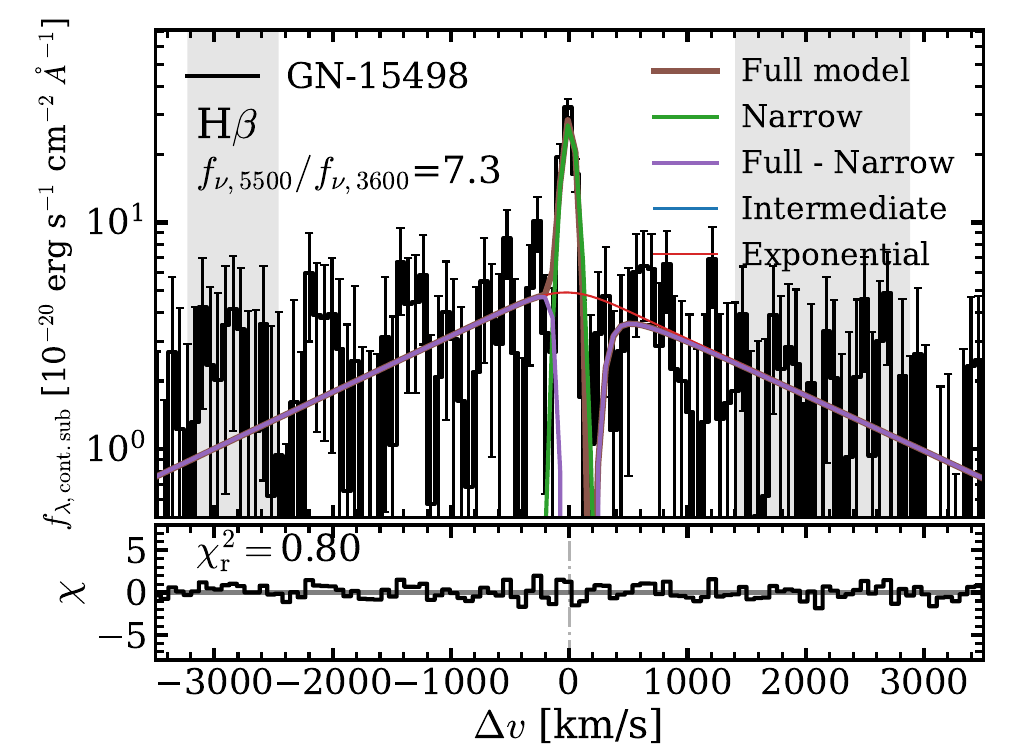} & 
        \includegraphics[width=0.43\linewidth]{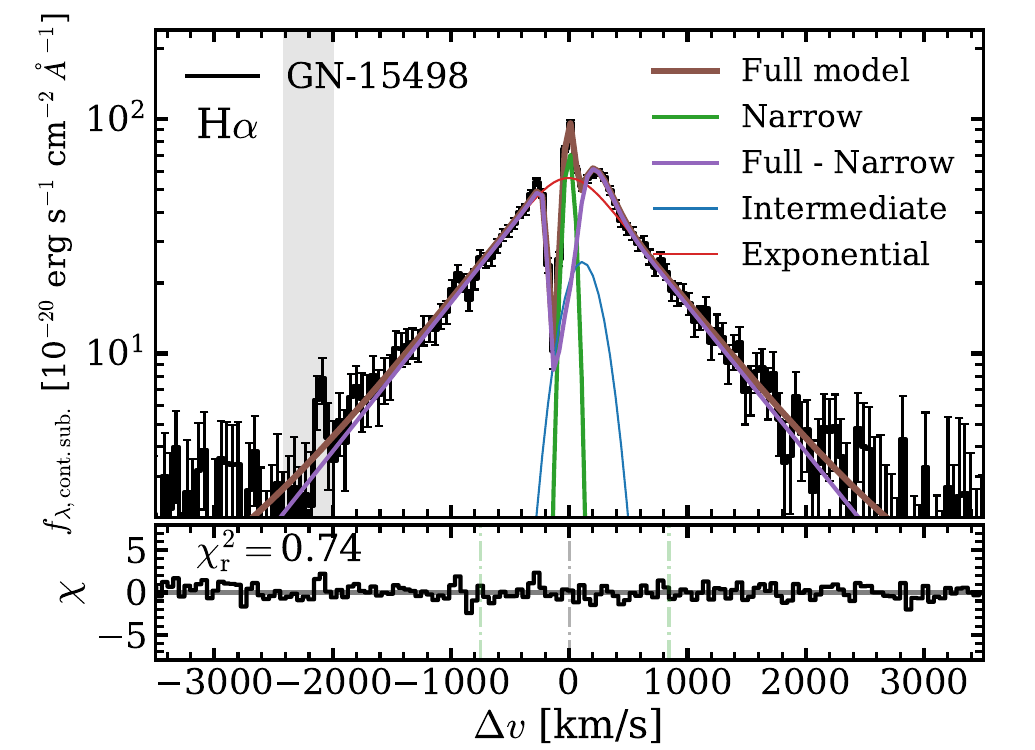}  \\
    \end{tabular}\vspace{-0.23cm}
    \caption{{\bf Example model fits to the H$\beta$ and H$\alpha$ line-profiles of four broad-line H$\alpha$ emitters with H-grating spectra, sorted by optical to UV color.} Continuum-subtracted H$\beta$ profiles are shown in the left column, while H$\alpha$ is shown in the right. Each panel shows the data (black) and the residual of the best-fit (brown shows the full model), with associated $\chi^2_{\rm r}$ below. For the H$\alpha$ residuals, green dash-dotted lines mark the locations of the [N{\sc ii}] doublet. The purple line shows the narrow-subtracted model. Blue shows the intermediate emission component and pink the exponential component. Grey regions are excluded from the fits as they are possibly affected by [Fe{\sc ii}] emission. While exponential wings are present in virtually all sources, the core of the line-profile displays most variation. The cores of the H$\beta$ lines show stronger absorption compared to the H$\alpha$ line, and narrow central emission that is likely associated with the host galaxy.  }
    \label{fig:profilefit}
\end{figure*}

\subsubsection{Model results}

Figure $\ref{fig:profilefit}$ shows our best fits to the Balmer line profiles of four sources with deep, high resolution data from our IFU program that span the range in UV to optical colors. Broad exponential components are identified in all H$\alpha$ profiles, with FWHMs $\approx1400$ km s$^{-1}$. In H$\beta$, the profile of the broad component is exponential for the sources with highest signal to noise ratio. The FWHM of the exponential components of the H$\beta$ lines are on average somewhat higher, which is at odds with expectations from Thomson scattering through a uniform medium, but we discuss modeling limitations in Section $\ref{sec:expwings}$ that could impact this comparison. Comparing the H$\alpha$/H$\beta$ ratios of the exponential components alone, we find H$\alpha$/H$\beta \approx 8.1$, ranging from $4.4-13.8$.

The central parts of the line profiles are complex and display significant variation. In the bluest source shown, GN-16813, the narrow and intermediate components show very similar H$\alpha$ profiles and there are no indications of an absorber, whereas the H$\beta$ profile has relatively low signal-to-noise. The H$\alpha$ lines of GS-13971 and GN-15498, on the other hand, illustrate the need for an absorber and a separate narrow component at the systemic redshift and a broader intermediate component that is redshifted. The modeling prefers similar absorbers in the H$\beta$ lines, although the BIC differences are $\Delta$BIC $<10$, i.e. these are not significantly detected. GN-9771 clearly demonstrates the need for an intermediate component, which is most prominently seen in H$\alpha$. The H$\beta$ profile of GN-9771 shows a little contribution from narrow emission, which in H$\alpha$ is outshone by the intermediate component. Blue-shifted absorption is clearly detected in both its lines. In the H$\alpha$ lines, the intermediate components are about a factor two broader than the narrow component and red-shifted by $\approx60$ km s$^{-1}$ (spanning the range $-40$ to 150 km s$^{-1}$, see Table $\ref{tab:fit_averages}$). The intermediate components appear to be much weaker in the H$\beta$ line. This can be seen in the differences in the narrow/total ratio of the lines. The narrow/total flux ratios tend to be higher for H$\beta$ (up to 100 \% for RUBIES-UDS-182791, although at somewhat low S/N) compared to H$\alpha$. For example, for GN-15498, our modeling attributes 4 \% of the H$\alpha$ flux to the narrow component, whereas this ratio is 22 \% for H$\beta$. As the total line-fluxes are dominated by emission in the exponential component, the H$\alpha$/H$\beta$ ratios of the total lines are similar (7.8, ranging from 3.6-14.0). Interestingly, the H$\alpha$/H$\beta$ ratios in the line cores (which is the exponential wing-subtracted line-profile) tend to be closer to the case B value (with average H$\alpha$/H$\beta$=3), but display a range from $-2.2$ to 4.5, where negative values imply that the core of one of the lines is dominated by absorption, such as is the case for the H$\beta$ core of GN-9771.

\subsubsection{Model variations and degeneracies} \label{sec:degeneracies}
Given the complexity of our model, an important question to ask is whether we need all these components. In order to test the relevance of each component of the model across our sample, we run the Balmer line fits with the following variations in the model: a) fiducial, b) a model without absorber, c) a model without narrow component, d) a model without intermediate component, and e) a model without [\ion{N}{ii}] emission -- the latter only applicable to \Halpha{}. Then, we compare the differences in the BIC value of these fits. The addition of the components is then warranted if their removal yields an increase in the BIC by $\Delta$BIC$>10$. 

For 11 sources (61 \%), we significantly detect an absorption feature in the H$\alpha$ line, with a minimum EW$_0$ of 3.9 {\AA}. As can be seen from the annotation in Fig. $\ref{fig:overview}$, absorbers are ubiquitous among the redder objects and absent in the bluest. Several of the bluest objects (e.g. GS-3073, GN-16813) are among the objects with the highest signal to noise and with H-grating spectra. Furthermore, as we show later (Section $\ref{sec:core_flows}$), the central velocity of absorbers tends to be closer to the systemic redshift for the reddest sources, which is thus most difficult to detect. This strongly suggests that the absence of absorption is not an observational limitation. For H$\beta$, absorption is detected in 9 sources (all of these also in H$\alpha$). The two sources that lack H$\beta$ absorption, while showing it in H$\alpha$, have very low H$\beta$ signal-to-noise ratio. The optical depth of the absorbers are degenerate with the emission near line center, but for H$\alpha$ span the range $\tau_0=0.8\pm0.1$ (GN-9771, absorption EW$_0$ of 3.9 {\AA}) to $\tau_0=7.2\pm1.3$ (GN-15498, EW$_0$=6.5 {\AA}), being $\tau_0\approx3.2$ on average. The H$\beta$ optical depth is usually higher, $\tau_0\approx4.5$, being $\tau_0=3.0\pm1.3$ for GN-9771 (EW$_0$=12.2 {\AA}) and $\tau_0=4.5\pm3.0$ for GN-15498 (EW$_0$=12.2 {\AA}), respectively. Since these values are strongly degenerate with other fitted parameters, we refrain from detailed quantitative analysis of these $\tau$'s or the EWs.

The narrow and intermediate components are the most degenerate components in our model. By analysing the BIC differences of the H$\alpha$ fits, we find that fits without narrow component perform better than fits without intermediate component. This is because the intermediate component is more flexible and can replace the narrow emission at systemic. Objects for which a narrow component impacts the fits the least have strong P~Cygni like profiles, such as GN-9771, RUBIES-EGS-42046 and RUBIES-EGS-55604. A narrow component is strongly preferred for the reddest objects with `flattened' top profiles, e.g. RUBIES-EGS-49140 or  {\it the Cliff}. For the bluest objects, such as ALT-69688 and GN-16813, we find that the intermediate components have negligible velocity offsets to the narrow components, causing the strongest degeneracies between the flux in the narrow and the intermediate component. For H$\beta$, BIC differences are generally lower due to the lower signal to noise. However, we find that our H$\beta$ fits consistently yield much higher narrow to intermediate flux ratios (7 times higher than in H$\alpha$ on average), i.e. the intermediate component is significantly less important in the H$\beta$ profiles, and most of the central emission can be explained with a narrow component alone, as illustrated most clearly in GN-9771 in Fig. $\ref{fig:profilefit}$, but see also GS-13971, GN-15498. 

By comparing our model fits without the inclusion of [N{\sc ii}] emission, we find that [N{\sc ii}] (whose profile is tied to the narrow component of H$\alpha$) is detected in four sources (A2744-45924, GS-3073, ALT-69688 and GN-16813), that are among the bluest sources in the sample. The [N{\sc ii}]/H$\alpha$ ratios are challenging to measure due to degeneracies in the narrow and intermediate H$\alpha$ flux, especially for these sources, and span the range 0.01 -- 0.08. No [N{\sc ii}] is detected significantly in any of the red sources, including those with the best quality spectra, showing that this is a real physical effect.

Finally, we explore the use of priors on the relative H$\alpha$/H$\beta$ and [O{\sc iii}]/H$\beta$ ratio of the narrow component. Assuming this component originates from the host galaxy, we adopt the following flat priors: H$\alpha$/H$\beta$=2.5-4.0 and [O{\sc iii}]$_{5008}$/H$\beta$=1-6 corresponding to typical ratios for low mass star-forming galaxies hosting LRDs \citep[e.g.][]{Cameron23,WSun26}. Generally, we find very similar fits with small changes to the fitting statistics. GN-9771's H$\alpha$ emission has a small narrow component, unlike the fiducial fit in Fig. $\ref{fig:profilefit}$, as it breaks a degeneracy with the intermediate emission component. The parameter that is mainly sensitive to these priors are the depths of the absorbers (i.e. $\tau_0$), which tend to increase by a factor 1.4 (2) for H$\alpha$ (H$\beta$). The qualitative result that the H$\alpha$ absorption is weaker than the H$\beta$ absorption would only be strengthened with these priors. The widths of the exponential components tend to be slightly larger with these priors, although this is primarily for sources in the lower S/N regime of our sample.

\begin{figure*}
    \centering
    \includegraphics[width=0.95\linewidth]{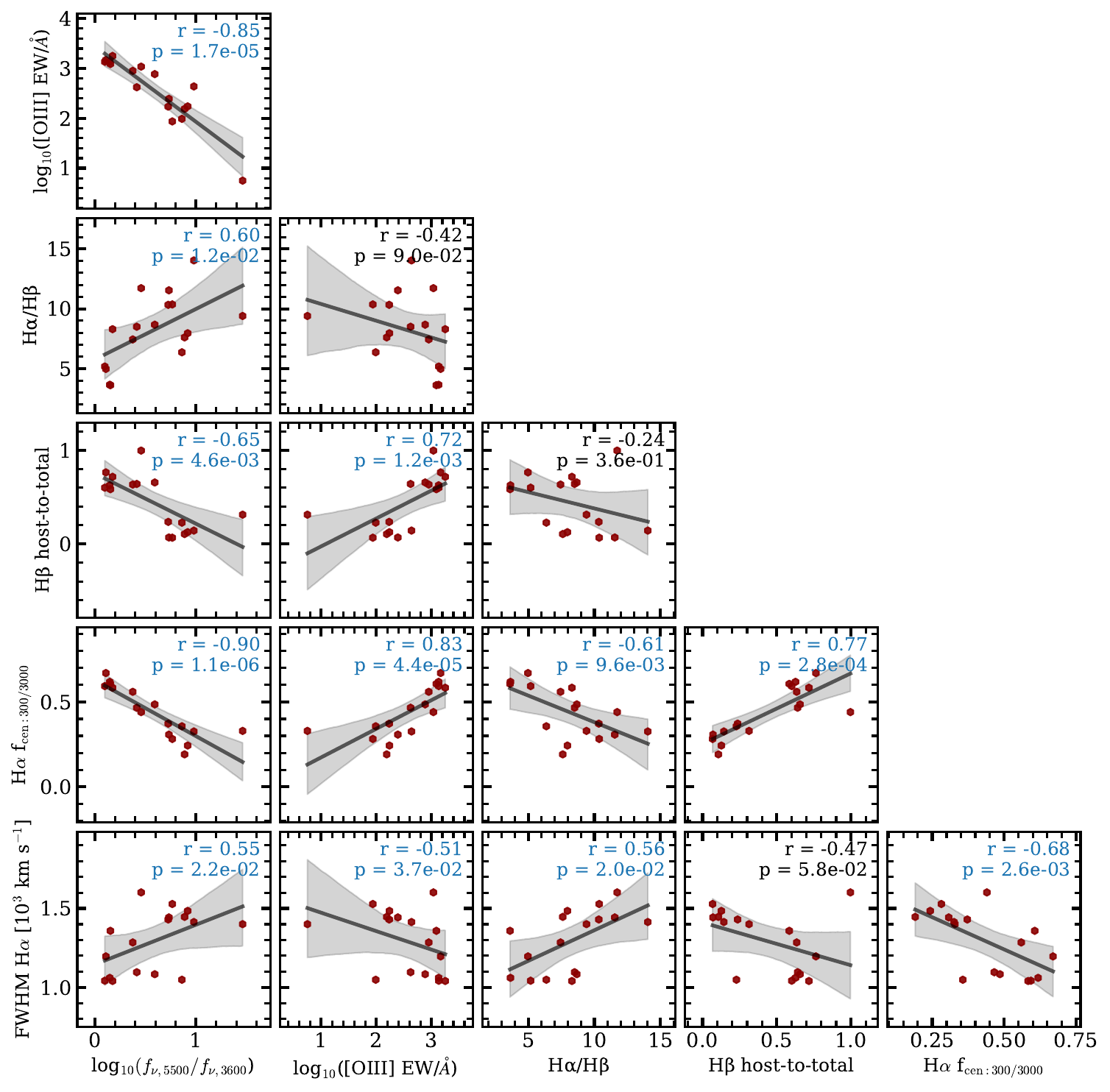}
    \caption{{\bf The correlation structure among various spectral properties and the H$\alpha$ and H$\beta$ line-profiles.} In each panel, we list the Spearman's rank correlation coefficient and associated p-value, highlighting significant correlations in blue. To guide the eye, we show linear correlations and their 95 \% confidence interval. The Balmer line profiles (quantified either in a model-independent way as the central H$\alpha$ fraction, or as the narrow-to-total flux ratio in the H$\beta$ line from our modeling) correlate most significantly with the UV to optical colors, the [O{\sc iii}] EW and the H$\alpha$/H$\beta$ ratio, such that objects with more centrally dominated Balmer emission have higher [O{\sc iii}] EWs, Balmer line ratios that are closer to case B and bluer SEDs.  }
    \label{fig:correlations_ratios}
\end{figure*}

\subsection{The overall correlation structure}\label{sec:definitions}
Here we explore the correlations illustrated in Fig. $\ref{fig:prism_sample}$ that we quantify with the empirically motivated properties as well as results from the fitting in Section $\ref{sec:profiles}$:

\begin{itemize}
    \item {\bf The UV to optical color:} As a measure of the broad-band color of the spectrum, we use the (continuum) flux ratio over a wide wavelength interval, $f_{\nu, 5500}/f_{\nu, 3600}$. The longer wavelength  range ($\lambda=5400-5700$ {\AA}) is chosen as this is free from strong features, whereas the shorter wavelength range ($\lambda=3500-3650$ {\AA}) is chosen to be just bluewards of the Balmer break wavelength. We note that the rank ordering with definitions of the Balmer break strength (e.g. in \citealt{deGraaff25}) are consistent with our measure. While we opt for an agnostic definition, we note that the UV to optical color mainly traces the Balmer break strength, which is sensitive to the H{\sc i}$_{n=2}$ column density around the central powering source \citep{Inayoshi24,Ji25}, but also to the contribution of the UV luminosity of a possible young, star-forming host galaxy \citep[e.g.][]{Naidu25}.
    
    \item {\bf [O{\sc iii}] EW:} With a critical density of $n_{\rm crit}\sim7\times10^5$ cm$^{-3}$, the strong [O{\sc iii}] doublet mainly traces emission regions with relatively low gas densities, such as those present in the ISM of host galaxies. This is further supported by its (very) narrow line-width and the absence of broad wings in the vast majority of objects. Given that the rest-frame optical continuum luminosity likely probed the luminosity from the compact engine, the [O{\sc iii}] EW could trace the relative emission originating from the engine and the host galaxy \citep{WSun26}.
    
    \item {\bf H$\alpha$/H$\beta$ ratio}. The Balmer decrement is usually associated with the dust attenuation. The H$\alpha$/H$\beta$ values in our sample are much higher (median: 7.8, up to 14.0, using the total line-flux) than the unattenuated value of $\approx2.8$ or 3.1 for galaxies and AGNs, respectively. This is similar to earlier stacking work on LRDs \citep[e.g.][]{Brooks24}. In addition to dust attenuation, the H$\alpha$/H$\beta$ ratio is sensitive to effects of collisional excitation \citep{Berg25,Torralba25b} and resonant scattering \citep{Chang26}, both present at high gas densities. 

    \item {\bf H$\beta$ narrow-to-total ratio}. Based on our emission-line modeling, we define the narrow-to-total ratio as the flux ratio of the narrow component to the total emission-line flux. We use the H$\beta$ line rather than H$\alpha$ as we find that these fits separate the narrow component with less degeneracies with the intermediate emission component.
    
    \item {\bf Central H$\alpha$ fraction, f$_{\rm cen: 300, 3000}$}. In addition to our model sensitive narrow-to-total ratio, we also define this quantity as an an empirical measure of the compactness of the line-profile. It is simply defined as the ratio of the integrated H$\alpha$ flux within $\pm300$ km s$^{-1}$ from the systemic, divided by that measured within $3000$ km s$^{-1}$. These windows were chosen to be insensitive to differences in grating resolutions and capture the majority of the total emission.

    \item {\bf The H$\alpha$ exponential FWHM}. Finally, we also include the line-width of the exponential component of our H$\alpha$ fit. If interpreted as being set by electron scattering, the width correlates with the effective temperature of the electrons, which is primarily sensitive to the temperature of the scattering medium in the $\tau_e\lesssim1$ regime, but also to the turbulence \citep{Chang26}.
\end{itemize}

In Fig. $\ref{fig:correlations_ratios}$ we show how the quantities defined above correlate with each other. The strongest correlation is found between the UV to optical color and the central H$\alpha$ fraction, with redder sources showing more wing-dominated H$\alpha$ emission. Further, we find that the UV to optical color correlates positively with the EW of the H$\alpha$ absorber. The H$\beta$ narrow-to-total ratio similarly correlates with the UV to optical color, but shows somewhat larger scatter, especially around the intermediate colors. 

Similar to the analysis by \cite{deGraaff25b}, we find that the [O{\sc iii}] EW decreases for increasingly red sources, which is stronger than the anti-correlation that the color has with [O{\sc iii}] luminosity ($r=-0.66$). This is probably because there is relatively more optical continuum emission emerging from the engine in redder sources, which lowers the EW \citep[e.g.][]{WSun26}. 

The H$\alpha$/H$\beta$ ratio of the total lines tends to be higher for redder sources \citep[similar to results on a larger samples from][]{Barro25,PerezGonzalez26}, but it does not correlate significantly with the [O{\sc iii}] EW. This suggests that the H$\alpha$/H$\beta$ ratio traces additional physics beyond the correlation that the [O{\sc iii}] EWs are lower in redder sources. A hint of this physics could be in the correlation between the H$\alpha$/H$\beta$ ratio with the central H$\alpha$ fraction, whereas its correlation with the H$\beta$ narrow-to-total ratio is weaker. As shown in \cite{Chang26}, in gas with a high column density of H{\sc i} in the n=2 state, a significant fraction of resonantly scattered H$\beta$ photons may branch into H$\alpha$ and Pa$\alpha$ emission. This would give rise to additional H$\alpha$ emission, relatively near the line-center (this could be interpreted as the intermediate components in our modeling), and would contribute to the high H$\alpha$/H$\beta$ ratios.

The width of the exponential wing of the H$\alpha$ line correlates with most properties discussed above, albeit with modest Spearman's rank correlation coefficients $|r|\approx0.5$. Broader wings are found for redder systems with stronger Balmer breaks \citep[see also][]{Sneppen26} and for sources with higher H$\alpha$/H$\beta$ ratios. The width of the exponential wings is determined by the effective temperature of the electrons if $\tau_e \lesssim 1$, and the broadening is enhanced by multiple scatterings in the $\tau > 1$ regime. The Balmer break strength instead correlates with the H{\sc i}$_{\rm n=2}$ column density. Therefore, this correlation suggests that the gas is partially ionised and that resonant and electron scattering zones are mixed. 

We note that the central H$\alpha$ fraction f$_{\rm cen: 300, 3000}$ anti-correlates strongly with the optical depth to Thomson scattering $\tau_e$ (see Eq.~\ref{eq:ha_model}) for the sources without Balmer absorption, where it ranges from $\tau_e \approx1 - 2.5$ (typically 1.2). For sources with absorption, $\tau_e$ is degenerate with the absorption depth, with values $\tau_e \approx1-5$ (typically 2.7) in our fiducial model.

\section{Physical Interpretation} \label{sec:interpretation}
In this section, we interpret the spectral shapes of the Balmer and [O{\sc iii}] lines, as well as their correlations with the broad-band spectra of LRDs. We will argue that these can be explained as caused by dense winds, with varying optical depths, around compact luminous engines that reside in low-mass host galaxies.

\subsection{Dense, flowing envelopes}  \label{sec:densewind} 

Empirically, Balmer line-shapes that are characterized by narrow central emission and absorption and exponential wings have been observed in the early weeks of Type IIn supernovae \citep[e.g.][]{Xu92,Chugai01,Dessart09,Smith10} and the `pseudo-envelope' phase of outbursts in luminous blue variable stars (LBVs; e.g. \citealt{Leitherer85,Humphreys94,Stahl01,Mehner13}), like P~Cygni. The main differences (apart from not being transients) are that LRDs are significantly more luminous (three orders of magnitude brighter than LBVs, e.g. \citealt{Jiang18}, comparable only to the brightest SNe; \citealt{Hiramatsu24}) and that LRDs show (strong) emission lines from forbidden lines as [Ne{\sc iii}] and [O{\sc iii}]. Both in SNe and in LBVs, the observed line profiles are associated with an optically thick envelope of dense gas around a source of energy (a supernova explosion or a massive star, respectively). Narrow Balmer emission is produced in these (slow) winds, where out-flowing optically thick gas causes the absorption features, whereas Thomson scattering on free electrons leads to exponential wings. 

Based on these empirical analogies and photo-ionization and line profile models of LRDs \citep{Torralba25b,Chang26}, we interpret the trends in a physical picture where a central, compact engine is surrounded by an envelope of dense partially ionised gas. In this picture, we can interpret the trends between the variation in the centers of the line profiles and UV to optical colors (probing the Balmer break) as highlighted in Fig. $\ref{fig:prism_sample}$ as primarily tracing the variation in the H{\sc i}$_{n=2}$ column density of the gas across our sample \citep[e.g.][]{Inayoshi24,Ji25}. A secondary, but not negligible effect is that variations in the contribution of host galaxy light also impact the UV to optical colors (via blue continuum emission) and the central parts of the line profiles (via narrow emission-lines from H{\sc ii} regions). Indeed, the [O{\sc iii}] luminosity -- that is primarily attributed to these regions -- correlates strongly with the UV luminosity ($r=0.77$), and this host emission may add scatter to correlations between the spectral properties and emission from the central engine.

\begin{figure}
    \centering
  \includegraphics[height=6.3cm]{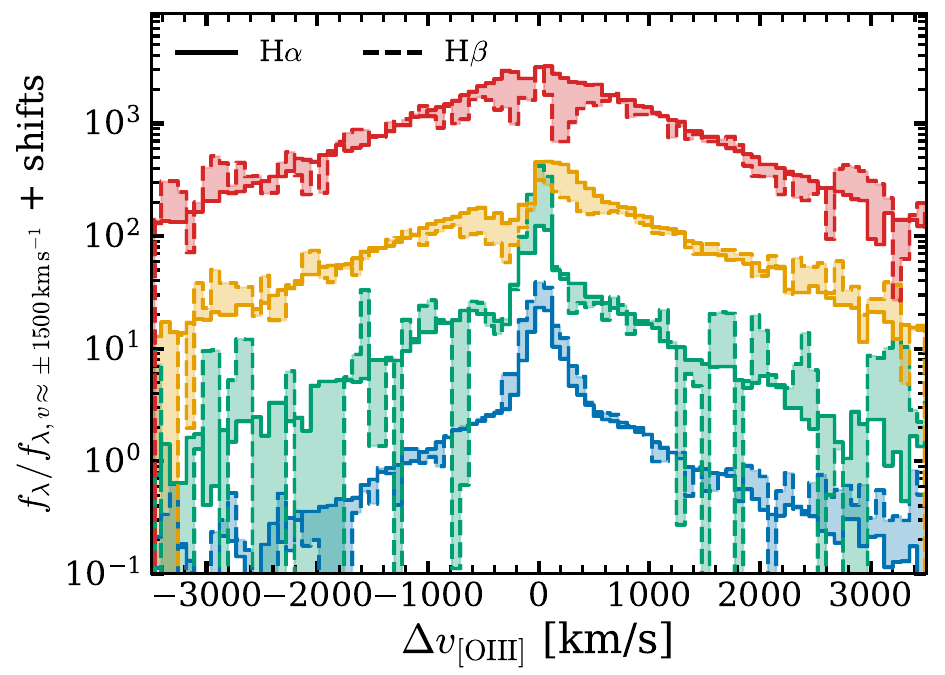} 
    \caption{{\bf The small differences in exponential wings of H$\alpha$ and H$\beta$ across UV to optical colors.} We show the continuum-subtracted, median stacked H$\alpha$ (solid) and H$\beta$ (dashed) line profiles, normalised at the average flux at $\pm1500$ km s$^{-1}$, i.e. where the exponential wings are dominant. Objects are binned by their UV to optical color as in Figure $\ref{fig:prism_sample}$. Bins are shifted vertically for clarity. Shaded regions highlight differences in the H$\alpha$ and H$\beta$ line profiles.   } 
    \label{fig:fwhm_exp}
\end{figure}

\subsection{Exponential wings in LRDs} \label{sec:expwings}
Figures $\ref{fig:prism_sample}$ and $\ref{fig:overview}$ show that exponential wings in the H$\alpha$ lines are ubiquitous in the sample (see also Appendix $\ref{app:fits}$), which we interpret as being due to electron scattering \citep[e.g.][]{Rusakov25}. In our fitting of the H$\alpha$ profiles, the FWHMs of the exponential components span the range $\sim1040-1600$ km s$^{-1}$ (Table $\ref{tab:fit_averages}$). We further identify marginal to weak asymmetries in the exponential wings, with the exception of some of the bluest sources that generally show a steeper blue wing. This indicates that the blue sources have outflowing scattering media \citep{Dessart09,Sneppen26}. The optical depth to electron scattering is challenging to measure in the majority of systems that show absorption near line centers, where resonant scattering effects make it challenging to estimate the unscattered component, but our modeling results indicate $\tau_e\approx1-5$. Nevertheless, it is qualitatively clear that the exponential components become increasingly more dominant in the redder objects. This suggests that redder LRDs have a higher optical depth to Thomson scattering $\tau_{e}$, hinting at a correlation between the column density of free electrons (setting $\tau_{e}$) and that of neutral hydrogen in the n=2 state (N$_{\rm HI, n=2}$), which sets the resonant scattering likelihood and Balmer break strength as well as the absorption EW. Such correlation could naturally arise in a partially ionised medium. Beyond $\tau_{e}$, redder sources show the flattest wings (see the bottom row in Fig. $\ref{fig:correlations_ratios}$), suggesting that they have a higher effective electron temperature or a higher Compton depth, or both.

\begin{figure*}
\begin{tabular}{ccc}
    \centering
    \includegraphics[width=0.31\linewidth]{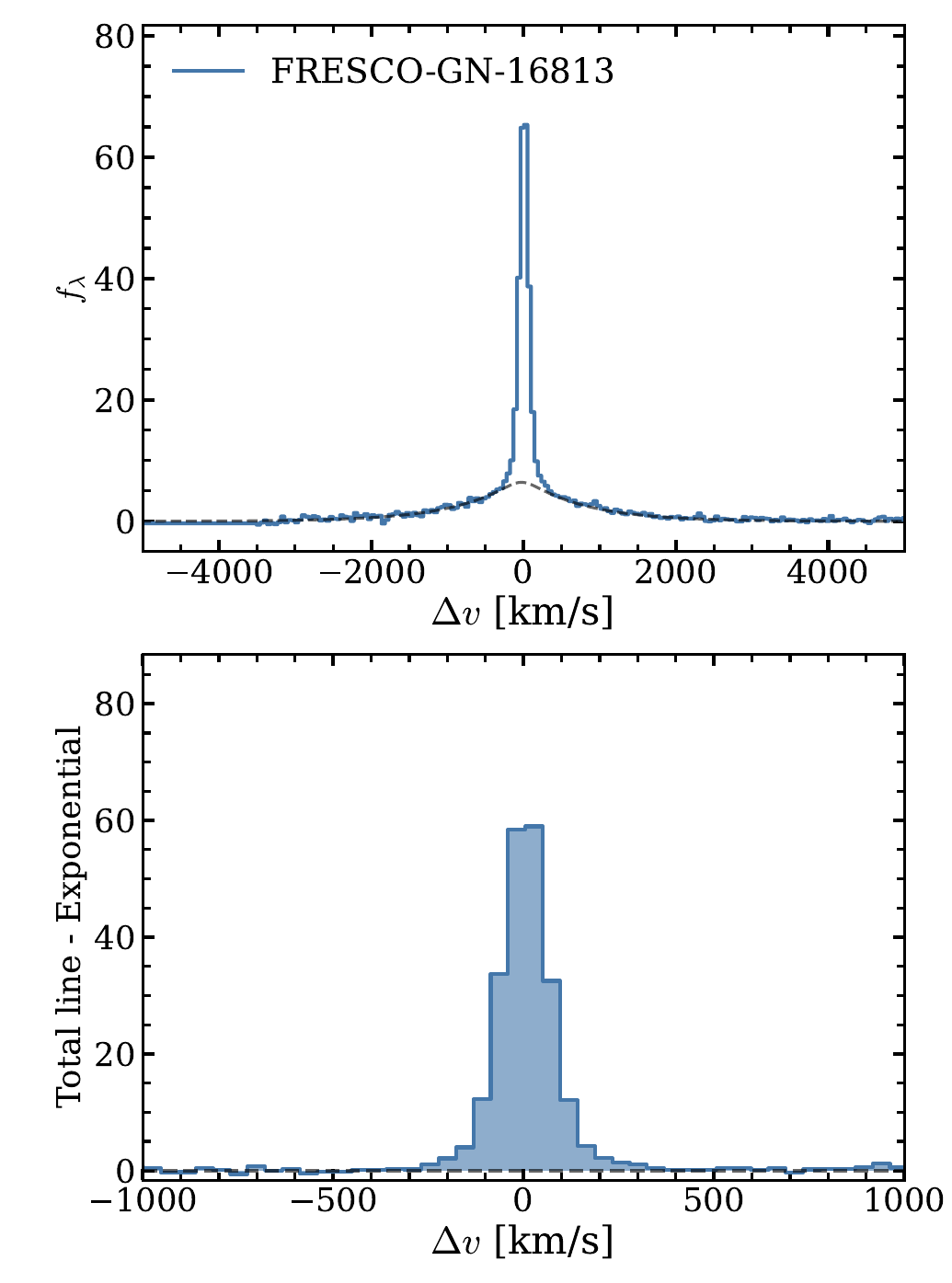} & 
    \includegraphics[width=0.31\linewidth]{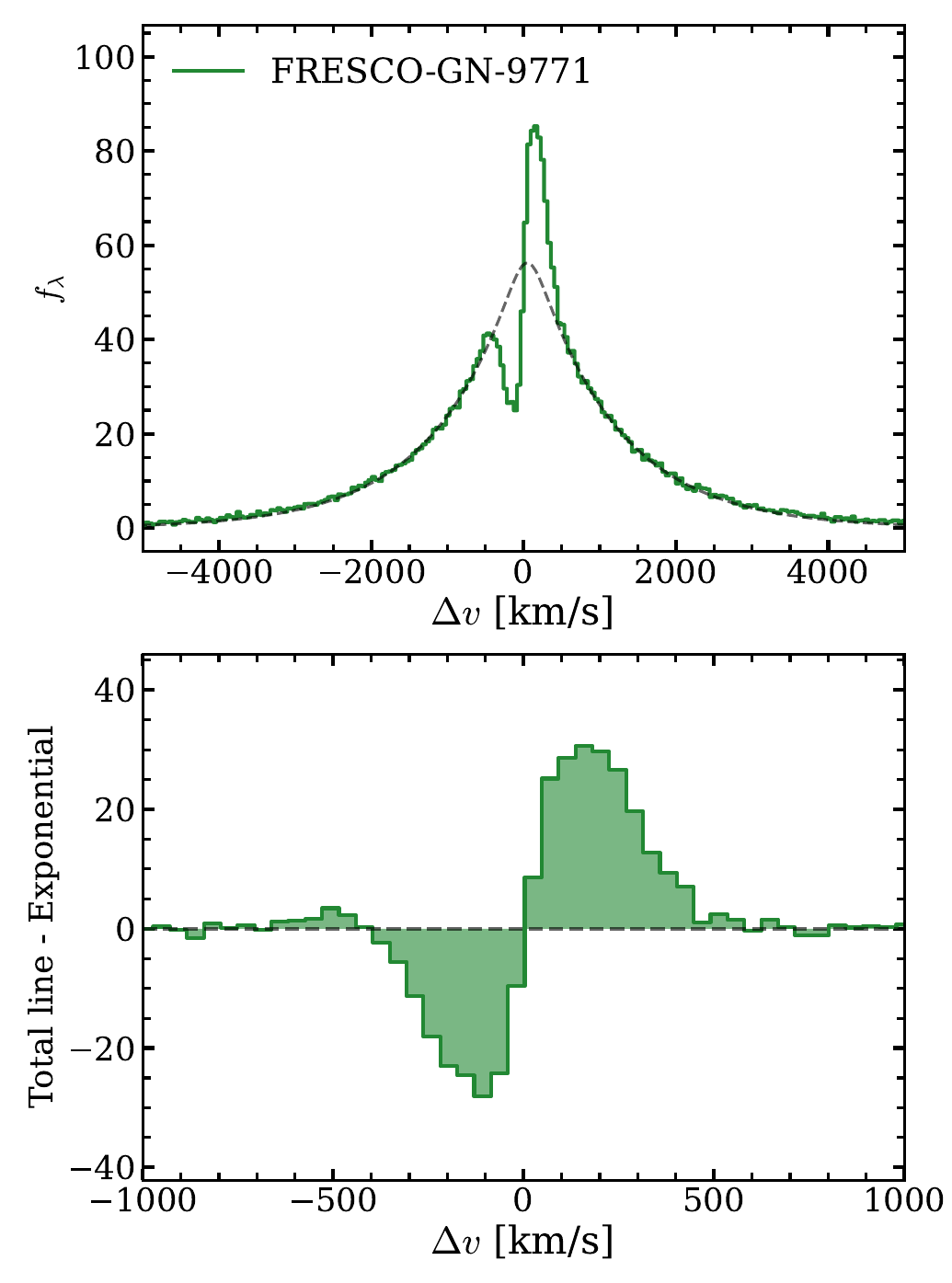} & 
    \includegraphics[width=0.31\linewidth]{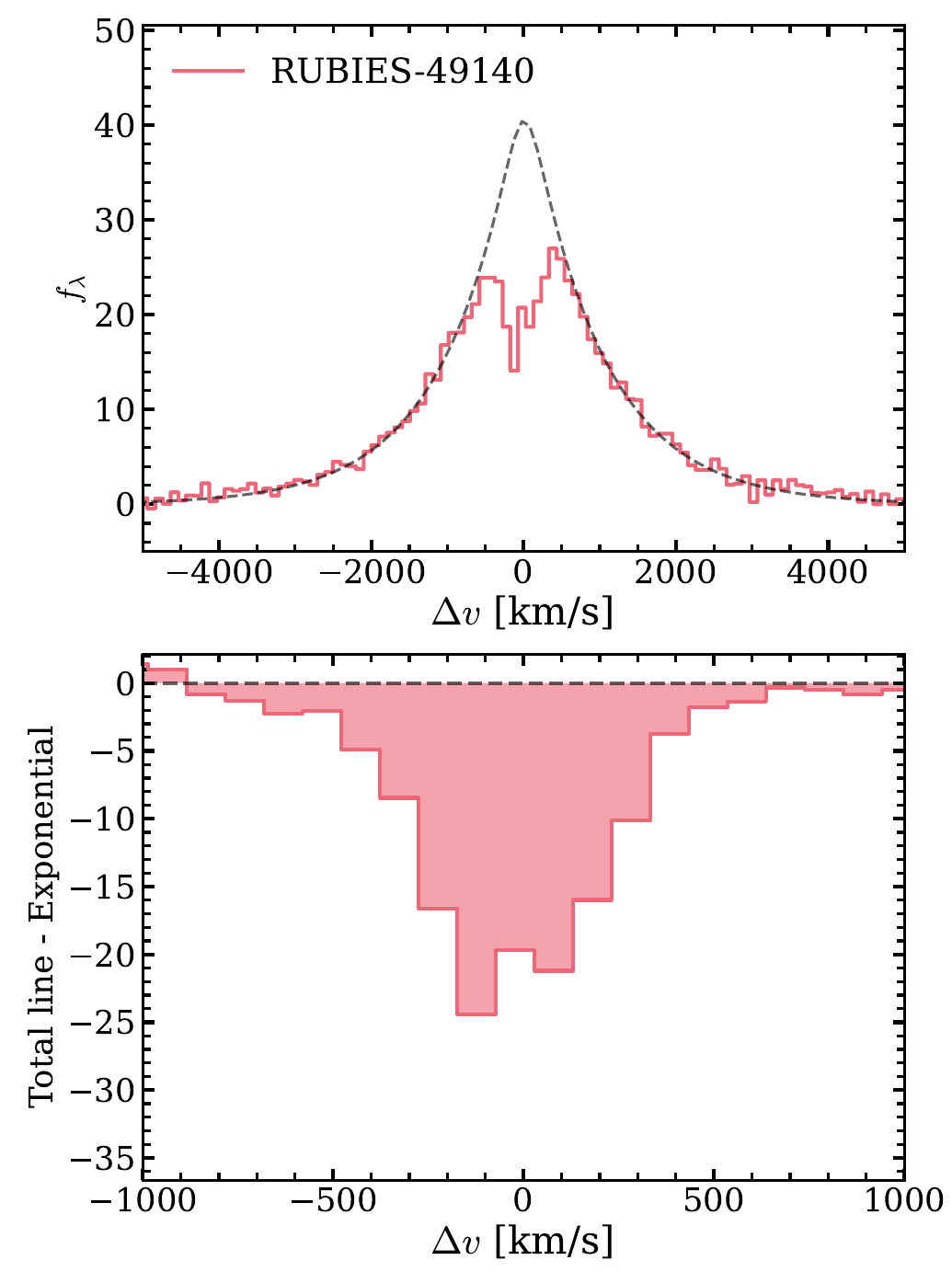} \\    
\end{tabular}
    \caption{{\bf Illustration of the H$\alpha$ ``core profiles'' of three sources with increasingly red UV to optical colors.} We define the ``core profile" as the total emission-line minus the exponential wing from our best-fit model. The three objects shown have increasingly red UV to optical color, highlighting how blue sources have centrally dominant narrow emission, whereas redder sources show P~Cygni profiles and the reddest sources show central Balmer absorption. }
    \label{fig:ProfileIllust}
\end{figure*} 

The relative widths of the wings of H$\alpha$ and H$\beta$ can constrain whether the scattering medium is outside the region that produces the emission-lines, or whether it is mixed with it \citep[see][for such an analysis]{Brazzini25}. If the scattering medium is outside the line emitting region, the exponential wings should have the same FWHM as $\tau_{e}$ is wavelength independent. In Fig. $\ref{fig:fwhm_exp}$ we show comparisons between the stacked H$\alpha$ and H$\beta$ line-profiles in the various subsets. The wings for H$\alpha$ are always best-fitted with exponentials (with very significant BIC differences $\Delta$BIC$>10$), but for H$\beta$ it remains challenging to distinguish exponentials from gaussian, even for stacks, especially considering the possible contribution of [Fe{\sc ii}] \citep[e.g.][]{Torralba25b}. Most of the differences between the H$\alpha$ and H$\beta$ profiles are near the line-center. For the bluest sources, H$\beta$ is stronger near the line-center, whereas in the reddest sources H$\beta$ shows a flattened central line profile compared to H$\alpha$. Resonant scattering of the Balmer lines, and H$\beta$ to H$\alpha$ branching decay leads to differences in the cores of the different line-profiles, but which can extend over several hundreds of km s$^{-1}$ \citep{Chang26}. Without a model that is flexible enough to capture these differences, we caution that the wings of H$\beta$ lines may be biased towards flatter values (i.e. higher FWHMs). At velocities beyond $\sim1000$ km s$^{-1}$ from line center, we find that the wings of the two lines have very similar slopes across our subsets.

\begin{figure}
    \centering
    \includegraphics[width=0.99\linewidth]{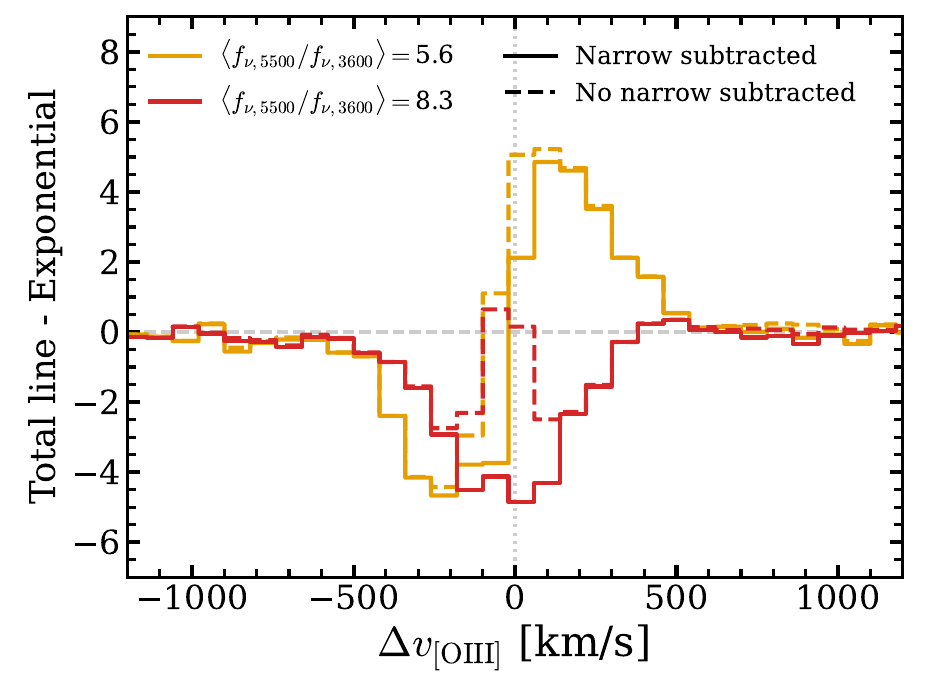}
    \caption{{\bf The impact of narrow emission on the core profiles.} We show the core profiles (as in Fig. $\ref{fig:ProfileIllust}$) of the median-stacked H$\alpha$ lines in the two reddest bins of UV to optical color in orange and red, respectively. In the solid lines we further also subtract the narrow emission component, whereas it is not subtracted for the dashed lines. The narrow emission (whose shape is fixed to the shape of the [O{\sc iii}] line) is clearly observed as a central peak in the H$\alpha$ profile of the reddest stack. Modeling the narrow component is also required to better characterize the P~Cygni shapes of intermediately red LRDs. }
    \label{fig:host}
\end{figure}

\subsection{Central emission: the contribution from the host galaxy} \label{sec:core_is_needed}
Given that all LRDs in our sample show narrow [O{\sc iii}]$_{4960,5008}$ line emission, it is expected that the Balmer line profiles contain a contribution from line emission originating from similar regions, that have a relatively typical ISM electron density to avoid collisional de-excitation of [O{\sc iii}]$_{4960,5008}$. This is therefore likely emission from the host galaxy \citep[e.g.][]{Barro25,WSun26}. Because we now mainly focus on variations in the central parts of the line, we define the ``core profile'' as the residual line profile after subtracting the contribution of our modeled exponential component. Figure $\ref{fig:ProfileIllust}$ illustrates this definition for three objects spanning the range of line profiles and UV to optical colors, ranging from narrow emission in the core, to P~Cygni and absorption-dominated profiles for increasingly red sources.

In the bluest sources, where the line centers are dominated by relatively narrow emission with no evidence for (strong) Balmer absorption, the flux in the narrow component is highly degenerate with the flux in the intermediate component (see Section $\ref{sec:degeneracies}$ for a detailed discussion of the degeneracies in our modeling). The central parts of the Balmer lines have a similar profile as [O{\sc iii}] (see Figs. $\ref{fig:prism_sample}$, $\ref{fig:overview}$), which suggests a common origin. While the [O{\sc iii}] EWs of the bluest broad-line sources are on average high ($\sim1500$ {\AA}), this is still well within the distribution of [O{\sc iii}] EW values found in high-redshift galaxies \citep[e.g.][]{Endsley23EW,Matthee23}. The [O{\sc iii}]/H$\beta$ values of the total line-profiles of $\approx5-7$ are also typical for galaxies at these redshifts \citep[e.g.][]{Kotiwale26}. These comparisons suggest that the [O{\sc iii}] lines in the blue broad-line emitters could plausibly originate from host galaxy emission. However, in some individual systems (e.g. GS-3073, ALT-69688, GN-16813), we note that faint broad (FWHM$\approx600$ km s$^{-1}$) components are visible in [O{\sc iii}] (accounting for $\sim7$ \% of the total flux), suggesting at least part of the [O{\sc iii}] emission in these blue systems originates from a wind, possibly powered by the engine or by the host galaxy.

\begin{figure*}
    \centering
\begin{tabular}{cc}
  \includegraphics[height=6.3cm]{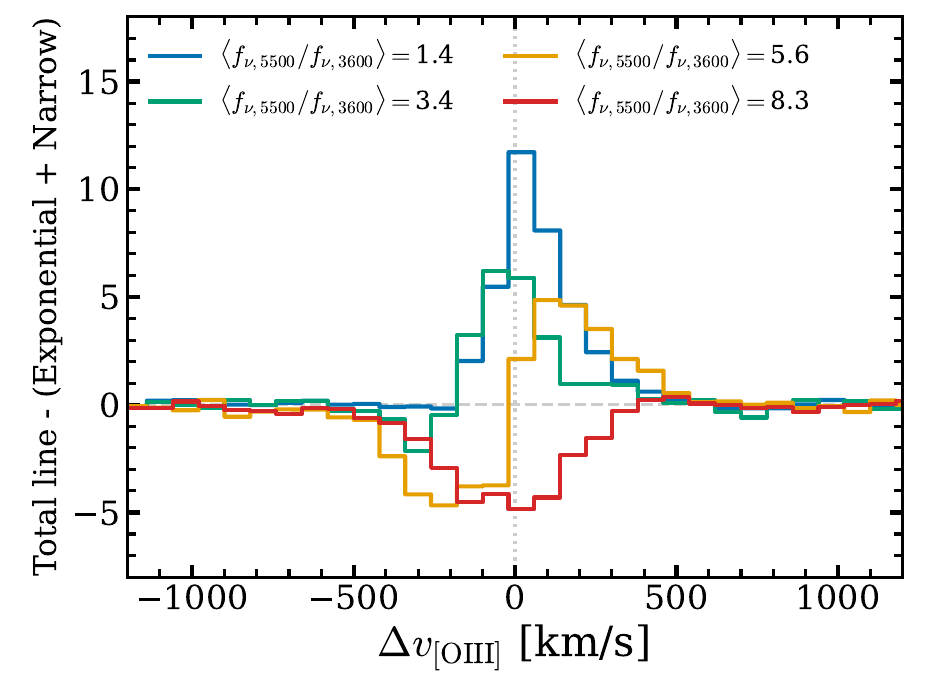}    &  \includegraphics[height=6.3cm]{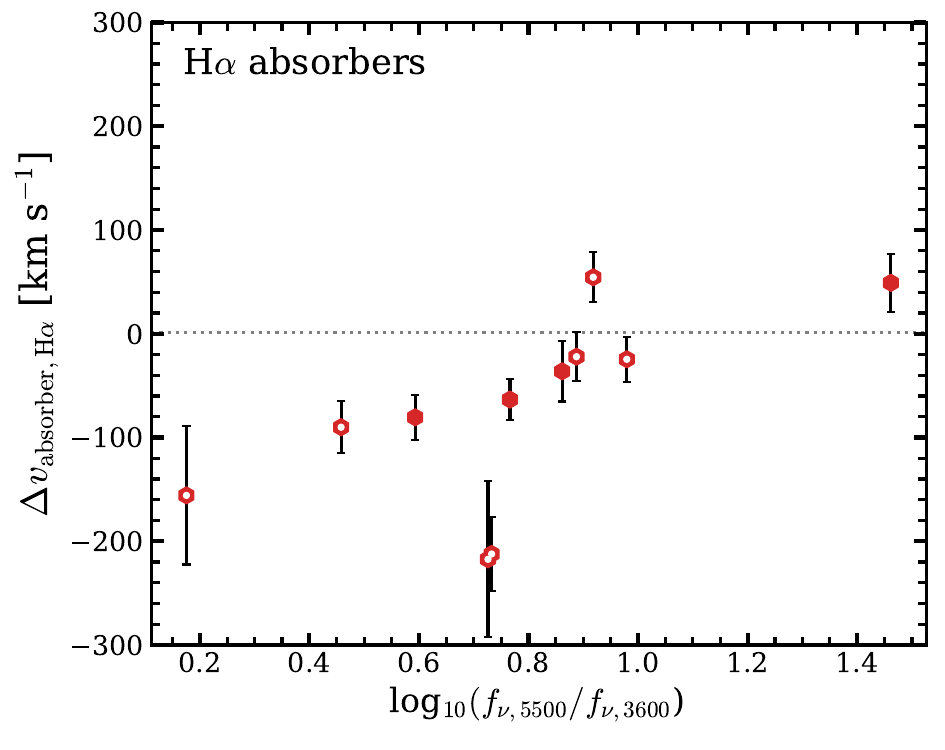}  \\
\end{tabular}
    \caption{{\bf The velocity of the H$\alpha$ absorber correlates with the UV to optical color, implying the winds' optical depth correlates with its velocity.} In the left panel, we show median stacked core profiles (see Fig. $\ref{fig:ProfileIllust}$) in bins of UV to optical color that primarily trace the Balmer brake strength. Line-profiles are normalised to total luminosity before stacking. In the right panel we show the central velocity of the H$\alpha$ absorbers in individual spectra with significantly detected H$\alpha$ absorption as a function of the UV to optical color. Sources with M-grating spectra are marked with an additional white dot. Both panels show a consistent trend that the H$\alpha$ absorption is increasingly less blue-shifted with redder UV to optical colors. The reddest sources have absorption that is redshifted with respect to the systemic redshift. The reddest sources do not show a central emission component, whereas intermediate sources show P~Cygni features. We note the two outliers in the right panel have M-grating spectra that challenge the identification of the absorption minimum.  }
    \label{fig:Hacores}
\end{figure*} 

In the red sources with P~Cygni or absorption-dominated core profiles (such as GS-13971 or GN-15498, respectively), narrow Balmer line emission can be distinguished near the line core, with a distinctively different velocity center and width compared to the intermediate component. In some cases with IFU data, the [O{\sc iii}] and narrow H$\alpha$ lines are also seen to be spatially extended (Ishikawa et al. in prep). These narrow central components are more clearly seen in H$\beta$ emission, because the Balmer decrement of the narrow emission is $\sim3$, whereas it is much larger for the intermediate component and the wings that are impacted by radiative transfer effects. Likewise, as shown in \cite{Deugenio2025-irony,Torralba25b}, narrow emission is relatively much stronger than the broad emission in the H$\gamma$ line. In Fig. $\ref{fig:host}$ we show how the contribution from the narrow emission -- here plausibly arising from the host galaxy -- impacts the cores of the H$\alpha$ profiles, with relatively stronger degeneracies in the P~Cygni dominated sources than in absorption-dominated sources. This figure clearly illustrates that ignoring the contribution of host galaxy H$\alpha$ emission impacts the absorption profiles, potentially causing over-estimated velocity blue-shifts, or the misinterpretation of a broad, central absorber as multiple absorption components.

In the reddest sources with absorption-dominated line cores, the deep absorption partially acts as a natural coronagraph of the strong emission of the powering engine, enabling (especially with H-grating spectra) a relatively clean view of the host's H$\alpha$ emission and therefore an estimate of the star formation rate \citep[e.g.][]{Kramarenko25}. Two of such sources are GN-15498 and GN-9771, for which we find star formation rates ranging from $\approx1$ to 6 M$_{\odot}$ yr$^{-1}$. If these galaxies have typical star formation rates for their stellar mass, these values correspond to $M_\star \approx(2-10)\times10^8$ M$_{\odot}$ \citep{diCesare25} that are consistent with SED decompositions \citep{WSun26} and the clustering strength of LRDs \citep{Matthee25clustering,Lin25}.

\subsection{Absorption velocities: a sequence of HI column densities}  \label{sec:core_flows}
Absorption features in the Balmer lines of LRDs have been interpreted as being due to dense gas clouds around the central engine \citep{Matthee24,Deugenio25,Lin25_Lowz}. Whether the absorber velocities and optical depths are somewhat randomly distributed or systematically vary across the sample and with the properties of the LRDs is yet to be explored in detail. Radiative transfer models \citep{Chang26} show that the resonant scattering of Balmer photons in a gas with a high H{\sc i}$_{n=2}$ column density leads to double peaked Balmer profiles, with the minimum tracing the outflow velocity (or inflow, in case the absorption is redshifted), which is typically on the order $\sim100$ km s$^{-1}$. The column density sets the strength of the absorber and is in the range N$_{\rm HI, n=2} \sim10^{15-16}$ cm$^{-2}$. Another effect is that a fraction of H$\beta$ photons are converted to H$\alpha$ and Pa$\alpha$ during the resonant scattering process, which leads to an increasing H$\alpha$/H$\beta$ ratio and differences in the central parts of the Balmer line profiles -- both as observed in our sample, spanning a range in H{\sc i} column density. As illustrated in Fig. $\ref{fig:prism_sample}$, it is primarily the central part of the Balmer line profiles that changes systematically with increasingly red colors, i.e. increasingly strong Balmer breaks. Thus, the spectral variations of broad H$\alpha$ line emitters that are blue to red from UV to optical can be explained by the sources probing a sequence of increasing column density of their dense envelopes.

The velocity of the absorber, as traced by the minimum in the core line profiles, appears to correlate with the UV to optical color, as hinted by the stacked spectra in Fig. $\ref{fig:prism_sample}$. In the left panel of Figure $\ref{fig:Hacores}$, we show median stacked core profiles in bins of UV to optical color, while the right panel shows the velocities of the modeled absorbers for the subset of sources with significantly detected H$\alpha$ absorption. The core H$\alpha$ profiles of the bluest sources in our sample are characterized by a single, relatively narrow peak, which we interpret as the envelope being optically thin (for Balmer photons) in these sources. The core profiles of redder objects tend to show P~Cygni features that become increasingly stronger with increasingly red colors. These profiles are suggestive of outflowing, optically thick winds with velocities $\approx100-200$ km s$^{-1}$. We note that iron (Fe{\sc ii}) absorption has been tentatively detected in LRDs \citep{Torralba25b,Deugenio2025-irony,PerezGonzalez26} potentially providing a significant source of opacity to drive these winds, especially in the outer layers of the envelope that are shielded of ionizing photons by the inner envelope. The reddest sources have core profiles that are characterized by absorption very close to the systemic redshift. These envelopes are the most optically thick and lack the P~Cygni profiles typical of less red objects (also reflected in weak intermediate components in our fits). From the model fits (right panel of Fig. $\ref{fig:Hacores}$), we unveil a strong correlation between the absorption velocity and the UV to optical color. \footnote{We note that the two outliers to the trend (RUBIES-EGS-55604 and RUBIES-EGS-42046, see Fig. $\ref{fig:overview}$) have M-Grating spectra and P~Cygni profiles that have the strongest degeneracies between the intermediate and narrow emission components, as the intrinsic width of the narrow component is less clearly defined. In H$\beta$, where we also detect absorption in these two sources, the absorbers are less blue-shifted, with $\Delta v_{\rm absorber, H\beta}\approx-30$ km s$^{-1}$, consistent with the trend in Fig. $\ref{fig:Hacores}$.} While the majority of absorber velocities are blue-shifted, indicative of outflowing envelopes, the absorption velocities in the reddest sources (A2744-45924 and {\it the Cliff}, see also J1148-18404 in \citealt{Matthee24}) are redshifted, suggesting that the densest envelopes are inflowing. 

A possible physical interpretation of the correlation between the absorber velocity and the UV to optical color (which traces the Balmer break strength and therefore H{\sc i}$_{n=2}$ column density, but is also a proxy for the effective temperature of  optically thick medium) is that we observe the reddest objects along the axis of accretion \citep[reminiscent of the quasar studied in][]{Zhou19}, whereas most of the solid angle is occupied by outflowing gas (similar to spectra emerging from stellar winds; \citealt[e.g.][]{Knigge95,Tampo24}, but with a somewhat lower column density). Alternatively, LRDs may evolve through different phases where their outer envelopes expand or contract with associated differences in column density. We discuss these scenarios further in Section $\ref{sec:discuss_variation}$.

\subsection{The differences in H$\alpha$ and H$\beta$ absorption} \label{sec:core_differenceHaHb}

As reported in various studies on high-quality LRD spectra \citep{Lin25_Lowz,Torralba25b,Deugenio2025-irony,Lambrides25}, the H$\alpha$ absorption usually shows a higher transmission than the H$\beta$ absorption. This is in conflict with the theoretical cross sections for the H$\alpha$ and H$\beta$ lines, which suggests H$\alpha$ absorption should be $\approx5$ stronger than the H$\beta$ absorber. Moreover, some studies report differences in the absorption velocities of different transitions (i.e. blue-shifted absorption in H$\alpha$ but red-shifted absorption for all other transitions in RUBIES-EGS-49140 reported in \citealt{Deugenio2025-irony}), which is at odds with a single absorption component (although this could be attributed to temperature and/or density gradients that lead to slight differences in the emission regions of the lines as has been seen in Type IIn supernovae; e.g. \citealt{Dessart09}). As can be seen by inspecting Fig. $\ref{fig:overview}$, the relatively stronger H$\beta$ absorption can also be noticed in various objects in our sample. Differences in absorption velocities have also been noticed in the high resolution spectrum of GN-9771, where H$\alpha$ appears more blue-shifted \citep[see also][]{Torralba25b}.

\begin{figure}
    \centering
    \includegraphics[width=0.99\linewidth]{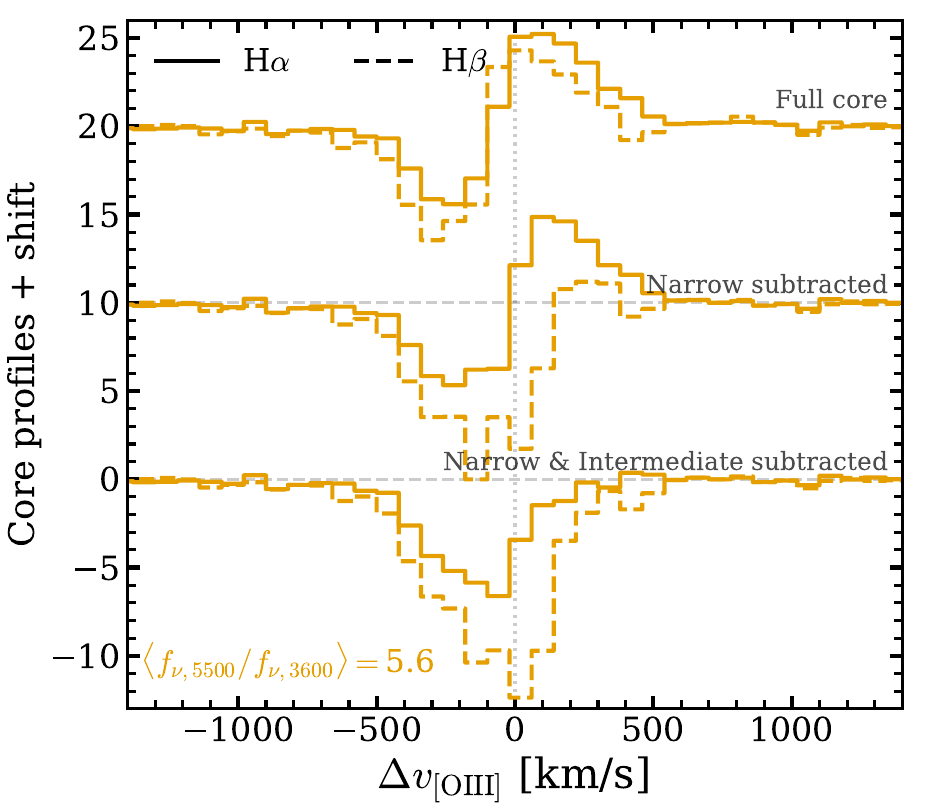} \\ 
        \includegraphics[width=0.99\linewidth]{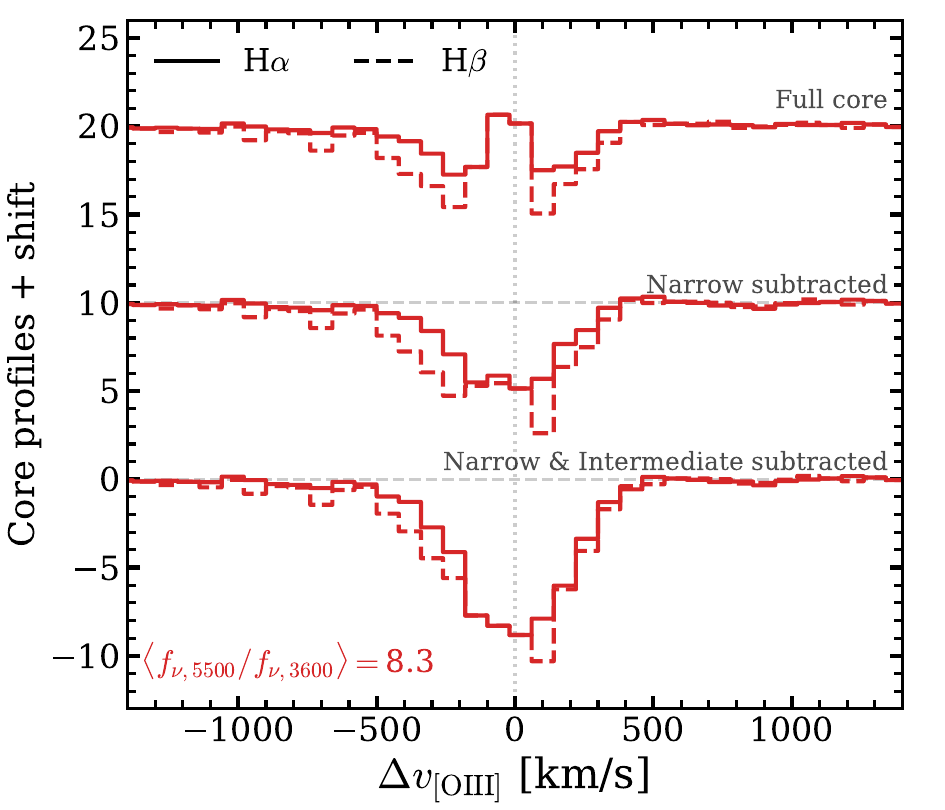} \\
    \caption{\textbf{The difficulties in measuring and comparing Balmer absorption profiles.} Median stacked core profiles of H$\alpha$ and H$\beta$ in the two bins with reddest UV to optical colors. We show the spectra where we only subtracted the exponential component (shifted to top), the spectra where we also subtracted the best-fit narrow component (middle shifts) and also the intermediate component (bottom shifts). As spectra were normalised to the H$\alpha$ and respective H$\beta$ luminosities before stacking, we caution against interpreting the depth of the absorption.}
    \label{fig:Hbcores}
\end{figure}

As discussed in Section $\ref{sec:degeneracies}$, properties of the absorbers are challenging to measure robustly due to degeneracies with other components near the line core. Figure $\ref{fig:Hbcores}$ illustrates the degeneracies of the shape of the absorption with emission components, both for H$\alpha$ and for H$\beta$, which were fit independently. In the reddest sources, the H$\alpha$ and H$\beta$ profiles show central narrow emission that we attribute to the host galaxy. Accounting for this in our modeling, we find that the H$\alpha$ and H$\beta$ absorption has a similar velocity center and width, with FWHM of about 600 km s$^{-1}$. The fitting degeneracies are stronger in the objects that are slightly less red, whose H$\alpha$ profiles have P~Cygni shapes. In these sources, we find distinct differences in the H$\beta$ and H$\alpha$ profiles, where our model attributes a significantly larger flux near the line center to the narrow emission in the case of H$\beta$. However, after accounting for the emission from the intermediate component, the profiles of the absorption appear more similar. The differences are attributed to additional H$\alpha$ photons that are emitted near the line-center in the resonant branching of H$\beta$ photons. Higher resolution data are required to assess the significance of the remaining differences between the narrow and intermediate-subtracted H$\alpha$ and H$\beta$ profiles that we see in the top panel of Figure $\ref{fig:Hbcores}$. These comparisons illustrate that differences in H$\alpha$ and H$\beta$ absorbers should be carefully interpreted, as emission-infilling can strongly impact the measurements and interpretation of H$\alpha$ absorption.

A possible explanation for these differences in absorption depth can be found in the scenario where H$\alpha$ scattering in the column density regime considered may behave similar to well known Lyman-$\alpha$ scattering in lower column density regimes. In Ly$\alpha$ resonant scattering in a uniform medium, the emerging profile is a double peak with the peak separation depending on column density and the line-ratio on gas dynamics \citep{Neufeld90}. Furthermore, in a clumpy medium photons might escape with less frequency diffusion leading to a larger flux at line center compared to the homogeneous case \citep{Neufeld91,Hansen06,Laursen13}. In a \textit{very} clumpy medium (i.e., when the number of clumps along the line of sight exceeds a critical threshold which depends on density and kinematics) this trend is reversed \citep{Gronke16,Chang23}\footnote{From some clumpiness threshold onwards the escape time of Ly$\alpha$ photons escaping via `excursion’ \citep{Adams72,Neufeld90} is shorter than via `random walking’ between the clumps leading to this reversal.}. Similar processes could impact H$\alpha$ and H$\beta$ transfer as well, but it would impact H$\beta$ differently because of the significant fraction of H$\beta$ photons that branch into H$\alpha$ and Pa$\alpha$ in this process. Detailed radiative transfer simulations are required to test this hypothesis. Alternatively, differences in optical depth that may occur when line-formation regions are somewhat different could also explain the stronger H$\beta$ absorption.

\section{Discussion} \label{sec:discussion}
\subsection{Implications for the engine and origin of LRDs} \label{sec:discuss_engine}

Since the discovery of LRDs in JWST images, the nature of the `engine' powering the emission of the compact, red sources has been debated \citep[e.g.][]{Labbe22,Barro23,Wang24_RUBIES,PerezGonzalez24}. The identification of ubiquitous broad Balmer emission has been decisive in shifting the consensus to an AGN as the powering mechanism \citep[e.g.][]{Kocevski23,Matthee24,Harikane2023,Greene24}, further supported by various spectral features typically associated with AGN activity \citep[see][for an overview]{InayoshiHo25}, although most LRDs lack emission in X-Rays or hard photons \citep[e.g.][]{Ananna24,Yue24,Wang25HeII}. However, both the specific shapes (exponential and P~Cygni like) as well as the correlation structure (line-profiles correlating with tracers of column density) suggest that the Balmer line profiles are primarily driven by radiative transfer effects in dense gas envelopes (see also \citealt{Rusakov25,Chang26,Sneppen26}), which do not require virial motions to broaden lines above $>1000$ km s$^{-1}$ \citep[e.g.][]{GreeneHo2005}. This naturally means we should reassess the evidence for AGNs as the powering source of LRDs \citep[see][for similar early discussion on the engine of quasars]{Oke65,Mathis70}. 

\begin{figure}
    \centering
    \includegraphics[width=0.99\linewidth]{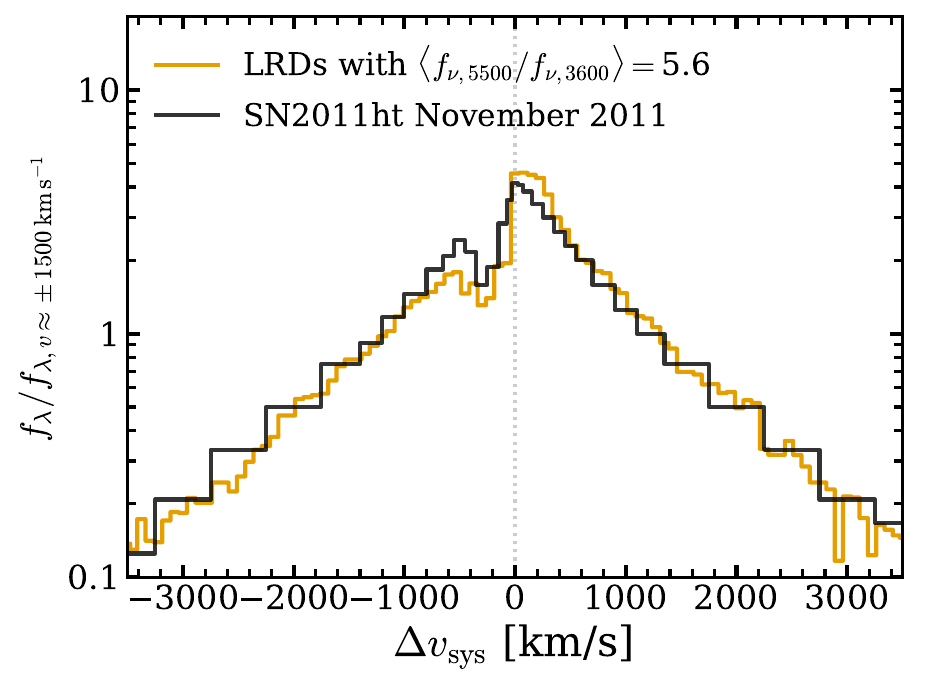} \\ 
    \caption{\textbf{The similarity between LRD spectra and the interaction phase of a Type IIn supernova.} We show the H$\alpha$ spectrum from SN2011ht obtained on 17 November 2011 by \cite{Humphreys12}, when the spectrum was characterized by P~Cygni features from interactions between the supernova shock-front and dense circum-stellar material. For comparison, we show the stacked H$\alpha$ profile of the subset of LRDs that most resembles this profile in terms of the exponential wings and the P~Cygni profile. For visualisation purposes, spectra are normalised at $\pm1500$ km s$^{-1}$.  }
    \label{fig:SN2011}
\end{figure}

We illustrate this point in Fig. $\ref{fig:SN2011}$, where we show the similarity of the H$\alpha$ profile of our second-most red subset of LRDs with the H$\alpha$ spectrum of a Type IIn supernova one month after explosion \citep{Humphreys12,Roming12}, with a similar line EW and luminosity as LRDs. While the analogy breaks at various comparisons such as the broad-band SED of supernovae that lack lines as [Ne{\sc iii}] and [O{\sc iii}], the unusually high H$\alpha$/H$\beta$ ratios of LRDs (but see \citealt{Filippenk97,Izotov09}) and the lack of variability of LRDs on similar time-scales \citep{Kokubo25}, the similarity in the spectra is informative of the geometry and physical conditions driving the line-profile formation and radiative transfer effects \citep[e.g.][]{Dessart09}. Specifically, the similarity suggests that the optical depths to electron and Balmer scattering are similar in LRDs. Outburst events in luminous blue variable stars that are associated with super-Eddington luminosities \citep[e.g.][]{Mauerhan15} also show similar line-profiles as LRDs. Here, dense stellar winds from massive stars lead to pseudo-photospheres that can last for decades and that may show pulsations \citep[e.g.][]{Genderen97,Weis20}. Indications of century-scale variability in an LRD \citep{Zhang25var} are reminiscent of the long time-scale variability reported in some LBV outbursts \citep{Lovekin14,Guseva22}, supporting similarities in the physical conditions based on empirical grounds. Some key differences between LBVs and LRDs are that LRDs are $\sim10^{2-4}$ times more luminous, that their Balmer lines have higher EWs and that they have steeper Balmer decrements compared to LBV outbursts.

\begin{figure*}
    \centering
    \includegraphics[width=0.98\linewidth]{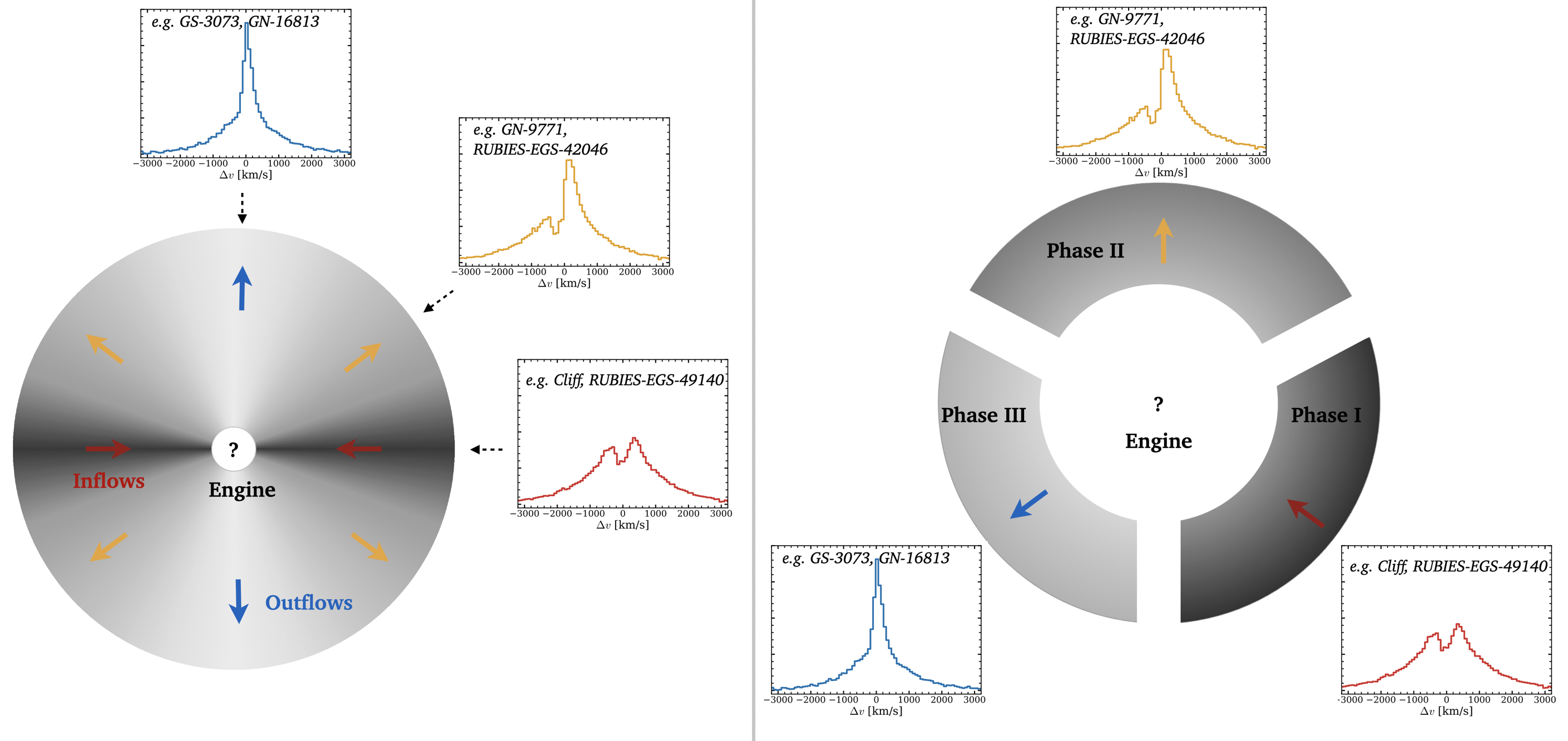}
    \caption{{\bf Illustration of possible scenarios explaining the correlation between the H{\sc i} column density and absorption velocity (Fig. $\ref{fig:Hacores}$) through a viewing angle effect on a geometry that is not spherically symmetric (left) or through evolutionary phases of the dense envelope (right).} We show the H$\alpha$ line profiles of our first, third and fourth sub-sets (in bins of UV to optical color) at different viewing angles or during different phases. Left: Here we show a bi-conical geometry with inflows along the equatorial direction, where the H{\sc i} column density decreases towards the poles. The highest H{\sc i} column densities are found along the equatorial direction, yielding the strongest Balmer breaks and velocities close to the systemic (in the most extreme cases inflowing). At lower inclination angles we may observe P~Cygni profiles, whereas sources at the lowest inclinations have relatively little H{\sc i} gas, though their broad wings are still strongly shaped by high columns of ionised gas. Right: Here we illustrate three distinct phases. In Phase I, very dense and highly optically-thick gas is inflowing as gravity overcomes saturated radiation pressure. In Phase II, when the column density of the gas layer has thinned sufficiently, radiation pressure overcomes gravity, causing an outflow. In Phase III, the column density has dropped enough that absorption features and the Balmer break are no longer visible.}
    \label{fig:sketch}
\end{figure*}

While these comparisons illustrate that the Balmer line profiles of LRDs do not necessarily require high gas velocities around a supermassive black hole, the high luminosities and compact sizes of LRDs strongly suggest that accretion onto a massive black hole is involved. An estimate of the mass scale associated to LRDs can be derived assuming the Eddington limit and a bolometric luminosity correction of $L_{\rm H\alpha} \approx 0.05 L_{\rm bol}$ \citep[e.g.][]{Greene26}. This implies bolometric luminosities of $10^{42.7-44.7}$ erg s$^{-1}$, typically L$_{\rm bol}\sim10^{44}$ erg s$^{-1}$, implying masses of $\approx10^{5.7}$ M$_{\odot}$ (ranging from $10^{4.6 - 6.6}$ M$_{\odot}$). An even lower mass scale of $\sim10^4$ M$_{\odot}$ is found consistent with data of a local LRD, based on the atmospheric modeling of surface gravity-sensitive absorption features \citep{Liu26}. Besides accreting black holes of these masses, we note that these bolometric luminosities are within the range achievable by recent models of supermassive or quasi stars \citep{Dotan12,Gieles18,Nandal26,Santarelli26}. Given the spectral similarity to stellar phenomena, especially those associated with super-Eddington episodes of massive stars, supermassive stars should be explored as an alternative engine, and new models of winds and accretion flows around such sources are highly warranted.

\subsection{Variation across the sample: the same central engine at different angles or in different phases?} \label{sec:discuss_variation}
So far, we primarily focused on red sources with strong Balmer absorption, but there are also broad Balmer line emitters without a strong Balmer break, but with otherwise similar weak X-ray emission, low temporal variability and similar rest-optical spectrum and broad exponential wings \citep[e.g.][]{Brazzini26}. Importantly, unlike classical Type II AGNs, these sources do not show strong [Ne{\sc v}] or Mg{\sc ii} emission. A possibility is that the powering mechanism in these blue sources are similar to the reddest LRDs, and the key difference is that their envelopes are optically thin for Balmer absorption due to a lower H{\sc i}$_{n=2}$ column density (at least along our line of sight), whereas the electron effective temperature and optical depth to Thomson scattering is more similar. Blue broad-line emitters therefore could enable a less obscured view of the powering engine, in particular through unveiling the rest-frame UV part of the spectrum that shows high excitation lines as He{\sc ii} and N{\sc iv}]. These emission lines require effective temperatures of the ionizing sources $\sim50,000$ K, but not power-law spectra as these sources lack lines with ionization energies $\gtrsim60$ eV. Such variation in column densities could be due to evolutionary differences (akin to the evolution of LBV outbursts), as for example indicated by the luminosity differences in the sample (with the reddest sources on average being more luminous in H$\alpha$, Fig. $\ref{fig:sample}$) or it could be due to different viewing angles in a geometry that is not spherically symmetric \citep[e.g.][]{Sneppen26,Madau26}.

The empirical relation between the absorber velocity and H{\sc i}$_{n=2}$ column density (Fig. $\ref{fig:Hacores}$) may provide a clue that points towards viewing angle differences as the primary cause of color variation in the sample. Specifically, we speculate that the objects with the highest column densities that show inflowing motions are seen along the equatorial direction, where gas inflows fuel the engine with a large covering angle (which is included in various models explaining LRDs; e.g. \citealt{Zwick25,Chen26} and which has been associated with super-Eddington accretion; e.g. \citealt{Skadowski16,Mayer19}). In this picture, the envelope is primarily outflowing in the polar direction, and objects seen with increasingly small angle towards the poles may have a lower column density of H{\sc i}$_{n=2}$ gas and a faster outflow velocity (see the left panel in Fig. $\ref{fig:sketch}$). The bluest objects may be seen close along the polar direction. Although the low column density of partially excited H{\sc i} gas in these sources prevents us from measuring the outflow velocity in absorption, the observed asymmetries in the wings (in e.g. ALT-69688) are qualitatively in agreement with this picture.

Such accretion flow + bi-conical outflow geometries are reminiscent of AGN winds \citep[e.g.][]{Proga04,Giustini19}, but they are also found in various types of accreting stars \citep[e.g.][]{Knigge95,Erkal22}. Specifically, comparing our core profiles to those modeled in \cite{Knigge95}, where the outflow originates from the surface of the accreting disk, we find that the LRD profiles resemble those with a classical \cite{Castor79} velocity law, with a monotonically increasing velocity with increasing radius. In a bi-conical outflow picture, we can estimate the opening angle $\theta$ of the outflow from our statistics where two sources out of 18 show inflows (and we assume sources without absorption have outflows and that our sample is representative), yielding $\cos(\theta)=2/18$, or $\theta\sim80$ degrees. We note that a similar viewing-angle scenario could also be the case in a (quasi-)spherical configuration where the envelope would have large coherent regions with inflow and outflow motions at the outer unstable layer of an atmosphere.

Nevertheless, variations in the viewing angle do not explain all trends in our data, nor are they a unique explanation. Sources with intermediate red colors in our sample are more luminous in H$\alpha$ than the bluest or reddest sources. This may imply correlations between the luminosity and the column density distribution, possibly reflecting different phases during the evolution of an LRD. In the right panel of Fig. $\ref{fig:sketch}$, we illustrate an alternative explanation for the correlation between the envelope velocity and column density that is linked to luminosity. Such scenario is as follows \citep[see also][for a similar idea]{DEugenio26}. The densest envelopes are optically thick, such that gravity overcomes the radiation pressure generated in the envelopes, causing inflows (Phase I). Optically-thick inflows could be associated with super-Eddington accretion and therefore (generally) high luminosities (see \citealt{Liu25} for a modeling perspective in the context of LRDs). Envelopes with lower column density will, on the other hand, be outflowing due to the relatively stronger impact of radiation pressure from the engine (Phase II). Such winds are reminiscent of radiation-driven winds around massive stars \citep[e.g.][]{Smith14,Jiang18}, where they could be powered by super-Eddington luminosities or dynamical events such as pulsations \citep[see][for discussions on pulsations in quasi-star models]{Cantiello25}. A third phase could correspond to an increasingly thinner envelope that is no longer optically thick, unveiling the inner blue UV spectrum of the engine. 

Finally, while we separately discuss evolutionary and viewing angle effects, there is of course the possibility that these are simultaneously relevant. Observationally, more detailed insights could be obtained from larger statistical samples with deep grating spectra that could be used to separate our identified trends in various luminosity bins, in order to isolate the relative role of luminosity variations. Further insight may be gained by searching for systematic differences as a function of redshift. From the modeling side, we caution that interpreting these trends purely in terms of column density or covering fraction is non-trivial, as radiative transfer effects are highly non-linear and depend sensitively on the full three-dimensional gas geometry, and our UV to optical colors may to some extent be impacted by host galaxy light. Robust interpretation therefore ultimately requires dedicated radiative transfer modeling in realistic environments that self-consistently accounts for scattering and photon redistribution, but our toy models may inspire the setup of such models.

\subsection{The fraction of Balmer absorbers among LRDs and faint AGNs at $z=4-7$} \label{sec:absorber_frac}
Narrow Balmer absorption lines are among the most unusual features among any class of galaxy or AGN \citep[e.g.][]{Schulze18,Zhang18}, and initially was only detected significantly in a $\approx10-20$ \% of broad-line selected LRDs observed with the NIRCam grism \citep{Matthee24,Lin24} and a handful of objects with deep R$\gtrsim1000$ resolution spectra from NIRSpec follow-up \citep[e.g.][]{Kocevski24,Juodzbalis24b,Labbe2024}. In this work, we find that two additional sources from the \cite{Matthee24} sample show Balmer absorption with higher sensitivity data, already increasing the fraction of broad line emitters with Balmer absorbers to $\geq25$ \%. Publicly available data from the BlackTHUNDER program (PID 5015, PIs: \"Ubler, Maiolino) show asymmetric broad emission in another source from the \cite{Matthee24} sample that could indicate absorption, although it could also be due to a superposition of two broad line regions \citep{Uebler25}.

As highlighted by the annotations in Fig. $\ref{fig:overview}$, our modeling (Section $\ref{sec:profiles}$) identifies H$\alpha$ absorption with H$\alpha$ EW $\gtrsim4$ {\AA} in 11/18 objects ($\approx$ 60 \%). It is challenging to assess whether our identification is robust to the sensitivity and resolution of the data. However, we find that there are no significant differences in the total H$\alpha$ signal-to-noise ratio of absorbers vs non-absorbers. The fraction of sources with H versus M grating spectra is also similar, and we measure absorbers with the lowest EW $\approx4$ {\AA} in data with both high and medium resolution. This suggests that our absorber fraction is not strongly limited by the data quality (unlike the earlier, shallower grism-based fractions).\footnote{It is challenging to confidently rule out an absorber in the H$\alpha$ spectrum of GN-12839 due to its closeness to the NIRSpec chip gap. } The stacked H$\alpha$ spectrum of sources without absorption also does not show indications of absorption, neither does its stacked H$\beta$ spectrum. We also note that there are no sources that show H$\beta$ absorption (usually stronger than H$\alpha$ absorption) without absorption in H$\alpha$, even if data quality would enable this.

Another source of uncertainty on the fraction of absorbers is the possibility that there are selection effects in the sources that were chosen for deep grating follow-up. The absorber fraction correlates with the UV to optical color, which is expected as it traces the Balmer break strength sensitive to the H{\sc i}$_{n=2}$ column density \citep[e.g.][]{Inayoshi24}. This provides a handle for correcting for such possible selection effect, as the UV to optical color measurements are available for large numbers of sources. For the 8 sources with $f_{\nu, 5500}/f_{\nu, 3600}>5$ ($f_{\nu, 4100}/f_{\nu, 3600}>2$ following the definition used in e.g. \citealt{Wang24_RUBIES,deGraaff25b}) we find that all of them show Balmer absorption. For bluer colors, the fraction increases from 20 \% (1/5) at $f_{\nu, 5500}/f_{\nu, 3600}=1-2$ to 40 \% (2/5) at $f_{\nu, 5500}/f_{\nu, 3600}=2-5$. Based on this, we predict H$\alpha$ EW$\gtrsim4$ {\AA} absorber fractions of $\approx65$ \% (for the \citealt{deGraaff25b} LRD sample) or $\approx30$ \% (for the \citealt{Juodzbalis25} faint AGN sample at $z=4-7$). The main uncertainty here is that the majority of these samples have intermediate UV to optical colors, where our statistics are poorest, stressing the need for more deep grating data of broad-line sources regardless of their broad-band colors.

\section{Summary} \label{sec:summary} 
In order to investigate the physical origin of the broad and complex Balmer line profiles in JWST's Little Red Dots, we assembled deep high resolution grating as well as PRISM spectra covering the rest-UV to optical of 18 broad-line emitters at $z=3-7$ identified in JWST fields. Section $\ref{sec:data}$ details how the sample generally is representative for the brighter end of the population and it details the new and archival data employed here. Throughout the paper, we explore empirical trends between the line-profiles and general spectral properties of the galaxies and we also describe the line-profiles with simple models to investigate variations in their detailed shapes. These are the main results:

\begin{itemize}
    \item We show that the [O{\sc iii}] lines are ubiquitously narrow, but the Balmer lines are composites of broad exponential wings around complex line-cores that (can) contain narrow emission, (narrow) absorption, and/or further emission with intermediate width that is most clearly seen in H$\alpha$. The Balmer line-profiles correlate with the UV to optical colors that probe the Balmer break strength. The bluest sources have narrow lines at the systemic redshift, broad exponential wings and they do not show absorption. P~Cygni profiles with blue-shifted absorption are seen in redder sources. The line-cores in the reddest sources are dominated by absorption near line-center. The prominence of the broad line increases from blue to red UV to optical colors. [Sections $\ref{sec:maintrends}$ and $\ref{sec:profiles}$, Figs. $\ref{fig:prism_sample}$, $\ref{fig:overview}$]

    \item Based on our modeling of the profiles, we find that LRDs have steep Balmer decrements (H$\alpha$/H$\beta \approx8$, from 3.6 to 14.0 for the total line-fluxes) that are higher in redder sources and are driven by the broad components. The broad wings of H$\alpha$ have a relatively narrow range in exponential FWHM $\approx1400$ (1040-1600) km s$^{-1}$, that is higher in the reddest sources. Additionally, the [O{\sc iii}] equivalent width is strongly anti-correlated with the Balmer break strength, but positively with the fraction of the Balmer line emission that emerges at the line-center, suggesting that (most of the) narrow [O{\sc iii}] emission emerges from H{\sc ii} regions in the host galaxy. [Section $\ref{sec:definitions}$, Figs. $\ref{fig:profilefit}$, $\ref{fig:correlations_ratios}$, Table $\ref{tab:fit_averages}$]
    
    \item Our results suggest that the Balmer line profiles trace radiative transfer effects in dense gas envelopes, with the diversity across the sample primarily tracing the column density. The envelopes are clumpy, partially ionised and the free electrons cause exponential wings through Thomson scattering with optical depths $\tau_e\approx1-5$. Excited H{\sc i}$_{n=2}$ causes a high opacity for Balmer photons, leading to resonant transfer effects. Branching of H$\beta$ photons to H$\alpha$ facilitates the high H$\alpha$/H$\beta$ ratios that are observed. The Balmer absorption tends to be stronger in H$\beta$ than in H$\alpha$, which could indicate that the medium is clumpy. [Sections $\ref{sec:densewind}$, $\ref{sec:expwings}$ and $\ref{sec:core_flows}$, Fig. $\ref{fig:fwhm_exp}$]

    \item We argue that part of the Balmer lines originate from star formation in the host galaxy, although a (faint) contribution from the main engine is not ruled out, especially in the bluer sources that show a faint broad base in [O{\sc iii}]. Degeneracies with the intermediate emission and absorption components from the envelopes exist but differ in the details across the sources. In the reddest sources characterized by central absorption, the absorption acts as a natural coronagraph of the engine near the line-center, allowing a clear view of the host galaxy emission. [Sections $\ref{sec:core_is_needed}$ and $\ref{sec:core_differenceHaHb}$, Figs. $\ref{fig:overview}$, $\ref{fig:host}$, $\ref{fig:Hbcores}$] 
        
    \item We unveil a correlation between the absorber velocity and the H{\sc i} column density (traced by UV to optical color). The objects with highest column densities are characterized by inflows, where increasingly fast outflows are associated with less dense absorbers. We discuss that a possibility is that this could indicate that the variety among the sample is due to differences in the viewing angle on geometry that is not spherically symmetric. The highest column density is found along the equatorial direction where gas is inflowing, but the majority of the solid angle is dominated by outflows with lower column densities towards the poles. Alternatively, we also discuss the possibility that this correlation traces different evolutionary phases, where the bulk motions of the outer envelope trace the competition between the gravity and the radiation pressure whose effectivity depends on the optical depth of the envelope. [Sections $\ref{sec:core_flows}$ and $\ref{sec:discuss_variation}$, Figs. $\ref{fig:Hacores}$ and $\ref{fig:sketch}$]

    \item As we find that line-profiles are tracing radiative transfer effects in dense envelopes, rather than dynamical broadening, there is less direct evidence that LRDs are powered by accretion onto supermassive black holes, except for the high luminosity and the compactness. We highlight physical analogies between LRDs and stellar phenomena such as outbursts in luminous blue variables and Type IIn supernovae, that yield similar line profiles, albeit typically at much fainter luminosity. These similarities, and the Eddington luminosity-based mass scale of $\approx10^{5-6}$ M$_{\odot}$ of our LRDs, warrant the exploration of possible alternative hot engines of LRDs. [Section $\ref{sec:discuss_engine}$, Fig. $\ref{fig:SN2011}$]

    \item Absorption in the H$\alpha$ line (with EW$\gtrsim4$ {\AA}) is detected in the majority of sources. The detection fraction increases with the UV to optical color, such that Balmer absorption is ubiquitous among the reddest LRDs. Based on the distribution of Balmer break strengths in LRD and JWST broad-line samples, we estimate that they have absorber fractions in the range $\sim30-60$ \% when observed with sufficient sensitivity and resolution. [Section $\ref{sec:absorber_frac}$]    
\end{itemize}

Our work demonstrates the potential of deep, high resolution spectroscopy of LRDs with JWST, but is limited by the small sample sizes that prevents more detailed statistical explorations of the multi-dimensional correlation structure. In particular, deep spectroscopic observations of intrinsically fainter systems will be helpful to extend the dynamic range in luminosities. Generally, larger samples would also allow us to control for luminosity as well as redshift differences, explore outliers to the general trends identified and more precisely test the predicted fraction of LRDs with absorbers. Our possible interpretation that the reddest sources could be seen along an equatorial axis along which dense gas is inflowing could be falsified in case extremely red sources with outflowing Balmer absorption would be detected. From the modeling side, our results suggest that successful models of LRDs -- whether it be with AGN or stellar engine -- should incorporate an engine hot enough to power emission-lines including those with ionization energies $\sim50$ eV (e.g. He{\sc ii}), as well as dense gas flows and their radiative imprint on observed spectra as a necessary ingredient.

\begin{acknowledgements}
We thank Zoltan Haiman, Harley Katz and Hannah \"Ubler for insightful discussions.

JM and AT acknowledge funding by the European Union (ERC, AGENTS,  101076224).
MG thanks the European Union for support through ERC-2024-STG 101165038 (ReMMU).
AdG acknowledges support from a Clay Fellowship awarded by the Smithsonian Astrophysical Observatory. IL acknowledges support by the Australian Research Council through Future Fellowship FT220100798.

This work is based in part on observations made with the NASA/ESA/CSA James Webb Space Telescope. The data were obtained from the Mikulski Archive for Space Telescopes at the Space Telescope Science Institute, which is operated by the Association of Universities for Research in Astronomy, Inc., under NASA contract NAS 5-03127 for JWST. These observations are associated with programs \#1181, 1216, 4233, 5664, 8204 and 9433. Support for program \#5664 was provided by NASA through a grant from the Space Telescope Science Institute. The authors acknowledge the UNCOVER team led by coPIs Labbe \& Bezanson, the RUBIES team led by coPIs De Graaff \& Brammer and the \#9433 team led by coPIs Maiolino \& D'Eugenio for developing their observing program with a zero-exclusive-access period.

Some of the data products presented herein were retrieved from the Dawn JWST Archive (DJA). DJA is an initiative of the Cosmic Dawn Center (DAWN), which is funded by the Danish National Research Foundation under grant DNRF140.

This research was supported by the International Space Science Institute (ISSI) in Bern, through ISSI International Team project \#25-659.
This work was supported by funding from the Swiss State Secretariat for Education, Research and Innovation (SERI) under contract number MB22.00072, as well as the Swiss National Science Foundation (SNSF) through project grant 200020\_207349

\end{acknowledgements}

%%%%%%%%%%%%%%%%%%%% REFERENCES %%%%%%%%%%%%%%%%%%
\bibliographystyle{aa}
\bibliography{my_bibliography_cleaned}

%%%%%%%%%%%%%%%%%%%% APPENDICES %%%%%%%%%%%%%%%%%%%%
\appendix
\section{Detailed line-profile fit results}\label{app:fits}
Here we list detailed fitting results to the H$\alpha$ and H$\beta$ line profiles of our LRD sample, as described in Section $\ref{sec:profiles}$. In Table $\ref{tab:bic_cat}$ we list the $\chi^2_{\rm reduced}$ values of the best H$\alpha$ and H$\beta$ fits of the fiducial model, as well as a model where the broad component has a Gaussian profile rather than an exponential. We also list the BIC differences between these two models, where an exponential profile is preferred significantly for $\Delta$BIC $<-10$. In Table $\ref{tab:lum_cat}$ we list the UV to optical colors, H$\alpha$ and [O{\sc iii}] equivalent widths and the total line luminosities of [O{\sc iii}], H$\beta$ and H$\alpha$. In Table $\ref{tab:profile_cat}$, we list the results from our line-profile fitting that were shown in Figs. $\ref{fig:correlations_ratios}$ and $\ref{fig:Hacores}$.

\begin{table*}
\centering
\caption{Goodness of fit statistics for H$\alpha$ and H$\beta$ line-profile modeling as described in Section $\ref{sec:profiles}$. The reduced $\chi^2$ values are listed for models with exponential broad wing (fiducial) and a Gaussian broad wing. The $\Delta$BIC is defined as BIC$_{\rm fiducial}$- BIC$_{\rm Gauss}$. A BIC difference $\lesssim-10$ suggests that an exponential wing is significantly preferred over a model with a broad gaussian wing.  }
\label{tab:bic_cat}
\begin{tabular}{lccccrr}
\hline
ID & $\chi^2_{\mathrm{H}\alpha,\,\mathrm{fiducial}}$ & $\chi^2_{\mathrm{H}\beta,\,\mathrm{fiducial}}$ & $\chi^2_{\mathrm{H}\alpha,\,\mathrm{Gauss}}$ & $\chi^2_{\mathrm{H}\beta,\,\mathrm{Gauss}}$ & $\Delta\mathrm{BIC}_{\mathrm{H}\alpha}$ & $\Delta\mathrm{BIC}_{\mathrm{H}\beta}$ \\
\hline
FRESCO-GN-9771    &  2.70 & 1.03  & 12.59  & 1.12 & $-2967.6$ & $ -12.8$ \\
FRESCO-GN-12839   &  0.99 & 0.75  &  1.27  & 0.73 & $ -36.6$  & $ -13.3$ \\
FRESCO-GN-15498   &  0.74 & 0.86  &  1.51  & 0.85 & $-152.2$  & $   1.0$ \\
FRESCO-GN-16813   &  1.09 & 1.07  &  1.31  & 1.05 & $ -44.3$  & $   2.9$ \\
FRESCO-GS-13971   &  0.99 & 0.86  &  1.38  & 0.82 & $-108.0$  & $   6.5$ \\
JADES-GN-68797    &  3.00 & 1.15  &  5.88  & 1.09 & $-256.6$  & $   2.3$ \\
JADES-GN-38147    &  1.06 & 0.74  &  5.80  & 0.80 & $-492.4$  & $ -20.4$ \\
JADES-GN-73488    &  1.06 & 1.22  &  1.09  & 1.09 & $ -20.3$  & $   4.1$ \\
RUBIES-EGS-42046  &  1.29 & 0.77  &  2.32  & 0.72 & $ -96.6$  & $   2.0$ \\
RUBIES-EGS-50052  &  1.09 & 2.18  &  1.43  & 2.29 & $ -27.7$  & $  -5.0$ \\
RUBIES-EGS-55604  &  1.02 & 0.63  &  1.45  & 0.61 & $ -49.8$  & $   1.2$ \\
RUBIES-EGS-49140  &  1.16 & 1.12  &  1.60  & 1.19 & $ -51.0$  & $  -3.7$ \\
RUBIES-UDS-47509  &  0.64 & 0.78  &  1.59  & 0.74 & $ -98.3$  & $ -16.3$ \\
RUBIES-UDS-182791 &  2.68 & 1.17  &  3.17  & 1.20 & $ -41.6$  & $ -14.2$ \\
{\it the Cliff} &  1.06 & 0.80  &  1.11  & 0.81 & $ -16.5$  & $  -0.7$ \\
UNCOVER-A2744-45924        & 19.24 & 5.88  & 60.21  & 8.97 & $-3036.2$ & $-213.4$ \\
GS-3073            &  3.48 & 2.85  &  6.09  & 3.20 & $-561.0$  & $ -36.5$ \\
ALT-69688          &  5.78 & 4.13  &  5.87  & 3.50 & $ -12.9$  & $  43.1$ \\
\hline
\end{tabular}
\end{table*}

\begin{table*}
\centering
\caption{The UV to optical colors, Balmer break strength, H$\alpha$ and [O{\sc iii}] equivalent widths (based on the PRISM spectra) and the total line luminosities of [O{\sc iii}], H$\beta$ and H$\alpha$ (based on line-profile fitting).}
\label{tab:lum_cat}
\begin{tabular}{lccccccc}
\hline
ID &  $f_{\nu, 5500}/f_{\nu, 3600}$ &  $f_{\nu, 4100}/f_{\nu, 3600}$ & EW$_{\mathrm{H}\alpha}$ & EW$_{\mathrm{[OIII]}}$ & $L_{\mathrm{[OIII]5008}}$ & $L_{\mathrm{H}\alpha, \rm tot}$ & $L_{\mathrm{H}\beta, \rm tot}$ \\
  & &                       & \AA                   & \AA                  & $10^{42}$\,erg\,s$^{-1}$ & $10^{42}$\,erg\,s$^{-1}$ & $10^{42}$\,erg\,s$^{-1}$ \\
\hline
FRESCO-GN-9771    & 5.84 & 3.28  & 1713 &   87 & $2.0\pm0.1$      & $64.4\pm0.1$    & $6.2\pm0.1$  \\
FRESCO-GN-12839   & 4.20 & 1.14  & 1718 &  628 & $8.7\pm0.1$      & $29.5\pm8.2$ & $3.7\pm0.2$  \\
FRESCO-GN-15498   & 7.27  & 2.34 &  536 &   98 & $0.7\pm0.1$      & $5.6\pm0.1$     & $0.9\pm0.1$  \\
FRESCO-GN-16813   & 1.41  & 1.66 & 1574 & 1354 & $8.7\pm0.1$      & $9.3\pm0.1$     & $2.5\pm0.1$  \\
FRESCO-GS-13971   & 3.92  & 1.04 &  894 &  771 & $5.7\pm0.1$      & $11.9\pm0.1$    & $1.4\pm0.1$  \\
JADES-GN-68797    & 9.54  & 2.94 & 1306 &  436 & $8.4\pm0.4$      & $60.0\pm0.3$    & $4.3\pm0.2$  \\
JADES-GN-38147    & 1.50  & 0.80 & 1483 & 1784 & $7.7\pm0.2$      & $13.8\pm0.3$    & $1.7\pm0.2$  \\
JADES-GN-73488    & 2.60  & 1.03 & 1265 &  420 & $1.9\pm0.2$      & $6.0\pm0.1$     & $0.7\pm0.1$  \\
RUBIES-EGS-42046  & 5.32  & 2.02 & 1772 &  172 & $2.5\pm0.1$      & $38.6\pm0.3$    & $3.7\pm0.2$  \\
RUBIES-EGS-50052  & 1.29  & 0.90 & 1556 & 1473 & $7.5\pm0.2$      & $9.2\pm0.2$     & $1.8\pm0.1$  \\
RUBIES-EGS-55604  & 5.40  & 2.30 & 137 &  246 & $6.1\pm0.2$      & $75.2\pm0.8$    & $6.5\pm0.3$  \\
RUBIES-EGS-49140  & 7.72  & 2.77 & 1505 &  155 & $5.5\pm0.1$      & $60.8\pm0.6$    & $8.0\pm0.3$  \\
RUBIES-UDS-47509  & 2.37  & 1.41 & 779 &  896 & $5.3\pm0.2$      & $7.9\pm0.2$     & $1.1\pm0.3$  \\
RUBIES-UDS-182791 & 2.87  & 1.04 & 1840 & 1088 & $6.5\pm0.2$      & $15.0\pm0.2$    & $1.3\pm0.4$  \\
{\it the Cliff} & 28.94 & 7.63 &  372 &    6 & $0.2\pm0.1$      & $0.7\pm0.1$     & $0.2\pm0.0$  \\
A2744-45924        & 8.28 & 3.06  & 1503 &  173 & $5.9\pm0.2$      & $66.0\pm0.1$    & $8.3\pm0.0$  \\
GS-3073            & 1.43 & 1.04  & 1054 & 1230 & $67.4\pm0.7$     & $58.5\pm0.2$    & $16.2\pm0.1$ \\
ALT-69688          & 1.26 & 1.05 & 1556 & 1362 & $7.7\pm0.3$      & $8.1\pm0.1$     & $1.6\pm0.0$  \\
\hline
\end{tabular}
\end{table*}

\begin{table*}
\centering
\caption{The best-fitted exponential FWHM of the H$\alpha$ line, the minimum absorption velocity and the rest-frame EW of the H$\alpha$ absorption (for the systems in which this is detected), \ldots otherwise. $f_{\mathrm{cen,300/3000}}$ is a non-parametric meaure of the compactness of the H$\alpha$ profile, being the fraction of H$\alpha$ emission within $\pm300$ km s$^{-1}$ divided by the emission within $\pm3000$ km s$^{-1}$. The narrow-to-total flux ratio of the H$\beta$ line is derived from our line-profile fitting. }
\label{tab:profile_cat}
\begin{tabular}{lccccc}
\hline
ID & FWHM$_{\mathrm{H}\alpha, exp.}$ & $\Delta v_{\mathrm{H}\alpha, \rm absorber}$ & EW$_{\mathrm{H}\alpha,\,\mathrm{abs}}$ & $f_{\mathrm{cen,300/3000}}$ & narrow-to-total H$\beta$ \\
   & km\,s$^{-1}$            & km\,s$^{-1}$             & \AA                                 &                        & \\
\hline
FRESCO-GN-9771    & $1528\pm9$    & $-63$  & $3.9$  & $0.28$ & $0.07$ \\
FRESCO-GN-12839   & $1361\pm35$   & \ldots & \ldots  & \ldots & $0.37$ \\
FRESCO-GN-15498   & $1048\pm9$    & $-36$  & $6.5$  & $0.36$ & $0.22$ \\
FRESCO-GN-16813   & $1060\pm46$   & \ldots & \ldots  & $0.62$ & $0.59$ \\
FRESCO-GS-13971   & $1083\pm35$   & $-81$  & $5.7$  & $0.49$ & $0.63$ \\
JADES-GN-68797    & $1414\pm22$   & $-25$  & $11.9$ & $0.33$ & $0.14$ \\
JADES-GN-38147    & $1040\pm63$   & $-156$ & $4.3$  & $0.58$ & $0.71$ \\
JADES-GN-73488    & $1096\pm53$   & \ldots & \ldots  & $0.47$ & $0.67$ \\
RUBIES-EGS-42046  & $1430\pm40$   & $-217$ & $11.9$ & $0.37$ & $0.21$ \\
RUBIES-EGS-50052  & $1196\pm73$   & \ldots & \ldots  & $0.67$ & $0.70$ \\
RUBIES-EGS-55604  & $1443\pm52$   & $-212$ & $6.7$  & $0.31$ & $0.07$ \\
RUBIES-EGS-49140  & $1446\pm43$   & $-22$  & $11.1$ & $0.19$ & $0.11$ \\
RUBIES-UDS-47509  & $1284\pm143$  & \ldots & \ldots  & $0.56$ & $0.67$ \\
RUBIES-UDS-182791 & $1602\pm50$   & $-90$  & $5.0$  & $0.44$ & $1.04$ \\
{\it the Cliff} & $1400\pm12$   & $49$   & $5.1$  & $0.33$ & $0.07$ \\
A2744-45924        & $1485\pm18$   & $54$   & $8.0$  & $0.24$ & $0.13$ \\
GS-3073            & $1358\pm15$   & \ldots & \ldots  & $0.61$ & $0.59$ \\
ALT-69688          & $1042\pm17$   & \ldots & \ldots  & $0.59$ & $0.58$ \\
\hline
\end{tabular}

\end{table*}

\end{document}